\documentclass[manuscript,screen]{acmart}
\usepackage{graphics}
\usepackage{hyperref}
\usepackage{multirow}
\usepackage{soul}
\usepackage{tabularx}
\usepackage{tcolorbox}
\usepackage[normalem]{ulem}
\usepackage{xcolor}

\usepackage{paralist}
\usepackage{enumitem}

\usepackage[outercaption]{sidecap}    
\usepackage[figuresright]{rotating}

\AtBeginDocument{%
  \providecommand\BibTeX{{%
    \normalfont B\kern-0.5em{\scshape i\kern-0.25em b}\kern-0.8em\TeX}}}

\graphicspath{{}{images/}{dia/}}
\DeclareGraphicsExtensions{.pdf,.png}

\newcounter{o}
\definecolor{1c1}{RGB}{188,162,6}
\definecolor{1c2}{RGB}{137,129,80}
\definecolor{1c3}{RGB}{239,167,31}
\definecolor{1c4}{RGB}{88,194,241}
\definecolor{1c5}{RGB}{6,180,188}

\def\bf{\textbf}
\def\eq {Equation~}

\def\fig {Figure~}

\def\tbl {Table~}

\def\sec {Section~}

\def\it{\textit}

\def\tt{\mct}

\newcommand{\mct}[1]{{\footnotesize {\texttt {#1}}}}

\newcommand{\nd}{\vspace{1mm}\noindent}
\newcommand{\urls}[1]{{\scriptsize\url{#1}}}
\newcommand{\emt}[1]{\emph{``#1''}}

\usepackage{appendix}

\begin{document}

\title[An Empirical Study of the Effectiveness of an Ensemble of Stand-alone SE Sentiment Detection Tools]{An Empirical Study of the Effectiveness of an Ensemble of Stand-alone Sentiment Detection Tools for Software Engineering Datasets}

\author{Gias Uddin}
\email{gias.uddin@ucalgary.ca}
\affiliation{%
  \institution{University of Calgary}
  \country{Canada}
}
\author{Yann-Ga\"{e}l Gu\'{e}h\'{e}neuc}
\email{yann-gael.gueheneuc@concordia.ca}
\affiliation{%
  \institution{Concordia University}
  \country{Canada}
}
\author{Foutse Khomh}
\email{foutse.khomh@polymtl.ca}
\affiliation{%
  \institution{Polytechnique Montr\'{e}al}
  \country{Canada}
}
\author{Chanchal K. Roy}
\affiliation{%
  \institution{University of Saskatchewan}
  \country{Canada}
}
\email{croy@cs.usask.ca}

\begin{abstract}
Sentiment analysis in software engineering (SE) has shown promise to analyze and support diverse development activities. 
Recently, several tools are
proposed to detect sentiments in software artifacts. While the tools improve accuracy over off-the-shelf tools, recent research shows that their
performance could still be unsatisfactory.
A more accurate sentiment detector for SE can help reduce noise in analysis of software scenarios where sentiment analysis is required. 
Recently, combinations, i.e., hybrids of stand-alone classifiers are found to  offer better performance than the 
stand-alone classifiers for fault detection. However, we are aware of no such approach for sentiment detection for software artifacts.
We report the results of an empirical study that we conducted to determine the feasibility of developing an ensemble 
engine by combining the polarity labels of stand-alone SE-specific sentiment detectors. Our study has two phases. In the first phase, 
we pick five SE-specific sentiment detection tools from two recently 
published papers by Lin et al.~\cite{Lin-SentimentDetectionNegativeResults-ICSE2018,Lin-PatternBasedOpinionMining-ICSE2019}, who first reported negative results with stand alone 
sentiment detectors and then proposed an improved SE-specific sentiment detector, POME~\cite{Lin-PatternBasedOpinionMining-ICSE2019}. 
We report the study results on 17,581 units (sentences/documents) coming from six currently 
available sentiment benchmarks for software engineering. We find that the existing tools can be complementary to each other in 85-95\% of the cases, i.e., one is wrong but another is right. However, a majority voting-based ensemble of those tools fails to improve the accuracy of sentiment detection. We develop Sentisead, a supervised tool by combining the polarity labels and bag of words as features. Sentisead improves the performance (F1-score) of the individual tools by 4\% (over Senti4SD~\cite{Calefato-Senti4SD-EMSE2017}) -- 100\% (over POME~\cite{Lin-PatternBasedOpinionMining-ICSE2019}). 
The initial development of Sentisead occurred before we observed the use of deep learning models for SE-specific sentiment detection. In particular, 
recent papers show the superiority of advanced language-based pre-trained transformer models (PTM) over rule-based and shallow learning models. Consequently, in a second phase, we compare and improve Sentisead infrastructure using the PTMs. 
We find that a Sentisead infrastructure with RoBERTa as the ensemble of the five stand-alone rule-based and shallow learning SE-specific tools from Lin et al.~\cite{Lin-SentimentDetectionNegativeResults-ICSE2018,Lin-PatternBasedOpinionMining-ICSE2019} offers the best F1-score of 0.805 across the six datasets, while 
a stand-alone RoBERTa shows an F1-score of 0.801. 
\end{abstract}





\maketitle

\section{Introduction}
\label{sec:intro}

According to Bing Liu, ``the textual information in the world can broadly be
categorized into two main types: facts and
opinions"~\cite{Liu-SentimentAnalysisOpinionMining-MorganClaypool2012}. Such
conjecture can also be applied to software-engineering (SE) artifacts, e.g., the
discussions in online developer forums, and so on. ``Facts'' are objective
expression (e.g., ``I use this tool''). Opinions are subjective expressions that
convey the sentiments towards entities (e.g., ``I like this tool''). While the
concept of opinion can be broad, the major focus of sentiment analysis is to
detect
polarity~\cite{Liu-SentimentAnalysisOpinionMining-MorganClaypool2012,Pang-SentimentClassificationMachineLearning-EMNLP2002},
i.e., given a sentence or a post as a unit of analysis (e.g., a sentence/post - denoted as \bf{\ul{``unit"}} from now on), whether
the unit exhibits positive, negative, or neutral sentiment (i.e., absence of
positivity or negativity).
 
Opinions are key determinants to many of the activities in SE~\cite{Uddin-OpinerReviewAlgo-ASE2017,Guzman-EmotionalAwarenessSoftwareDevelopmentTeams-FSE2013,Mika-MiningValenceBurnout-MSR2016}.
The productivity in a team may depend on developers' sentiments in/of their
diverse development
activities~\cite{Ortu-BulliesMoreProductive-MSR2015,Guzman-EmotionalAwarenessSoftwareDevelopmentTeams-FSE2013,Pletea-SecurityEmotionSE-MSR2014},
while the analysis of sentiments towards software artifacts may lead to better
selections of software
artifacts~\cite{Uddin-OpinerReviewAlgo-ASE2017,Uddin-OpinerReviewToolDemo-ASE2017}.
Therefore, it is necessary to accurately detect sentiments in
software-engineering
artifacts~\cite{Novielli-ChallengesSentimentDetectionProgrammerEcosystem-SSE2015}.

Most prior work on sentiment analysis in SE used cross-domain
sentiment-detection tools. This ``off-the-shelf'' usage is insufficiently
accurate for SE, due to the difference in the domain (see
\sec\ref{sec:related-work}). Many cross-domain sentiment-detection tools were
developed using movie reviews (e.g.,
Stanford~\cite{Socher-Sentiment-EMNLP2013}). The context of sentiments expressed
in movie reviews is different from that in SE. For example,
the `API is simple' is a positive API
review~\cite{Uddin-OpinerReviewAlgo-ASE2017}, but `The movie is simple' is
generally a negative movie
review~\cite{Liu-SentimentAnalysisOpinionMining-MorganClaypool2012}.
Over the last few years, several tools have been developed to detect sentiments
in software artifacts, e.g.,
SentistrengthSE~\cite{Islam-SentistrengthSE-MSR2017},
Senti4SD~\cite{Calefato-Senti4SD-EMSE2017},
SentiCR~\cite{Ahmed-SentiCRNIER-ASE2017}. These tools offer better accuracy than
off-the-shelf tools. However, recently Lin et
al.~\cite{Lin-SentimentDetectionNegativeResults-ICSE2018} reported that the
tools can still be inaccurate significantly.

Recent research in other domains has seen a rise in combining multiple sentiment
detectors, such as hybrids of rule-based and supervised classifiers observed to
be better detectors of sentiments in Twitter
messages~\cite{Balage-HybridSentimentDetectorTwitter-SemEval2014,Goncalves-ComparingCombiningSentimentAnalysisTools-COSN2013}.
In software engineering, the choice of classification methods and the
combination of diverse classifiers is effective to detect
defects~\cite{Ghotra-RevisitingImpactClassificationDefectPrediction-ICSE2015,Nucci-DynamicSelectionOfClassifiersBugPrediction-IEEETranEmergingTopics2017}.
Intuitively, an ensemble of existing sentiment detection tools for SE can  
learn from the strengths and weaknesses of each tool for a given dataset, which then may offer 
it better insight to determine the correct sentiment polarity label of the different units in the dataset. 
This is because we find that different SE specific sentiment detection tools are developed using diverse 
rules and algorithms. We are particularly interested to study how a hybrid of existing sentiment detection tools for SE 
performs, because of the superiority of hybrid tools for sentiment detection other domains like Twitter. For example, 
Balage et al.~\cite{Balage-HybridSentimentDetectorTwitter-SemEval2014} found that a hybrid of rule-based (i.e., that uses sentiment lexicons as cues) and supervised tools (i.e., that 
uses features to learn)  offered better sentiment detection performance for Twitter messages, because the tools could completement each other. The hybrid tool 
in Balage et al.~\cite{Balage-HybridSentimentDetectorTwitter-SemEval2014} was superior to the stand-alone tools, because each stand-alone tool (e.g., the rule-based tool) focused 
on specific estimators (e.g., sentiment lexicons). In SE, sentiment detection tools like SentistrengthSE are rule-based while other tools like Senti4SD are supervised. 
Therefore, the tools can also exhibit strengths and weaknesses on particular features/estimators in a dataset, which a hybrid tool can find as complementary. However, 
we are aware of no such hybrid sentiment detection tool developed for SE so far.

To determine the feasibility of developing a hybrid of stand-alone
SE-specific sentiment tools, our empirical study in this paper started in 2019 with an aim to gain more
positive outlook on the state of sentiment detection than the negative results
reported by Lin et al.~\cite{Lin-SentimentDetectionNegativeResults-ICSE2018}. Therefore, we
initially intended to study all the sentiment detection tools developed for software
engineering included in the study of Lin et
al.~\cite{Lin-SentimentDetectionNegativeResults-ICSE2018} and that are currently
available through GitHub or other online sources, including the latest such
tool, POME, by also Lin et al.~\cite{Lin-PatternBasedOpinionMining-ICSE2019}.
These tools are either rule-based or shallow machine learning
models. Looking forward in 2020, we observed new uses of deep learning models to detect sentiment in
software engineering~\cite{Zhang-SEBERTSentiment-ICSME2020,Biswas-ReliableSentiSEBERT-ICSME2020,Biswas-SentiSEWordEmbedding-MSR2019,Chen-SentiEmoji-FSE2019}. 
The most recent works are by Zhang et al.~\cite{Zhang-SEBERTSentiment-ICSME2020}
and Biswas et al.~\cite{Biswas-ReliableSentiSEBERT-ICSME2020}, who investigated
the effectiveness of advanced pre-trained language based transformer models
(PTMs) like BERT~\cite{Delvin-BERTArch-Arxiv2018}. They found that
the PTMs outperform rule-based and shallow learning models.
Therefore, we also study the effectiveness of those
PTMs to support/outperform our hybrid engine of SE-specific sentiment detectors.
Thus, we divided our empirical study into two phases:

\begin{itemize}
\item \bf{Phase 1.} We studied the feasibility of developing a hybrid of stand-alone SE-specific sentiment tool, based on the tools studied by Lin et al.~\cite{Lin-PatternBasedOpinionMining-ICSE2019,Lin-SentimentDetectionNegativeResults-ICSE2018}. 

\item \bf{Phase 2.} We studied how the hybrid tool that we developed in Phase 1 compares (and can be improved) with the recent, advanced pre-trained language-based models by Zhang et al.~\cite{Zhang-SEBERTSentiment-ICSME2020}.
\end{itemize}

We studied a total of nine available sentiment detection tools for software engineering 
based on the studies of Lin et al.~\cite{Lin-SentimentDetectionNegativeResults-ICSE2018, Lin-PatternBasedOpinionMining-ICSE2019} 
and Zhang et al.~\cite{Zhang-SEBERTSentiment-ICSME2020}: five from Lin et al.~\cite{Lin-SentimentDetectionNegativeResults-ICSE2018, Lin-PatternBasedOpinionMining-ICSE2019} 
and four from Zhang et al.~\cite{Zhang-SEBERTSentiment-ICSME2020}. Through our empirical study, we answer total eight research questions as discussed below.

\bf{In Phase 1}, our focus was the feasibility of developing an ensemble of the five tools that could potentially offer better performance than each of the stand-alone tools from Lin et
al.~\cite{Lin-SentimentDetectionNegativeResults-ICSE2018, Lin-PatternBasedOpinionMining-ICSE2019}. In Phase 1, we answered a total of five research questions as follows:

\nd\bf{RQ$_1$. How frequently do we get at least one correct classification
while using multiple SE-specific tools?} On a total of 17,581 units from six
software specific sentiment benchmarks from the literature, we ran five
stand-alone SE-specific sentiment tools:
Senti4SD~\cite{Calefato-Senti4SD-EMSE2017},
SentiCR~\cite{Ahmed-SentiCRNIER-ASE2017},
SentistrengthSE~\cite{Islam-SentistrengthSE-MSR2017},
Opiner~\cite{Uddin-OpinionValue-TSE2019}, and
POME~\cite{Lin-PatternBasedOpinionMining-ICSE2019}. We find that 85\% of
misclassification by the tools related to non-neutral polarity (positive,
negative) and 99\% neutral classes can be potentially corrected by at least one
other tool. This finding motivated us to investigate whether a hybrid sentiment
detector can be developed by a simple majority analysis as follows.

\nd\bf{RQ$_2$. Can a majority voting-based classifier perform better than
individual classifiers?} For each of the 17,581 units, we assign a polarity
label by simply taking the majority voting on the unit by the tools, i.e., if
three tools say `positive' and two say `negative', we assign a final polarity as
`positive'. Unfortunately, this classifier only achieves a F1-score (Macro) of
0.746, which was outperformed by both supervised stand-alone detectors, Senti4SD
(0.750) and SentiCR (0.756). This finding necessitates towards a more
sophisticated design of the hybrid tool. To determine that, we analyzed why one
tool misclassifies when another does not.

\nd\bf{RQ$_3$. What are the error categories that could potentially be corrected
by combining multiple SE-specific tools?} We manually label 153 units where at
least one tool was right and at least one another was wrong. For the labeling,
we follow error categories similar
to~\cite{Novielli-BenchmarkStudySentiSE-MSR2018}, i.e., we determine the
specific reason for misclassification by a tool. We find five major error
categories. The most prevalent category is the failure of a tool to understand
the underlying context (related to an expressed sentiment). Such contexts can be
better understood by analyzing the textual contents in the unit. This motivates
us to investigate the possibility of leveraging ensemble learning to capture the
strength of multiple sentiment detectors. We therefore answer the following
research questions.

\nd\bf{RQ$_4$. Can a supervised ensemble of classifiers offer better performance
than individual classifiers?} For each of the 17,581 units, we combine the
polarity labels of the five individual sentiment tools with the bag of words of
the unit as features. We train a supervised machine learning classifier based on shallow machine learning Random Forest (RF) model using
the features.
We name the classifier Sentisead$_{RF}$. Sentisead$_{RF}$ outperforms each stand-alone
sentiment detectors from 4\% (Senti4SD) to $>$100\% (POME). We investigate the
possibility to further improve Sentisead$_{RF}$ with additional features through the
following research question.

\nd\bf{RQ$_5$. Can the addition of more features improve the performance of the
hybrid tool?} We introduce two new types of features in Sentisead$_{RF}$ based on the
analysis of polarity words and their positions in a given unit (e.g., first vs.\
last) and \it{entropy} measures using information theory. The enhanced hybrid
engine, i.e., Sentisead$_{RF}+$, however, shows negligible performance increase over
Sentisead$_{RF}$ (increase of 0.001 in F1-score Macro).

\bf{In Phase 2}, our focus was to study the effectiveness of Sentisead$_{RF}$ against
the most recently introduced advanced language-based models for SE-specific
sentiment tools, which were studied in multiple papers like Zhang et
al.~\cite{Zhang-SEBERTSentiment-ICSME2020}. We answer a total of three research
questions as follows:

\nd\bf{RQ$_6$. Can Sentisead$_{RF}$ outperform the BERT-based advanced pre-trained
language based models?} Given that pre-trained advanced language-based
transformers (PTMs for short) offer better performance than shallow learning
models for sentiment detection in SE, we want to know whether the PTMs can also
outperform Sentisead$_{RF}$, our hybrid tool of rule-based and shallow learning
models based on the Random Forest model. We investigated four PTMs from Zhang et
al.~\cite{Zhang-SEBERTSentiment-ICSME2020}, ALBERT, BERT, RoBERTa, and XLNet, on
our six datasets. We found that RoBERTa is the best performer among the four
stand-alone PTMs with a Macro F1-score of 0.801, which is slightly better than
Sentisead$_{RF}$ by 1.8\%. Following this minor improvement, we decided to investigate
the following research question.

\nd\bf{RQ$_7$. Can a deep learning model as ensembler for Sentisead outperform the stand-alone BERT-based
PTMs?} Motivated by superior performance of stand-alone PTMs over Sentisead$_{RF}$
that uses a shallow learning Random Forest model, we replaced the Random Forest
model in Sentisead$_{RF}$ by the PTMs: we trained and tested four versions of Sentisead
where features are the unit textual contents and the polarity labels from the
five rule-based and shallow learning models from Lin et
al.~\cite{Lin-SentimentDetectionNegativeResults-ICSE2018,
Lin-PatternBasedOpinionMining-ICSE2019}. However, the ensembler was a PTM
instead of Random Forest. We found that the best performing PTM-based ensembler
of the stand-alone rule-based and shallow learning sentiment detectors is based
on RoBERTa, which we named Sentisead$_{RoBERTa}$. Across the six datasets (and
using the same 10-fold cross validation), Sentisead$_{RoBERTa}$ shows a Macro
F1-score of 0.805: it offers 0.5\% increase over a stand-alone RoBERTa model and
a 2.3\% increase in performance over Sentisead$_{RF}$. This
slight performance increase motivated us to include the stand-alone PTMs into
the hybrid architecture as follows.

\nd\bf{RQ$_8$. Can Sentisead based on an ensemble of all the models (i.e., PTMs + non-PTMs)
offer the best performance of all tools?} We investigated whether the
performance of the ensembler could further increase using the sentiment polarity
labels of a unit from all available stand-alone models (i.e., rule-based,
shallow, and deep learning models) in addition to the textual content of the
unit. We trained and tested four versions of Sentisead by taking as input the
above polarity labels and textual content of a unit as features. The ensemblers
are the four PTMs, which yield four ensemblers: Sentisead$_{ALBERT}+$,
Sentisead$_{BERT}+$, Sentisead$_{RoBERTa}+$, and Sentisead$_{XLNet}+$. The best
performing model is Sentisead$_{RoBERTa}+$ with a Macro F1-score of 0.8.
However, Sentisead$_{RoBERTa}+$ is not more accurate than Sentisead$_{RoBERTa}$
from RQ$_7$. Therefore, the inclusion of polarity labels of individual PTMs does
not contribute to any performance improvement in Sentisead.

Based on our empirical study findings, we make following major observations and suggestions:
\begin{enumerate}
\item Our hybrid classifier, Sentisead$_{RoBERTa}$, which is based on the PTM
RoBERTa, offers the best Macro F1-score of 0.805. This classifier can be easily
built by taking as input the sentiment polarity of the five stand-alone
sentiment detectors (Opiner, POME, Senti4SD, SentiCR, and SentistrengthSE) and
the unit textual contents. However, Sentisead$_{RoBERTa}$ shows only 0.5\%
performance improvement over the stand-alone RoBERTa model. Therefore, when 
a mere increase of 0.5\% is
not warranted, a stand-alone RoBERTa model should suffice.
\item The hybrid classifier using Random Forest achieves 
the best Macro F1-score of 0.787 out of all shallow learning-based hybrid models. 
Thus, this Random Forest model shows a mere decrease of 0.018 in F1-score compared to 
the overall best hybrid model, i.e., the RoBERTa based PTM which shows a Macro F1-score of 0.805. 
Therefore, for systems where GPU-based servers are not readily available, the Random Forest-based hybrid model can suffice.
%
%
\item On the six benchmarks, Sentisead$_{RF}$ or Sentisead$_{RoBERTa}$ show more than 0.8-0.9 F1-score (for
positive and negative classes) for four datasets but only 0.3-0.5 for the other
two datasets. Further improvement in the latter two datasets require the
analysis of contextual information \it{surrounding} a given unit (e.g.,
preceding \& following units); an aspect that current tools cannot handle. Thus,
future improvement of stand-alone SE-tools should focus on developing algorithms
to analyze surrounding contexts.
\end{enumerate}

Our study benefited from the availability of open source code, tools, and
datasets that were shared by software engineering researchers working on
sentiment detection. In fact, the code base of Sentisead uses the
SentiCR~\cite{Ahmed-SentiCRNIER-ASE2017} code base to bootstrap. Given that our
study shows promise to advance the state-of-the-art research in software
engineering, we have open sourced the entire codebase of Sentisead in our online
appendix~\cite{website:senisead-online-appendix-ase2020}.

\section{Studied Benchmarks and Tools}\label{sec:background}
In this section, we describe the sentiment detection benchmarks and tools that 
we studied to develop our hybrid engine.

\subsection{Studied Sentiment Benchmarks}\label{sec:benchmarks}
We analyze the following six benchmarks available from software-engineering
research. Each unit in a benchmark is labeled as a polarity (positive, negative,
or neutral). A unit can be a sentence or a document (i.e., a list of sentences).
\tbl\ref{tab:benchmark-dataset-stat} provides descriptive summary statistics of
the benchmarks. Overall, 46\% of the units in the benchmarks are labeled as
positive or negative and the rest (64\%) are labeled as neutral (see last column
in \tbl\ref{tab:benchmark-dataset-stat}).

\begin{inparaenum}
\item\bf{Stack Overflow: Calefato et al.~\cite{Calefato-Senti4SD-EMSE2017}.} The
benchmark is based on 4,423 randomly-sampled units from Stack Overflow. It was
annotated for polarity by 12 coders. Each unit was annotated by three coders
using majority voting. The benchmark was used to train and test Senti4SD~\cite{Calefato-Senti4SD-EMSE2017}. It is available in the Senti4SD
GitHub repository.

\item\bf{Stack Overflow: Lin et al.~\cite{Lin-SentimentDetectionNegativeResults-ICSE2018}.} The benchmark contains 1,500 
units, each manually assigned a sentiment score ranging from -2 (strong negative) to +2 (strong positive) by two coders. Each unit was divided into a number of nodes (e.g., clauses), which were assigned a sentiment score. We use the final sentiment scores available in the replication package of~\cite{Lin-SentimentDetectionNegativeResults-ICSE2018} to assign a polarity: +1 positive, -1 negative, and 0 neutral.

\item\bf{Stack Overflow: Uddin et al.~\cite{Uddin-OpinionValue-TSE2019}.} The benchmark consists of the 4,522 units in 71 randomly-sampled Stack Overflow threads. Each unit was manually labeled for polarity by at least two coders with a third coder consulted in case of disagreements. Unlike the above four benchmarks, this benchmark includes all the units from the threads, i.e., contextual information. We obtained this benchmark from the authors of Opiner~\cite{Uddin-OpinerReviewAlgo-ASE2017}, who used it to develop an online opinion summarization engine.

\item\bf{Jira: Ortu et al.~\cite{Ortu-BulliesMoreProductive-MSR2015}.} The benchmark consists of 6,000 units (2,000 issue comments and 
4,000 units contributed by developers using Jira). The issue comments were collected from four popular open-source 
frameworks/repositories: Apache, Codehaus, JBoss, and Spring. The original benchmark was annotated for emotions, such as love, joy, surprise, anger, fear, and sadness and later 
labeled with three polarities by Novielli et al.~\cite{Novielli-BenchmarkStudySentiSE-MSR2018}, which we obtained from the authors for this paper. In their translation, 
Novielli et al. coded ``joy'' and ``love'' as positive, while ``sadness'', ``fear'' and ``anger'' as negative. An absence of any emotion for a unit was considered neutral. They discarded 131 units labeled as ``surprise'' because of the lack of contextual information needed to identify the underlying polarity.

\item\bf{Jira: Lin et al.~\cite{Lin-SentimentDetectionNegativeResults-ICSE2018}.} Lin et al.~\cite{Lin-SentimentDetectionNegativeResults-ICSE2018} 
took 926 units from Ortu et al.~\cite{Ortu-BulliesMoreProductive-MSR2015} dataset that was originally labeled as six different types of emotions: love, joy, surprise, anger, fear, and sadness. 
Lin et al.~\cite{Ortu-BulliesMoreProductive-MSR2015} considered the units labeled as joy and love as positive and the units 
labeled as anger and sadness as negative.

\item\bf{Apps Reviews: Lin et al.~\cite{Lin-SentimentDetectionNegativeResults-ICSE2018}.} It contains 
341 mobile-app reviews from Villarroel et al.~\cite{Villarroel-ReleasePlanningMobileAppReviews-ICSE2016}, who labeled each review to identify its type (e.g., bug reporting, request for enhancement, etc.). Each review was labeled for polarity by two of the authors of \cite{Lin-SentimentDetectionNegativeResults-ICSE2018}.  
\end{inparaenum}
\begin{table}[t]
  \centering
  \caption{Benchmarks used in the study (SO = Stack Overflow)}
    \begin{tabular}{lr|rr|r}\toprule
    \textbf{Dataset} & \multicolumn{1}{l}{\textbf{\#Units}} & \multicolumn{1}{l}{\textbf{+VE}} & \multicolumn{1}{l}{\textbf{-VE}} 
     & \multicolumn{1}{l}{\textbf{$\pm$/\#Units}} \\
    \midrule
    SO Calefato et al.~\cite{Calefato-Senti4SD-EMSE2017} & 4423  & 1527  & 1202  & 62\% \\
    SO Lin et al.~\cite{Lin-SentimentDetectionNegativeResults-ICSE2018} &      1,500  &        131  &      178  &      21\% \\
    SO Uddin et al.~\cite{Uddin-OpinionValue-TSE2019}  &      4,522  &      1,048  &      839  &      42\% \\
    JIRA Ortu et al..~\cite{Ortu-BulliesMoreProductive-MSR2015} &      5,869  &      1,128  &      786  &      33\% \\
    JIRA Lin et al.~\cite{Lin-SentimentDetectionNegativeResults-ICSE2018} &        926  &        290  &      636  &       100\% \\
    Mobile App Lin et al.~\cite{Lin-SentimentDetectionNegativeResults-ICSE2018} &        341  &        186  &      130  &      93\% \\
    \midrule
    Total &                     17,581  &                        4,310  &                 3,771  &                        46\% \\
    \bottomrule
    \end{tabular}%
  \label{tab:benchmark-dataset-stat}%
  \vspace{-2mm}
\end{table}%

\subsection{Studied Sentiment Detection Tools in Phase 1}\label{sec:senti-tools} 
The observed shortcomings in cross-domain sentiment detection tools motivated
the development of several recent sentiment detection tools for software
engineering. In this paper, we have studied five sentiment detection tools developed for software engineering: SentistrengthSE~\cite{Islam-SentistrengthSE-MSR2017},
Senti4SD~\cite{Calefato-Senti4SD-EMSE2017}, 
SentiCR~\cite{Ahmed-SentiCRNIER-ASE2017}, Opiner~\cite{Uddin-OpinionValue-TSE2019}, and POME~\cite{Lin-PatternBasedOpinionMining-ICSE2019}.
While Senti4SD and SentiCR are supervised, the other three tools are rule-based classifiers. As we noted in \sec\ref{sec:intro}, these tools were the subject of two papers 
by Lin et al.~\cite{Lin-PatternBasedOpinionMining-ICSE2019,Lin-SentimentDetectionNegativeResults-ICSE2018}. From the two studies of Lin et al.~\cite{Lin-PatternBasedOpinionMining-ICSE2019,Lin-SentimentDetectionNegativeResults-ICSE2018}, 
we picked the tools whose trained model/source code were available online during the time of analysis. We also picked the tools that were developed 
specifically for software engineering artifacts. These five tools are also the five most cited sentiment detection tools 
developed for software engineering as of now.  

\begin{inparaenum}
\item\bf{Senti4SD~\cite{Calefato-Senti4SD-EMSE2017}.} Given as input a short text, Senti4SD 
detects the polarity of the text as positive, negative or neutral.  
Senti4SD~\cite{Calefato-Senti4SD-EMSE2017} is trained on a dataset of 4000 posts
(questions, answers, and comments) from Stack Overflow. The classifier leverages a suite of textual features 
both based on input text (e.g., ngrams) as well as a pre-defined list of sentiment lexicons and 
word embeddings. The word embeddings are compiled from an entire Stack Overflow dump to 
offer domain specific information, such as semantic similarity between the input text and 
Stack Overflow texts that are closely similar to the 4000 posts used in the benchmark. The supervised 
classifier is a trained support vector machine (SVM) model.

\item\bf{SentiCR~\cite{Ahmed-SentiCRNIER-ASE2017}.} 
SentiCR~\cite{Ahmed-SentiCRNIER-ASE2017}  was trained on a dataset of 1600 code
reviews from Gerrit. 
While Senti4SD uses a SVM classifier, the currently distributed version of
SentiCR leverages the Gradient Boosting Tree (GBT) algorithm. Unlike Senti4SD,
SentiCR handles the under-representation (i.e., class imbalance) of the polarity
labels (i.e., positive and negative) compared to the neutral classes by using
the SMOTE algorithm (synthetic minority over-sampling
technique)~\cite{Chawla-SMOTE-JournalAIResearch2002}. Note that we train SentiCR to detect three polarity classes following Novielli et al.~\cite{Novielli-BenchmarkStudySentiSE-MSR2018} by using numeric identifiers: positive (+1), 
negative (-1), and neutral (0).

\item\bf{SentistrengthSE~\cite{Islam-SentistrengthSE-MSR2017}.}
SentiStrengthSE~\cite{Islam-SentistrengthSE-MSR2017} was developed on top of
Sentistrength~\cite{Thelwall-Sentistrength-ASICT2010} by introducing rules and
sentiment words specific to the domain of software
engineering~\cite{Islam-SentistrengthSE-MSR2017}. Each negative word has a score
ranging from -2 to -5, positive word has a score ranging from +2 to +5. The
polarity scores are \it{a priori}, i.e., they do not depend on the contextual
nature of the sentiment expressed in a unit.
Similar to Sentistrength, SentistrengthSE outputs both positive and negative
scores for an input text. Following state of
art~\cite{Novielli-BenchmarkStudySentiSE-MSR2018}, the overall polarity score of
an input text can be calculated by taking the algebraic sum of the positive and
negative scores. The text is labeled as `positive' if the sum of scores is
greater than 0, negative if the sum is less than 0, and neutral otherwise.

\item\bf{Opiner~\cite{Uddin-OpinionValue-TSE2019}}. The tool adapts 
the Domain Sentiment Orientation (DSO) algorithm 
originally proposed by Hu and Liu~\cite{Hu-MiningSummarizingCustomerReviews-KDD2004} to collect online customer reviews. 
DSO assigns a polarity to a unit based on
the presence of sentiment words. 
The SE-specific adaptation in Opiner was used 
to mine and summarize API reviews from Stack Overflow~\cite{Uddin-OpinerReviewAlgo-ASE2017,Uddin-OpinerReviewToolDemo-ASE2017,Uddin-OpinionValue-TSE2019}. 
Opiner assigns polarity to an input text in three major steps: 
\begin{inparaenum}
\item\it{Detect potential sentiment words.} by identifying adjectives in the unit
that match the sentiment words in a list of predefined sentiment vocabularies.
Opiner gives a score of +1 to a recorded
adjective with positive polarity and a score of -1 to an adjective
of negative polarity. \item\it{Detect Negations.} The sign for an adjective alternates if a
negation word is found close to the word. 
\item\it{Label unit.} If the sum of all polarity scores is greater than 0, Opiner labels the unit as `positive'.
If the sum is less than 0, Opiner labels it as `negative'. 
\end{inparaenum} 

\item\bf{POME~\cite{Lin-PatternBasedOpinionMining-ICSE2019}}. 
In 2019, Lin et al.~\cite{Lin-PatternBasedOpinionMining-ICSE2019} proposed POME,
a pattern-based tool to mine opinions about software aspects from Stack
Overflow. The tool serves two purposes simultaneously: detection of the aspects
and of sentiment polarity in units (i.e., sentences). The patterns are derived
by manually assessing 1,662 units from Stack Overflow. The manual assessment
produced 157 distinct patterns. For example, a pattern can check the presence of
specific API aspect (e.g., performance) and sentiment lexicons related to the
API aspect in an input text. The pattern follows strict matching rules, such as
subject (in a unit) followed by API aspect (e.g., performance, usability),
followed/preceded by a list of predefined sentiment lexicons.

In a benchmark-based study against six state-of-the-art SE-specific sentiment
detection tools including all four stand-alone rule-based and shallow sentiment
detection tools that we have studied in this paper (i.e., Opiner, Senti4SD,
SentiCR, StrengthSE), POME outperformed each tool in the benchmark. While POME
serves two features (sentiment and aspect detection) compared to the other four
studied tools that serve one feature (sentiment detection), POME does not limit
itself to a specific purpose (e.g., only aspect detection or only sentiment
detection). Indeed, in their evaluation of POME, Lin et
al.~\cite{Lin-PatternBasedOpinionMining-ICSE2019} considered other sentiment
detection tools like Opiner as baselines (see RQ$_2$ and RQ$_3$ in Lin et
al.~\cite{Lin-PatternBasedOpinionMining-ICSE2019}). We, therefore, include POME
as a stand-alone sentiment detection tool in our study.
\end{inparaenum}

\subsection{Studied Deep Learning Tools for Sentiment Detection in Phase 2}
\label{sec:senti-tools-bert} 

Starting from mid 2019, the SE community started using deep learning models to
detect sentiment in SE artifacts. Most recently, several research results showed
the superior performance of pre-trained advanced language-based transformers for
sentiment detection in SE. For the sake of brevity, we refer to those models as
PTMs (Pre-trained Transformer Models) in the following. Among the PTMs, four
models, ALBERT, BERT, RoBERTa, and XLNet, were studied in the most recent papers
by Zhang et al.~\cite{Zhang-SEBERTSentiment-ICSME2020} and Biswas et
al.~\cite{Biswas-ReliableSentiSEBERT-ICSME2020}, who found that these PTMs offer
better performance than shallow learning models, like Senti4SD, SentiCR, etc. In
the second phase of our study, we thus investigate the effectiveness of the PTMs
with our ensemble sentiment detection model. We discuss briefly the PTMs below.

\begin{inparaenum}[(1)]
\item \bf{BERT~\cite{Delvin-BERTArch-Arxiv2018}} stands for Bidirectional Encoder Representation from Transformers. Google developed this transformer-based ML technique for textual analysis. BERT is a transformer model that processes words in a unit in relation to all words in the unit, which is different from sequence to sequence models (e.g., LSTM), which utilizes the forward/backward sequences of words one-by-one in a unit. BERT models thus are capable of considering the full context of a given word in a unit by looking at the preceding and following words. Contextual word representations are learned based on two tasks:

\begin{inparaenum}[(a)]
\item Masking some words from the input unit based on a Masked Language Model (MLM) and predicting the masked words based on the context. 

\item Next unit prediction (NSP) to determine whether one unit follows another. 
\end{inparaenum} 

Google provided two architecture of BERT, both based on a multi-layer bidirectional transformer model: BERT$_{BASE}$ and BERT$_{LARGE}$. BERT$_{BASE}$ has 12 bidirectional self-attention heads while BERT$_{LARGE}$ has 24 heads. Both models are trained using unlabeled data from English Wikipedia and a  corpus of books. BERT achieved state-of-the-art performance on text understanding tasks in several benchmarks (e.g., GLUE).

\item \bf{RoBERTa~\cite{Liu-Roberta-Arxiv2019}} is a \bf{Ro}bustly Optimized \bf{BERT} \bf{a}pprach developed by Facebook to offer improvements over the BERT model developed by Google. RoBERTa follows BERT's strategy of masking, i.e., the system learns to predict intentionally masked words within a unit. In addition, the following modifications led to better performance than BERT for downstream tasks, e.g., classification:

\begin{inparaenum}[(a)]
\item RoBERTa removes next unit prediction in BERT, which is not important for text classification (e.g., for our sentiement detection). This removal offers more flexibility to improve on the MLM of BERT. 

\item RoBERTa is trained and hyper-tuned on more data than BERT, for a longer amount of time.
\end{inparaenum}

\item \bf{XLNet~\cite{Yang-Xlnet-Arxiv2020}} is designed to address the concern in BERT that by ``relying on corrupting the input with masks, BERT neglects dependency between the masked positions and suffers from a pretrain-finetune discrepancy''. XLNet is a generalized auto-regressive pre-training method based on Transformer-XL, which learns bidirectional contexts by maximizing the expected likelihood of words over all permutations of the order. XLNet also provides a modified language model training objective over BERT to learn conditional distributions for all permutations of word tokens. In empirical evaluations, XLNet outperformed BERT in 20 tasks (e.g., sentiment analysis).

\item \bf{ALBERT~\cite{Lan-Albert-Arxiv2020}} offers an upgrade on BERT on 12 textual processing tasks (e.g., question answering) by allocating the model more efficiently for input-level embeddings (words, sub-tokens) that are needed to learn context independent representations and hidden-layer embeddings that are needed to refine the input-level embeddings into context-dependent representations. ALBERT factorizes the embedding parameterization, where the embedding matrix is split into input-level embedding with relative-lower dimension than then hidden-layer embeddings (e.g., 128 in input-level vs.\ 768 or more in hidden-layer), which allows ALBERT to achieve an 80\% reduction in the parameters. Another design decision in ALBERT is to apply the same layer on top of each other, while BERT creates separate independent layers on top of each other. This decision in ALBERT is based on the observation that often multiple layers in BERT learn almost similar operations. Thus, ALBERT reduces the architectural complexity of BERT, but at the cost of a slight decrease in performance than BERT. As such, ALBERT is considered as ``lite'' BERT.  
\end{inparaenum}

\begin{table}[ht]
\centering
\caption{Architecture details of variants of BERT Model (L = Layers, H = Hidden, D = Heads, P = Parameters)}
\begin{tabular}{l|p{3.2cm}rrrr}
\toprule
\textbf{Architecture}& \bf{Used Model} & \bf{L}  & \bf{H} & \bf{D} & \bf{P}\\ 
\midrule
 BERT & bert-base-uncased & 12& 768&12&110M\\
 XLNet & xlnet-base-cased & 12& 768& 12& 110M                  \\ 
 RoBERTa& roberta-base & 12&768&12&125M                               \\ 
 ALBERT & albert-base-v1 & 12 &  768 & 12 & 11M \\
\bottomrule
\end{tabular}
\label{tab:architecture_details} 
\end{table}

In our study, we use the implementations of the above four models from the Hugging Face transformer library, which is an open-source community around the PTM libraries. \tbl\ref{tab:architecture_details} shows the architecture of the four PTMs.

\section{Study Phase 1: Ensemble of Rule-Based and Shallow Machine Learning Models}
\label{sec:results}
We offer insights into the feasibility of developing a hybrid sentiment detection engine for the domain of software engineering. 
Our study has two phases. In the first phase, we answer five research questions we introduced in \sec\ref{sec:intro}.
\begin{enumerate}[leftmargin=12pt]
  \item How frequently can we correct the misclassification of a tool by another tool? (\sec\ref{subsec:complementarity})
  \item Can a majority voting-based classifier perform better than individual classifiers? (\sec\ref{subsec:ensemble-majority})
  \item What are the error categories that could be potentially corrected by combining multiple SE-specific tools? (\sec\ref{subsec:misclassification-cats})
  \item Can an ensemble of classifiers offer better performance than individual classifiers? (\sec\ref{subsec:sentisead})
  \item Can the addition of more features improve the performance of the hybrid classifier? (\sec\ref{subsec:sentiseadplus})
\end{enumerate}

\subsection{\texorpdfstring{RQ$_1$}{RQ1} How frequently can we correct the misclassification of a tool by another tool?}
\label{subsec:complementarity}

\subsubsection{Motivation} The five SE-specific tools (SentiCR, Senti4SD, Opiner, SentistrengthSE, and POME) 
in our study can complement each other, if the misclassification of one tool can be \it{potentially} corrected 
by the correct polarity label from another tool. If all the tools are wrong for a given unit, the proper processing of the unit may require the development of a new tool. 
As noted in \sec\ref{sec:intro}, our focus is to improve the accurate polarity classification probability of a given unit by analyzing the polarity 
labels of the units from the existing SE-specific tools. This concept is similar to the bagging or boosting combination of classifiers in 
machine learning~\cite{Misrili-IndustrialCaseStudyEnsembleSoftwareDefects-SQJ2011,
Petric-EnsembleDefectDiversity-ESEM2016,Zhang-CombinedClassifierCrossDefect-FrontierComputerScience2018,Yang-TwoLayerEnsembleJustInTimeDefect-IST2017}. However, for the 
combination to perform better than the stand-alone SE-specific classifier, we first need to confirm whether another tool can offer correct polarity classification for a given 
unit when a tool is wrong on that unit.


\subsubsection{Approach}
\label{sec:approach-rq1-complementarity}

We have six datasets in our study. For each dataset, we collect sentiment
polarity labels on each unit from the five sentiment detection tools as follows.
We run each of the three unsupervised classifiers (i.e., SentistrengthSE,
Opiner, and POME) on the entire benchmark dataset. The output is a polarity
label for each unit. The other two tools, Senti4SD and SentiCR, are supervised
sentiment detectors so, for each dataset, we trained and tested as follows.

Literature on the evaluation of SE-specific supervised sentiment detection tools
has so far split a given dataset into 70\% training set and 30\% testing set (e.g., Novielli et al.~\cite{Novielli-BenchmarkStudySentiSE-MSR2018} Calefato et al.~\cite{Calefato-Senti4SD-EMSE2017}, 
Zhang et al.~\cite{Zhang-SEBERTSentiment-ICSME2020}. The
split is done based on a stratified sampling, i.e., 70\% of each polarity class
will stay in the training set and 30\% of that class will be in the testing set.
The split is done once, i.e., before a classifier is trained or tested. In our
paper, we improved on this approach as follows.

We split each dataset into 10-folds using a stratified sampling\footnote{We use  the StratifiedKFold method from Python Scikit-learn to create the $k = 10$ folds. The folds preserve the percentage of samples for each of the polarity  classes.}. 
Thus, each fold contains the same proportion of a polarity class as in the entire dataset.
Also, a unit cannot be present in two different folds, because otherwise the
classifier would see the same unit in both training and testing sets, which
would lead to overfitting. Then, we train and test a classifier 10 times. For
example, in the first run, we train the classifier using folds 1-9 and then test
it on Fold 10. In the second run, we train the classifier using folds 2-10 and
test it on Fold 1. Once all the 10 runs are completed, we have test results on
each of the 10 folds (e.g., test result of Fold 10 coming from Run 1, test
result of Fold 1 coming from Run 2, etc.). We can then compute the performance
of the classifier on all test results.

Therefore, our approach is similar to the existing 70–30 evaluation approach
used in the literature but our training vs. testing split is 90-10. In addition,
unlike current approaches in SE-specific sentiment detection, because we do the
testing on the entire dataset by running a tool 10 times, our test results are
more robust than a 70-30 split, which is run only once and is tested only on 30\%
of all data.

Because we are training and testing each tool 10 times compared to only once in
the current literature, our approach also better addresses any variance in the
machine learning classification model due to its stochastic nature. An
alternative to our approach would have been to do a repeated sampling of
training and testing sample from a given dataset for N times (e.g., N = 10 or N
= 100) and then report the performance of the model by taking the average of the
performance across all test samples. Our approach slightly differs from this
alternative in that we created the 10 folds once, before we started all the
classification tasks. This means that we were able to track and compare each
classification model performance on each fold, where the training and testing for each model followed the same pattern (i.e., train using folds 1-9 and testing
using fold 10 in run 1, train using folds 2-10 and test using fold 1 in run 2,
etc.). This process then allowed us to also check for tool complementarity,
i.e., see when a tool is wrong while another can be right for a given unit.
Moreover, given that we train and test a classification model 10 times instead of one
time, we also reduce any variance in the classification results. Nevertheless,
for each model, we pre-set the random variables by using seeds, which is useful
to get similar results for a model for a given dataset, even if we run the model
10, 20, or 100 times in the dataset.

Given that we have six datasets and five SE-specific sentiment detection tools,
we analyze the performance of each tool across the six datasets as follows.
First, we get the test results for each dataset following the steps listed
above. Second, we compute the performance metrics of each model in each dataset
by analyzing the polarity labels in the 10 test folds of the dataset. Third, we
compute the performance metrics of each model across the six datasets by
analyzing the polarity labels across all the test folds in the six datasets,
i.e., across the 60 test folds (10 each for each dataset). While we report the
consolidated performance metric for a given model across the datasets to save
space in the paper, we include the performance metric for each dataset in our
online replication package~\cite{website:senisead-online-appendix-ase2020}.

In summary, given as input the benchmark datasets and the five sentiment detectors, we produce the final output after all the training and testing for each dataset as a matrix $M = U \times F \times L$, where $U = \{U_1, U_2, \ldots U_n\}$ ($n = 17,581$) is each unit, $F = \{F_1, F_2, \ldots F_k\}$ ($k = 6$) corresponds to the benchmark dataset where unit $U_i$ is found, and $L = \{L_1, L_2, \ldots, L_m\}$ ($m = 5$) and $L_j$ corresponds to the polarity label of unit $U_i$ by a given sentiment detector $j$ in our study.

We analyze the complementarity of the tools on the generated matrix $M$ as follows. First, for each pair (e.g., $(i, j)$) of tools in our dataset (e.g., Senti4SD, SentiCR), we collect all the records $R_{i,j}$ in $M$ where at least one of the tools is right (i.e., matches with the ground truth). For each tool in $M$, we then compute the fraction of misclassification of each tool, for which one or more of the other tools offer correct classifications.

\begin{table}[ht]
  \centering
  \caption{How the misclassification of a tool can be corrected by one tool or at least one (i.e., last column $\geq 1$) tool. O = Neutral.}
    \begin{tabular}{lrrrrrr|r}\toprule
    {\textbf{Tool}} & \textbf{Polarity} & \multicolumn{6}{c}{\textbf{Tools Could Correct Wrong Polarity}} \\
    \cmidrule{3-8}
    {\textbf{Wrong}} & \textbf{Wrong} & \textbf{Senti4SD} & \textbf{SentiCR} & \textbf{SentistrengthSE} & \textbf{Opiner} & {\textbf{POME}} & \textbf{$\geq 1$} \\
    \midrule
    {\textbf{Senti4SD}} & $\pm$ (2685) &       & 24\%  & 29\%  & 29\%  & 7\%   & 54\% \\
          & O (1281) &       & 54\%  & 47\%  & 41\%  & 88\%  & 97\% \\
    {\textbf{SentiCR}} & $\pm$ (2948) & 31\%  &       & 28\%  & 37\%  & 7\%   & 59\% \\
          & O (917) & 36\%  &       & 37\%  & 43\%  & 88\%  & 0.96 \\
    {\textbf{SentistrengthSE}} & $\pm$ (2970) & 36\%  & 29\%  &       & 32\%  & 7\%   & 59\% \\
          & O (1458) & 53\%  & 61\%  &       & 40\%  & 91\%  & 98\% \\
    {\textbf{Opiner}} & $\pm$ (4087) & 54\%  & 55\%  & 51\%  &       & 4\%   & 70\% \\
          & O (2999) & 75\%  & 83\%  & 71\%  &       & 89\%  & 99\% \\
    {\textbf{POME}} & $\pm$ (7621) & 67\%  & 64\%  & 64\%  & 48\%  &       & 84\% \\
          & O (471) & 69\%  & 77\%  & 71\%  & 31\%  &       & 92\% \\
    \textbf{$\geq 1$}& $\pm$ (7938) & 66\%  & 63\%  & 63\%  & 49\%  & 4\%   & 85\% \\
          & O (4049) & 68\%  & 77\%  & 64\%  & 26\%  & 88\%  & 99\% \\
          \bottomrule
    \end{tabular}%
 \label{tab:toolMisclassificationCanBeCorrectedByAnother}%
\end{table}%

\subsubsection{Results} 

In \tbl\ref{tab:toolMisclassificationCanBeCorrectedByAnother}, we show the percentage of misclassified textual units 
for each tool that can be potentially corrected by another tool. The second column (`Polarity Wrong') shows the polarity that is misclassified by the 
tool listed in the first column. The third to seventh column show how frequently each of the other tools can correct the misclassification of a tool 
listed in the first column. For example, for Senti4SD (i.e., S4SD in \tbl\ref{tab:toolMisclassificationCanBeCorrectedByAnother}) the two unsupervised classifiers 
(SentistrengthSE and Opiner) offer the most complemenarity (31\% and 34\%) to correct the positive polarity as well as the negative polarity. 
The last column ($\geq 1$) shows how the misclassification of a tool can be corrected by the correct classification of at least one of the tools. For example, 60\% of the misclassified 
positive polarity by Senti4SD can be corrected by at least one of the other tools. The last row ($\geq 1$) shows the overall distribution of the positive, negative, and neutral 
units for which at least one tool is wrong. Around 87\% of such misclassified positive polarity and around 82\% of such negative polarity can be corrected by at least one of the other tools.
The tool POME seems to be the most effective to correct the misclassifications of neutral units, i.e., when other tools incorrectly label a unit as non-neutral. 
Other tools (i.e., except POME) can complement more each other to correct the misclassifications for the  positive and negative polarities. The primary challenge for a hybrid engine would be then to make a final decision from the different polarity labels on a given unit.
\begin{table}[t]
  \centering
  \caption{Confusion matrix for 3 class sentiment detection}
    \begin{tabular}{crr|r|r}\toprule
          &       & \multicolumn{3}{c}{\textbf{Predicted}} \\
          \cmidrule{3-5}
          &       & \textbf{Positive (P)} & \textbf{Negative (N)} & \textbf{Neutral (O)} \\
          \cmidrule{2-5}
    \multirow{3}[0]{*}{\textbf{Actual}} & \textbf{Positive (P)} & TP$_P$ & FP$_N$, FN$_P$ & FP$_O$, FN$_P$ \\
          & \textbf{Negative (N)} & FP$_P$, FN$_N$ & TP$_N$ & FP$_O$, FN$_N$ \\
          & \textbf{Neutral (O)} & FP$_P$, FN$_O$ & FP$_N$, FN$_O$ & TP$_O$ \\
          \bottomrule
    \end{tabular}%
  \label{tab:conmat-sentiment}%
\end{table}%

\subsection{\texorpdfstring{RQ$_2$}{RQ2} Can a majority voting-based classifier perform better than individual classifiers?}
\label{subsec:ensemble-majority}

\subsubsection{Motivation} 

The findings from RQ$_1$ confirm that the sentiment detectors can complement each other to correct their polarity misclassifications. However, 
the challenge is to determine the right polarity labels when multiple tools offer different labels. Given that we analyze five SE-specific sentiment detectors, 
the majority of the classifiers may agree on the correct classification for a given unit. 

\subsubsection{Approach} We take as input the matrix $M = U \times F \times L$ that we created for RQ$_1$, where for each unit $U_i$ we have polarity labels from each 
of the five sentiment detectors, i.e., $L_i = \{L_{i}^{1}, L_{i}^{2}, L_{i}^{3}, L_{i}^{4}, L_{i}^{5}\}$. As we noted in \sec\ref{subsec:complementarity}, this matrix 
was created by retraining the supervised classifiers on the benchmark datasets using a 10-fold cross validation. 
Out of the five polarity labels in $L_i$, we assign $U_i$ the 
label that was observed the most in $L_i$. We report performance using three standard metrics of information
retrieval: precision ($P$), recall ($R$), and F1-score ($F1$).

{
\begin{eqnarray*}
P  = \frac{TP}{TP+FP},~
R = \frac{TP}{TP+FN},~
F1 = 2*\frac{P*R}{P+R}
\end{eqnarray*}
}
{$TP = $ \# of true positives, $FN =$ \# of false
negatives, $TN = $ \# of true negatives, and $FP = $\# of false positives.} 
In addition, we also report the weighted $\kappa$ value for each tool by computing the agreement between the tool and the manual labels. We use the Python ml\_metrics to compute the weighted $\kappa$ values. 

We do not report accuracy as a metric while reporting the performance of the tools due to the following reason.
As we presented the descriptive statistics of our six datasets in \tbl\ref{tab:benchmark-dataset-stat} (\sec\ref{sec:background}), 
our datasets are imbalanced. This means that we do not have each polarity class present in equal proportion in a given dataset. 
For example, for the SO Lin et al.~\cite{Lin-SentimentDetectionNegativeResults-ICSE2018} dataset, only 21\% of the units are either positive or negative 
polarity and the rest (i.e., 79\%) are neutral. Overall, only 46\% of all units across the six datasets are either positive or negative and the rest (i.e., 54\%) 
units are neutral. Therefore, we are analyzing highly skewed (i.e., imbalanced) datasets. Metrics like accuracy can give misleading information for 
imbalanced datasets. For example, if we label each unit as neutral, the overall accuracy across the datasets would have been 0.46. However, given 
the accurate detection of negative/positive polarity is more important in SE, the accuracy would have given us a false impression of the effectiveness of the model. 
Indeed, in machine learning literature and discussions, it is warned to not use accuracy as a measure for imbalanced datasets\footnote{see \url{https://machinelearningmastery.com/failure-of-accuracy-for-imbalanced-class-distributions/}}.  

To assess performance, we use the confusion matrix in \tbl\ref{tab:conmat-sentiment}, 
which is consistent with the state of the art~\cite{Novielli-BenchmarkStudySentiSE-MSR2018,Calefato-Senti4SD-EMSE2017,Sebastiani-MachineLearningTextCategorization-ACMSurveys2002,Uddin-OpinionValue-TSE2019}. 
We report overall performance using both micro and macro averages, as well as by each polarity class.
Macro average is useful to emphasize the performance on polarities with few instances (e.g., positive/negative in our case). In contrast, 
micro average is influenced mainly by the performance of the majority polarity (e.g., neutral), because it merely takes the average of each polarity. We use the F1-score to report the best tool, following standard practices~\cite{Novielli-BenchmarkStudySentiSE-MSR2018,Manning-IRIntroBook-Cambridge2009}.

\begin{table}
  \centering
  \caption{Performance comparison of majority voting-based hybrid classifiers with the baselines. 
  Performance analysis for each polarity class is provided in \tbl\ref{tab:majorityVoting} of Appendix \ref{sec:detailsPerformanceAnalysis}.}
    \begin{tabular}{lrrrr|rrr}\toprule
         &  & \multicolumn{3}{c}{\textbf{Macro}} & \multicolumn{3}{c}{\textbf{Micro}}  \\
          \cmidrule{3-8}
         & \bf{$\kappa$} & \textbf{F1} & \textbf{P} & \textbf{R} & \textbf{F1} & \textbf{P} & \textbf{R} \\
    \midrule
    \textbf{Majority-All} & 0.56 & 0.746 & 0.806 & 0.694 & 0.768 & 0.768 & 0.768  \\
    \textbf{Majority-Supervised} & 0.61 & 0.745 & 0.811 & 0.690 & 0.769 & 0.769 & 0.769  \\
    \textbf{Majority-Unsupervised} & 0.42 & 0.630 & 0.715 & 0.563 & 0.671 & 0.671 & 0.671 \\
     \midrule
    Baseline - \textbf{Senti4SD} & 0.63 & 0.750 & 0.768 & 0.732 & 0.774 & 0.774 & 0.774  \\
    Baseline - \textbf{SentiCR} & 0.63 & 0.756 & 0.791 & 0.723 & 0.780 & 0.780 & 0.780  \\
    Baseline - \textbf{SentistrengthSE} & 0.60 & 0.723 & 0.746 & 0.701 & 0.748 & 0.748 & 0.748 \\
    Baseline - \textbf{Opiner} & 0.35 & 0.563 & 0.571 & 0.556 & 0.597 & 0.597 & 0.597\\
    Baseline - \textbf{POME} & 0.04 & 0.397 & 0.451 & 0.354 & 0.540 & 0.540 & 0.540 \\
    \bottomrule
    \end{tabular}%
  \label{tab:majorityVotingShort}%
\end{table}
\subsubsection{Results} In \tbl\ref{tab:majorityVotingShort}, we compare the performance of the developed majority-based voting detector against the stand-alone sentiment detectors. 
We show three versions of the majority-based detectors: \begin{inparaenum}
\item Majority-All: uses polarity outputs from all five sentiment detectors.
\item Majority-Supervised: uses polarity outputs from only the supervised sentiment detectors (i.e., Senti4SD and SentiCR), and 
\item Majority-Unsupervised: uses polarity outputs from only the unsupervised sentiment detectors (i.e., POME, Opiner, and SentistrensghtSE).
\end{inparaenum} None the of above two majority-based detectors outperform the individual supervised detectors. The Majority-All detector shows an F1-score of 0.746 (Marco), 
while Senti4SD shows 0.750 and SentiCR 0.756. With regards to the Micro average, Majority-All shows 0.768, while Senti4SD shows 0.774 and SentiCR shows 0.780. The breakdown of the Majority-All into sub-components, i.e., Majority-Supervised and Majority-Unsupervised shows that the major contributors of success in the 
Majority-All are the supervised classifiers. For example, Majority-Supervised shows a Macro F1-score of 0.745 (Majority-All shows 0.746), while Majority-Unsupervised shows 
an F1-score of 0.630. The Majority-Unsupervised performs better than the two unsupervised detectors (Opiner and POME), but it is outperformed by the SentistrengthSE (Macro F1-score = 0.723).
The results indicate a non-linear relationship between the correct polarity labels and the agreement among the tools when at least one of the tools can be correct. 

\subsection{\texorpdfstring{RQ$_3$}{RQ3} What are the misclassification categories in the complementary cases?}
\label{subsec:misclassification-cats}

\subsubsection{Motivation}

The findings from RQ$_1$ (\sec\ref{subsec:complementarity}) confirm the potential of developing a hybrid sentiment detector, 
RQ$_2$ (\sec\ref{subsec:ensemble-majority}) shows that a simple majority-based voting does not perform better than the individual detectors. 
To better design a hybrid sentiment detector, we thus need to understand why one tool misclassifies when 
another tool can offer the correct label. This can be achieved by analyzing the error categories in the complementary cases. 
 
\subsubsection{Approach} We study misclassification types in 159 
units from our complementarity dataset $R$ (\sec\ref{subsec:complementarity}). 
This sample is statistically significant with a 99\% confidence level (10 interval).
Given that the majority of units are neutrals, a purely random sample would contain mostly neutral units. We thus randomly pick 
equal number (i.e., 53) of units from each polarity class. We manually label the misclassification reasons as follows.  
\begin{inparaenum}
  \item We label the misclassification in the unit to one of the seven error categories from Novielli et al.~\cite{Novielli-BenchmarkStudySentiSE-MSR2018}. The categories are: 
\begin{inparaenum}
\item Polar facts by neutral,
\item General error,
\item Politeness,
\item Implicit polarity,
\item Subjectivity in annotation,
\item Inability to deal with context information, and 
\item Figurative language.
\end{inparaenum}
  \item Novielli et al.~\cite{Novielli-BenchmarkStudySentiSE-MSR2018} 
observed seven error categories in four sentiment detectors when they are all wrong on a given unit. 
In contrast, we aim to study the textual 
units of the misclassifications where at least one tool is correct. If we cannot match the misclassification to one of the seven categories, we assign it a new error category based on the analysis of the textual contents.  
\end{inparaenum}
\begin{table}[t]
  \centering
  \caption{Error Categories in complementarity records}
    \begin{tabular}{l|lr}\toprule
    \textbf{Category} & \textbf{Description} & \textbf{\#} \\
    \midrule
    Context  & Tool fails to understand underlying context required to determine the overall polarity & 62 \\
    Polarity Diversity & The text contains multiple polarity cues with diverse interpretations & 35\\
    Domain & Polarity cues are SE-specific & 23\\
    General & Tool cannot process linguistic cues/typos. & 22 \\
    Politeness & A neutral text is labeled polar due to the existence of politeness & 17 \\
    \bottomrule
    \end{tabular}%
  \label{tab:misclassificationCategories159Units}%
\end{table}%
\begin{figure}[t]
  \vspace{-0.3in}
  \centering
  \hspace*{-.6cm}
  \includegraphics[scale=.76]{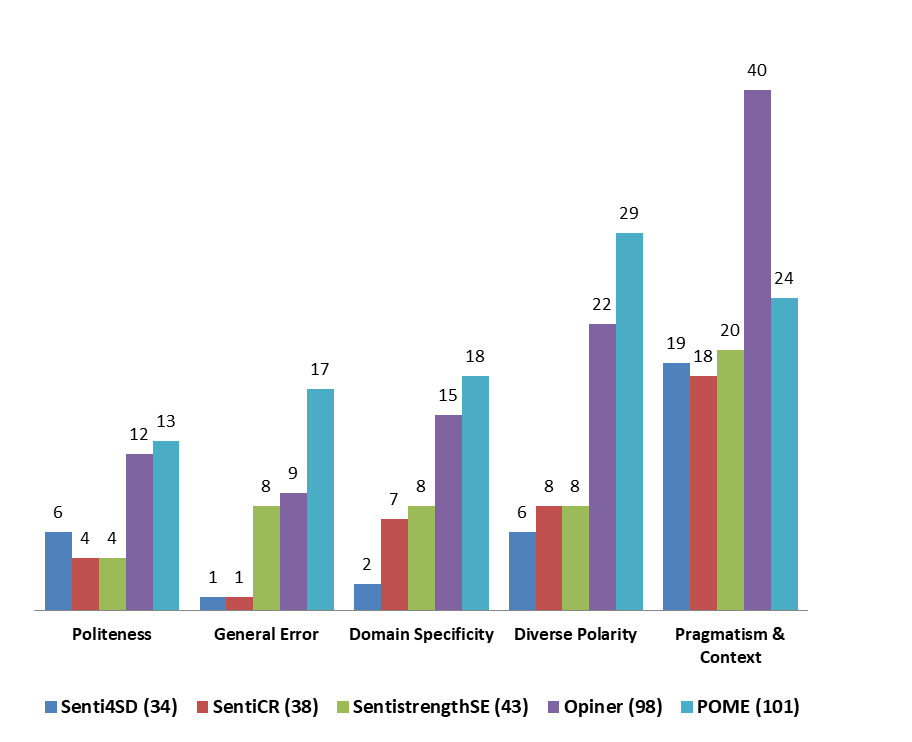}
  \vspace{-0.2in}
  \caption{Error categories in the complementary cases}
  \label{fig:sentimentToolComplementarityErrorCategories}
  \vspace{-.1in}
\end{figure}

\subsubsection{Results} We observed five error categories: three 
from \cite{Novielli-BenchmarkStudySentiSE-MSR2018} (Context/Pragmatics, General Error, and Politeness) and  
two new: Domain-specific polarity and presence of diverse polarity in the textual units. \tbl\ref{tab:misclassificationCategories159Units} and 
\fig\ref{fig:sentimentToolComplementarityErrorCategories} summarize the categories and their distributions in our analysis.
We discuss the categories below.

\begin{inparaenum}
\item \bf{Inability of classifiers to deal with pragmatics or context information.} We observe that the positive/negative polarity signals are mostly implicit 
or expressed through the status of specific event (e.g., the task cannot be completed or the underlying API worked). The two supervised classifiers (Senti4SD and SentiCR) and 
SentistrengthSE performed better to detect such context-specific pragmatics. For example, only Senti4SD and SentiCR 
correctly assigned the following unit as negative in Jira (Ortu et al.~\cite{Ortu-BulliesMoreProductive-MSR2015}), because the developer mentions the absence of certain 
features in their development context, \emt{Maurice I don't have such option or maybe I don't know where it is.} Only Senti4SD, SentiCR, and SentistrengthSE correctly 
labeled the following unit as negative in SO Calefato et al.~\cite{Calefato-Senti4SD-EMSE2017}, because the tools detect that the developer is reporting that a tool 
feature is not working: \emt{I need to pass a regex substitution as a variable: This, of course, doesn't work \ldots}
Opiner shows the most number of misclassifications followed by POME. Both tools are rule-based. 
POME relies on specific linguistic patterns, while Opiner counts the polar words. 
While SentistrengthSE is rule-based, it 
uses domain-specific contextual cues. 

\item \bf{Diverse Polarity.} The tools can get confused when diverse polarized statements are made in a textual unit. For example, all the tools except Senti4SD 
wrongly classified this unit as neutral, \emt{Seeing all the information and activity, it does look very constructive and perhaps 
it shouldn't have been closed so eagerly!} The presence of diversity of polarity in a textual unit can be confusing to the tools for two reasons:
\begin{inparaenum}
\item The tools may consider all the polarized words to determine polarity, while the final polarity of the unit may be decided based on a subset of those words. For example, consider the 
following post which starts with criticism towards a tool, but ends with praises towards the tool, \emt{this app is terrible i am in a good relationship and we are happy together}. 
\item The underlying design of the tool does not allow the processing such diversity. For example, only POME misclassifies the following post, because it does not 
conform to any of its linguistic patterns~\cite{Lin-PatternBasedOpinionMining-ICSE2019}, \emt{Hey, I'm new to this site. I think it is great! Okay, here's the deal. I just downloaded Smule Ocarina. I was wondering how they made it so you can upload a song to the cloud. I might have an app idea that might incorporate this. How would I do this? What would I need?}
\end{inparaenum} 

\item\bf{Domain specific polarity.} The expression of polarity can be unique to the domain of software engineering (e.g., the API is threadsafe.)
or differ from other domains (e.g., the API is simple to use vs the movie is too simple)~\cite{Novielli-BenchmarkStudySentiSE-MSR2018,Lin-SentimentDetectionNegativeResults-ICSE2018,Uddin-OpinionValue-TSE2019}. 
The tools in our study are designed for different SE-domains, making them suitable to complement to each other. Both Senti4SD and SentiCR label the following unit correctly as negative, \emt{alright its okay but it freezes all the time
}.  Both Opiner and Senti4SD algorithms correctly label the negative polarity for the unit, \emt{worried about IE compatibility!!}.

\item \bf{General Error.} The tools also differ from each other in how they process the textual contents and thus can 
misclassify a unit when the contents are not properly processed. For example, POME could not process the emoticons in the following unit while other tools were 
able to, \emt{@BoristheSpider :( :( That is really sad. :(}. Opiner wrongly labeled the following unit as positive, because it could 
not properly determine the presence of negation \emt{so I'm not happy with it.
}. Opiner incorrectly labels the following unit as positive, 
\emt{I also would like to see an answer to this..}. The reason Opiner labels it as positive is due to the presence of `like' in the unit, which it considered 
as positive. However, the presence of `would like to' makes the expression more of an intention than of an opinion.

\item \bf{Politeness.} In two datasets (Jira Ortu et al.~\cite{Ortu-BulliesMoreProductive-MSR2015} and 
SO Calefato et al.~\cite{Calefato-Senti4SD-EMSE2017}), the tools confused over the appropriate 
labeling of a unit that contained politeness. For example, Senti4SD and SentistrengthSE correctly label the following unit as positive, while other tools label it as neutral \emt{Thanks Arvind}. 
\end{inparaenum}

In summary, the supervised detectors show lower error coverage across all the five categories. 
When the misclassification of a unit was due to the presence of diverse polarity, Senti4SD showed lower misclassification than SentiCR. 
However, when the underlying contextual information is 
required, SentiCR had lower misclassification rates. The underlying design of the two supervised classifiers offers some insights into this: 
Senti4SD computes polarity labels based on their presence in the selected parts of the unit, e.g., the first vs last sentiment word 
in a unit may carry different weights. In contrast, SentiCR relies specifically on a preprocessed Bag of Words (BoW) produced from a unit, e.g., 
negations are detected and explicitly annotated in the bag of words. 

\subsection{\texorpdfstring{RQ$_4$}{RQ4} Can a supervised ensemble of detectors perform better than individual detectors?}
\label{subsec:sentisead}

\subsubsection{Motivation} The findings from RQ$_1$ (\sec\ref{subsec:complementarity}) and RQ$_2$ (\sec\ref{subsec:ensemble-majority}) 
show that the SE-specific sentiment detectors can complement each other to correct the 
misclassfications, but an unsupervised majority voting-based detector fails to properly capture the complementarity cases. 
The analysis of misclassifications in RQ$_3$ (\sec\ref{subsec:misclassification-cats}) shows that 
the failure of a tool to properly detect the underlying contextual information caused the most misclassifications. 
We also see that SentiCR offered better support for such cases. 
Therefore, an ensemble of the individual detectors in a supervised setting similar to SentiCR may provide more contextual information to a hybrid engine. 

\begin{table}[t]
  \centering
  \caption{Studied features to design the hybrid classifier}
    \begin{tabular}{p{3cm}p{10cm}}\toprule
    \textbf{Features} & \textbf{Rationale} \\
    \midrule
    Polarity Labels from SE-specific detectors & The supervised classifiers can learn the underlying context and domain-specific 
    pragmatics. 
    The rule-based classifiers can employ domain-specific rules.\\
    \midrule
    Bag of words & The vectorized representation of the bag words offers contextual information\\
    \bottomrule
    \end{tabular}%
  \label{tab:sentiseadStudiedFeatures}%
\end{table}%

\subsubsection{Approach} 

We developed a suite of supervised sentiment detectors using the features summarized in \tbl\ref{tab:sentiseadStudiedFeatures}. These features are divided into two categories:

\begin{itemize} 
\item Polarity labels from individual detectors: for each unit, we have five polarity labels, one from each of the five SE-specific sentiment detectors.

\item Bag of words: for each unit, we produce a bag of words by applying preprocessing steps similar to SentiCR~\cite{Ahmed-SentiCRNIER-ASE2017}, which are as follows:
\begin{itemize}
\item Removal of stop words (as SentiCR, we used stopwords from NLTK and those provided by SentiCR).

\item Identification and annotation of negation. For example, ``This isn't good'' will be preprocessed as ``This is NOT\_good''.

\item Identification and expansion of contractions. For example, ``let's'' will be expanded to ``let us''.

\item Detection of emojis. For example, `\%-(' will be replaced by a placeholder `NegativeSentiment'.
\end{itemize}
\end{itemize}

For our analysis, we used the entire dataset used in RQ$_1$, i.e., all the 17,581 units. Also similar to RQ$_1$, we used a 10-fold cross validation.  
We investigated a total of nine algorithms, both ensemble and non-ensemble. From
the ensemble algorithms, we trained and tested four classifiers:
\begin{inparaenum} \item Random Forest (RF), \item Gradient Boosting Classifier
(GBT), \item Adaboost, and \item XGBoost. \end{inparaenum} Ensemble algorithms,
like bagging and boosting, are designed to combine the predictions of several
base estimators. For example, an RF classifier creates multiple decision-trees
based on random samples of training data and then makes a final decision by
consulting the decisions of the individual trees. Each algorithm is widely used
in machine learning. Models, like RF, are robust against overfitting and they
work well with both categorical and numerical data, which is useful in our case
because, as features, we consider both vectorized textual contents from an input
unit and the sentiment polarity labels about the unit from stand-alone tools. We use the 
Python Sklearn API to experiment with the models. Following the Sklearn documentation, we obtain 
deterministic behavior during model fitting by setting a fixed integer to define the random state of a model. 
For example, we set random\_state = 45 for the Random Forest model\footnote{https://scikit-learn.org/stable/modules/generated/sklearn.ensemble.RandomForestClassifier.html}.  

From the non-ensemble algorithms, we trained and tested five classifiers:
\begin{inparaenum} \item Logistic Regression (family = `Multinomial'), \item
SVM, \item LASSO, \item Na\"{i}ve Bayes (Bernoulli), and \item From neural network
category, the Multi-layer Perceptron (MLP). \end{inparaenum} Algorithms like SVM
are used to design hybrid sentiment detection tools in other domains, like
sentiment analysis in Twitter
data~\cite{Balage-HybridSentimentDetectorTwitter-SemEval2014}. Logistic
Regression, LASSO, MLP, and Na\"{i}ve Bayes are used in almost all kinds of
regression and classification analysis. Na\"{i}ve Bayes is found to offer reasonably
good performance in sentiment detection in other domains (e.g., movie
reviews)~\cite{Wang-BaselinesBigrams-ACL2012}.

Following standard principles of machine learning, for each algorithm, we hyper tuned it on the six benchmark datasets under 
different configurations and picked the configurations that offered the best performance. We use the python scikit-learn GridSearchCV function for our hypertuning, which 
takes a list of configuration parameters and a range a values for each parameter to determine the best parameter values. For each algorithm, we studied all possible 
parameters. For example, we ran Random Forest algorithm 
by varying parameter values, such as number of decision trees, maximum tree depth, minimum number of leafs, etc. Similarly, for GBT we also tried changing the 
learning rate, and so on.

\begin{table}
  \centering
 \caption{Performance comparison of the developed hybrid classifier (Sentisead) with the baselines. 
 Performance analysis for each polarity class is provided in \tbl\ref{tab:ensembleBoW} of Appendix \ref{sec:detailsPerformanceAnalysis}.}
    \begin{tabular}{lrrrr|rrr}\toprule
          & & \multicolumn{3}{c}{\textbf{Macro}} & \multicolumn{3}{c}{\textbf{Micro}} \\
          \cmidrule{3-8}
         & $\kappa$ &{\textbf{F1}} &{\textbf{P}} &{\textbf{R}} &{\textbf{F1}} &{\textbf{P}} &{\textbf{R}}\\
     \midrule
    \textbf{Sentisead$_{RF-B}$} & 0.69 &{0.787} &{0.816} &{0.759} &{0.805} &{0.805} &{0.805} \\
    \bf{Sentisead$_{RF-N}$} & 0.67 & {0.774}  & {0.792} & {0.757} & {0.795} & {0.795} & {0.795}\\

     \midrule
    \multicolumn{8}{c}{\textbf{Macro/Micro Performance Increase of  Sentisead$_{RF-B}$ (i.e., Hybrid with Bag of}} \\
    \multicolumn{8}{c}{\textbf{Words) Over Baseline (SentiSE = SentistrengthSE)}} \\
     \midrule
    \textbf{Majority} & 0.61 & 5.5\% & 1.2\% & 9.4\% & 4.8\% & 4.8\% & 4.8\%\\
    \textbf{Senti4SD} & 0.63 & 4.9\% & 6.2\% & 3.7\% & 4.0\% & 4.0\% & 4.0\% \\
    \textbf{SentiCR} & 0.63 & 4.1\% & 3.2\% & 5.0\% & 3.2\% & 3.2\% & 3.2\%\\
    \textbf{SentiSE} & 0.60 & 8.9\% & 9.4\% & 8.3\% & 7.6\% & 7.6\% & 7.6\% \\
    \textbf{Opiner} & 0.35 & 39.8\% & 42.9\% & 36.5\% & 34.8\% & 34.8\% & 34.8\%\\
    \textbf{POME} & 0.04 & 98.2\% & 80.9\% & 114.4\% & 49.1\% & 49.1\% & 49.1\% \\
    \bottomrule
    \end{tabular}%
  \label{tab:ensembleBoWShort}%
\end{table}
\begin{table}[t]
  \centering
  \caption{Distribution of Misclassification Error Categories by Sentisead$_B$ on Dataset and Categories from \tbl\ref{tab:misclassificationCategories159Units}}
    \begin{tabular}{lrrrrrr}\toprule
    \textbf{Category} & \textbf{Sentisead$_B$} & \textbf{SentiCR} & \textbf{Senti4SD} & \textbf{SentistrengthSE} & \textbf{Opiner} & \textbf{POME} \\
    \midrule
    \bf{Context (62)} & 20\%  & 25\%  & 34\%  & 30\%  & 64\%  & 38\% \\
    \bf{Polarity (35)} & 23\%  & 26\%  & 20\%  & 20\%  & 63\%  & 83\% \\
    \bf{Domain (23)} & 9\%   & 35\%  & 4\%   & 30\%  & 65\%  & 78\% \\
    \bf{General (22)} & 5\%   & 9\%   & 5\%   & 14\%  & 41\%  & 77\% \\
    \bf{Politeness (17)} & 24\%  & 18\%  & 35\%  & 18\%  & 76\%  & 82\% \\
    \bottomrule
    \end{tabular}%
  \label{tab:misclassificationCatCovSentiseadB}%
\end{table}%

\subsubsection{Results} Out of the nine supervised algorithms we trained and tested in our datasets, 
the Random Forest-based classifier performed the best based on both Macro and Micro F1-scores. This finding is not surprising, given 
Random Forest is by design an ensemble algorithm and it is found to be a good predictor in other SE-specific tasks (e.g., defect prediction~\cite{Misrili-IndustrialCaseStudyEnsembleSoftwareDefects-SQJ2011}). 
We name the best performing Random Forest (RF) model in our analysis `Sentisead$_{RF}$'. We ran Sentisead on two settings: 
\begin{itemize}
\item Sentisead$_{RF-B}$: Using all the features described in \tbl\ref{tab:sentiseadStudiedFeatures}.
\item Sentisead$_{RF-N}$: Without using the bag of words in \tbl\ref{tab:sentiseadStudiedFeatures}. That means, this classifier only uses five feature values for a given unit, each corresponding to 
a polarity label for the unit from an individual detector.  
\end{itemize}  In \tbl\ref{tab:ensembleBoWShort}, we compare the performance of Sentisead$_{RF-B}$ and Sentisead$_{RF-N}$ 
against the individual sentiment detectors as well as the 
Majority-All classifier from RQ$_2$ (\sec\ref{subsec:ensemble-majority}). The comparison shows the percentage increase in performance 
in Sentisead for a given metric over a baseline. For example, Sentisead$_{RF-B}$ shows a 5.5\% increase in F1-score Macro over Majority-All 
classifier, which is computed as $\frac{\text{F1 (Sentisead)}_{RF-B} - \text{F1 (Majority-All)}}{\text{F1 (Majority-All)}}$. 

Overall, Sentisead$_{RF-B}$ perfoms better than Sentisead$_{RF-N}$. Therefore, the incorporation of bag of words as additional features in Sentisead$_{RF-B}$ 
improves the performance of sentiment detection in the hybrid engine. The Sentisead$_{RF-B}$ also outperforms the stand-alone sentiment 
detectors, which it combines. The increase in performance varies across the tools and the metrics. For F1-score Macro, Sentisead$_{RF-B}$ 
offers 4.1-98.2\% increase in performance over the individual detectors. For F1-score Micro, the hybrid engine offers 4-49\% increase 
in performance over those detectors. The increase in performance is significant over two unsupervised detectors (Opiner, POME). 
Acroos the supervised and unsupervised individual detectors, Sentisead$_{RF-B}$ offers better increase in performance to detect positive and negative 
polarity instead of neutral polarities. For example, excluding POME, Sentisead$_{RF-B}$ offers 4.5-54.7\% increase in performance to detect 
positive polarity (F1-score) and 7-44\% increase in performance to detect negative polarity. For POME, Sentisead$_{RF-B}$ offers more than 500\% increase 
in performance, due mainly to the fact that POME classifies more than 90\% of units as neutral. POME is designed to 
rely on specific linguistic rules detect polarity aspects (e.g., performance), which when the tool found it was very accurate. However, 
due to its focus on such precision, it also missed correct polarity classes where those patterns were not present.

In \tbl\ref{tab:misclassificationCatCovSentiseadB}, we show the coverage of the misclassification 
categories on the datasets of \tbl\ref{tab:misclassificationCategories159Units} from 
RQ$_3$ (\sec\ref{subsec:misclassification-cats}). Our hybrid engine Sentisead performs the best to reduce the mislassifications related to 
contextual information (20\% in Sentisead$_{RF-B}$ vs 25\% in SentiCR). Given that more than one third of the misclassifications are due to the failure of the tools 
to detect the contextual/pragmatics in the sentiment carrying expression, this particular focus of Sentisead$_{RF-B}$ is thus encouraging. Sentisead$_{RF-B}$ also offers 
performance boost over SentiCR while correcting errors related to `Domain specificity' (9\% vs 35\%). 
Thus, the incorporation of bag of words as features helped Sentisead$_{RF-B}$ to disentagle the contextual information 
during polarity labeling. Overall, Sentisead$_{RF-B}$ shows lower coverage of the misclassifications over the unsupervised (except for `Diverse Polarity' in SentistrengthSE). 
However, for the two misclassfication categories (`Diverse Polarity' and `Domain Specificity') Sentisead still has slightly higher coverage than Senti4SD. For example, 
for units misclassified due to `Diverse Polarity', Senti4SD shows 20\% coverage while Sentisead$_{RF-B}$ shows 23\% coverage. The cases related to `Politeness' are due mainly to 
the heterogeneity in the labeling of the datasets, which we discuss in \sec\ref{sec:discussion}.  
  

\begin{table}[t]
  \centering
  \caption{Features used in Sentisead+ in addition to \tbl\ref{tab:sentiseadStudiedFeatures}}
    \begin{tabular}{p{2.5cm}l}\toprule
    \textbf{Features} & \textbf{Rationale} \\
    \midrule
    Partial polarity & The polarity label may depend more on the first/last sentence of a unit. \\
    \midrule
    Polarity entropy & The information entropy of polarity may reflect diverse polarity. \\
    \bottomrule
    \end{tabular}%
    
  \label{tab:sentiseadPlusStudiedFeatures}%
\end{table}%

\subsection{\texorpdfstring{RQ$_5$}{RQ5} Can more features further improve the performance of the hybrid detector?}
\label{subsec:sentiseadplus}

\subsubsection{Motivation} 

The findings from RQ$_4$ (\sec\ref{subsec:sentisead}) confirm that Sentisead$_{RF-B}$, a supervised mode using Random Forest (RF) as an ensembler of stand-alone sentiment detectors 
can offer improved performance over the individual detectors. A closer look at the corrected misclassifications shows that the improvement occurred mainly on the 
understanding of underlying contextual information (see \tbl\ref{tab:misclassificationCatCovSentiseadB}). This is not surprising, given features used in the hybrid 
engine were the polarity labels from individual detectors and the vectorized representation of units as bag of words. As shown in \tbl\ref{tab:misclassificationCatCovSentiseadB}, 
Sentisead$_{RF-B}$ offers improved performance across the tools, but Senti4SD still slightly outperforms Sentisead$_{RF-B}$ while 
classifying units with diverse polarity and domain specificity. Given that the misclassifications due to the diverse polarity were second most in our manual analysis, 
it would be important to see if Sentisead$_{RF-B}$ (denoted as Sentisead$_{RF}$ hereon) can be further improved to handle those cases\footnote{For brevity, we will refer to Sentisead$_{RF-B}$ as Sentisead$_{RF}$ in the rest of the paper.}.    

\subsubsection{Approach} In addition to the features in \tbl\ref{tab:sentiseadStudiedFeatures} (i.e., polarity labels from SE-tools + Bag of words), we investigate two new 
categories of features: \begin{itemize}[leftmargin=10pt]
\item \it{Partial Polarity.} We label a unit based on the polarity information we find in the first or last clause/sentence of the unit. We decided to use this feature based on the observations 
that sometimes the final polarity label of a given input unit depends more on the first or the last polarity expression in the unit. Similar features were also 
previously used to develop Senti4SD~\cite{Calefato-Senti4SD-EMSE2017}.
\item \it{Polarity Entropy.} We compute entropy of a given unit following principles of information theory. We were motivated to use entropy as a feature, because entropy measures 
were found to be useful in various software engineering classification tasks previously, such as  predict/triage software faults~\cite{Hassan-PredictingFaultsUsingCodeComplexity-ICSE2009, 
Khomh-EntropyEvaluationTriageMozillaCrash-WCRE2011}. Intuitively, entropy can inform us of the underlying diversity of polarity/words in the input unit, which we hoped could be useful to 
make a better decision on the overall polarity label of the unit. 
\end{itemize} The new features are summarized in \tbl\ref{tab:sentiseadPlusStudiedFeatures}. We explain those below. 

\begin{inparaenum}
\item\bf{Partial Polarity.} The relative position of a polarity words in a unit can 
be important to correct a misclassification. 
For example, `I like this tool. But it is slow' could convey a negative polarity even though the first sentence is positive and the second sentence is negative. Given 
as input a unit, we first detect sentences using NLTK sentence detector. We then determine two polarity labels: \begin{inparaenum}
\item First Polarity. We label the unit the polarity of the first sentence.
\item Last Polarity. We label the unit the polarity of the last sentence.
\end{inparaenum}  

\item\bf{Polarity Entropy.} If one tool is wrong but another is right, then there is some \it{uncertainty} in the detection process. 
To quantify such uncertainty, we use \it{Shannon Entropy} from {Information Theory}~\cite{Shannon-MathematicTheoryOfCommunication-BellJournal1948}, 
which was previously used to predict/triage software faults~\cite{Hassan-PredictingFaultsUsingCodeComplexity-ICSE2009, 
Khomh-EntropyEvaluationTriageMozillaCrash-WCRE2011}. {Shannon Entropy} measures 
the information associated with a given input:
\begin{equation}\label{eq:shannon}
H_{n}(P) = -\sum_{k=1}^{n}(p_{k}*\log_{e}{p_k})
\end{equation}where $p_k$ is the probability of a data value in a given unit and $\sum_{i=1}^{k} p_k = 1$. 
Suppose we are monitoring an input text to categorize (e.g., polarity). The text has two words A and B. Using \eq\ref{eq:shannon}, the entropy value of the 
text would be 0.69. However, if the next word of text is again B, the entropy value drops to 0.63 because the uncertainty level 
has decreased: we see more similar contents (i.e., two Bs, one A). For each unit, we compute following metrics.

\begin{inparaenum}
\item\it{Polarity $_{Entropy}$}: We 
identify all the polarity lexicons in the unit and develop a data structure of the 
form $\{(w_1,f_1), ... (w_n, f_n)\}$, where each $w_i$ is a word in the unit that corresponds to a sentiment word and 
$f_i$ corresponds to its occurrence frequency in the unit. We then use \eq\ref{eq:shannon} to compute the entropy. 
For example, with the sentence ``The API is great, but it's slow'', if we only use the occurrence of polarity words in 
the sentence into \eq\ref{eq:shannon}, we have the following data values: \{'great': 1, `slow': 1\}. 
The entropy value would be 0.69. 

\item \it{Adjective $_{Entropy}$ }: We compute the \it{diversity} of all adjectives in a given unit. 
A polarity word may be missing in our sentiment lexicons, but may still be found in the adjectives. 

\item \it{Verb $_{Entropy}$}: We compute the \it{diversity} of all verbs in a unit using \eq\ref{eq:shannon} similarly as we did with 
adjectives. Ko et al.~\cite{Ko-LinguisticAnalysis-VLHCC2005a} observed that developers use verbs to describe software problems 
(e.g., ``This tool does not work''). 

\end{inparaenum} 
\end{inparaenum}

\begin{table}
  \centering
  
  \caption{Performance comparison of the designed Hybrid detector (Sentisead) with added features from \tbl\ref{tab:sentiseadPlusStudiedFeatures}. 
   Performance analysis for each polarity class is provided in \tbl\ref{tab:performanceSentiseadPlus} of Appendix \ref{sec:detailsPerformanceAnalysis}.}
    \begin{tabular}{lrrrr|rrr}\toprule
         &  & \multicolumn{3}{c}{\textbf{Macro}} & \multicolumn{3}{c}{\textbf{Micro}}\\
           \cmidrule{3-8}
          & \bf{$\kappa$} & {\textbf{F1}} & {\textbf{P}} & {\textbf{R}} & {\textbf{F1}} & {\textbf{P}} & {\textbf{R}}\\
          \midrule
      \textbf{Sentisead$_{RF-B}$+} & 0.69 & 0.788 & 0.815 & 0.763 & 0.806 & 0.806 & 0.806 \\
  
    \textbf{Sentisead$_{RF-BNE}$+} & 0.69 & 0.786 & 0.817 & 0.757 & 0.804 & 0.804 & 0.804 \\
    \textbf{Sentisead$_{RF-BNP}$+} & 0.68 & 0.786 & 0.812 & 0.761 & 0.804 & 0.804 & 0.804 \\
    \midrule
    \textbf{SentiSead$_{RF-N}$+} & 0.65 & 0.757 & 0.77  & 0.744 & 0.779 & 0.779 & 0.779 \\
    \textbf{Sentisead$_{RF-NNE}$+} & 0.66 & 0.768 & 0.786 & 0.751 & 0.791 & 0.791 & 0.791\\
    \textbf{Sentisead$_{RF-NNP}$+} & 0.65 & 0.756 & 0.769 & 0.744 & 0.779 & 0.779 & 0.779\\
    \midrule
    \multicolumn{8}{l}{Sentisead$_{RF-B}*$: With bag of words. Sentisead$_{RF-N}*$: No bag of words.}
    \\
    \multicolumn{8}{l}{NP = No partial polarity. NE = No polarity entropy}
    \\
    \bottomrule
    \end{tabular}%
  \label{tab:performanceSentiseadPlusShort}%
  
\end{table}
\begin{table}[t]
  \centering
  \caption{Error categories coverage by Sentisead+ variations}
    \begin{tabular}{lrrr|rrr}\toprule
    \textbf{Category} & \textbf{B+} & \textbf{BNE+} & \textbf{BNP+} & \textbf{N+} & \textbf{NNE+} & \textbf{NNP+} \\
    \midrule
    \textbf{Context (62)} & 18\%  & 21\%  & 18\%  & 26\%  & 25\%  & 26\% \\
    \textbf{Polarity (35)} & 23\%  & 26\%  & 26\%  & 23\%  & 17\%  & 29\% \\
    \textbf{Domain (23)} & 9\%   & 9\%   & 9\%   & 9\%   & 9\%   & 13\% \\
    \textbf{General (22)} & 5\%   & 5\%   & 5\%   & 9\%   & 9\%   & 9\% \\
    \textbf{Politeness (17)} & 24\%  & 24\%  & 24\%  & 24\%  & 24\%  & 24\% \\
    \midrule
    \multicolumn{7}{l}{B+: With bag of words. N+: No bag of words.}\\
    \multicolumn{7}{l}{NP = No partial polarity. NE = No polarity entropy}\\
    \bottomrule
    \end{tabular}%
  \label{tab:errorCatSentiseadPlus}%
\end{table}%

\subsubsection{Results} Sentisead$_{RF-B+}$ offers a very marginal increase in performance over Sentisead$_{RF}$ (0.001 increase in Macro F1-score). Therefore, the inclusion 
of the new features into Sentisead$_{RF}$ does not seem effective to further improve its performance over the stand alone detectors. To further understand the impact of the 
new features, in \tbl\ref{tab:performanceSentiseadPlusShort}, we report the performance of Sentisead$_{RF}$ with the additional features in six different settings:
\begin{enumerate}
\item Sentisead$_{RF-B}+$: with features bag of words + polarity labels from SE-tools + partial polarity + polarity entropy. 
\item Sentisead$_{RF-BNP}$+: with features bag of words + polarity labels from SE-tools + polarity entropy.
\item Sentisead$_{RF-BNE}$+: with features bag of words + polarity labels from SE-tools + partial polarity.
\item Sentisead$_{RF-N}$+:with features polarity labels from SE-tools + partial polarity + polarity entropy.
\item Sentisead$_{RF-NNP}$+:with features polarity labels from SE-tools + polarity entropy.
\item Sentisead$_{RF-NNE}$+:with features polarity labels from SE-tools + partial polarity. 
\end{enumerate} The models with bag of words perform better than those without bag of words as features. Both Sentisead$_{RF-BNP}$+ (i.e., without partial polarity) 
and Sentisead$_{RF-BNE}$+ (i.e., without polarity entropy)
show a Marco F1-score 0.786, but they differ from each other based on precisions and recalls. The precision is higher for Sentisead$_{RF-BNE}$+ 
between the settings, but the recall is higher for Sentisead$_{RF-BNP}$+. Therefore, the two features while combined with bag of words features, were able to slightly 
improve the overall performance than Sentisead$_{RF}$. 

In \tbl\ref{tab:errorCatSentiseadPlus}, we show the distribution of error categories across the six settings. Overall, the inclusion of the new features has reduced the coverage 
of errors due to contextual information from 20\% (Sentisead$_{RF}$) to 18\% (Sentisead$_{RF-B}$+). There is no impact of the new features on the other four error categories. 
A closer look at the settings, however, shows that the partial polarity features were successful to reduce errors related to diverse polarity. For example, Sentisead$_{RF-NNE}$+ shows 
a coverage of 17\% for errors related diverse polarity. This setting is trained using only the polarity labels from the five SE-tools and the partial polarity labels. However, 
those corrections were ignored in Sentisead$_B$+ tool, which shows a coverage of 23\% of errors related to diverse polarity. 
Thus, Sentisead$_{RF-B}$+ gave more preference over bag of words features than partial polarity features while making the final decisions on a polarity label. For example, the 
following unit was correctly labeled as positive by Sentisead$_{RF-NNE}$+ (due to first sentence being positive), but incorrectly labeled as neutral by Sentisead$_{RF-B}$+: 
\emt{Looks excellent.  
Really sweet.  \ldots [4 more sentences] \ldots + Can you add a note on how you've changed how REST works high-level?  Thats all for now.}



\section{Study Phase 2: Ensemble with Deep Learning Models}
\label{sec:resultsDeep}

In a second phase, we answer three more research questions introduced in \sec\ref{sec:intro}:

\begin{enumerate}[start = 6]
\item Can Sentisead outperform the BERT-based advanced pre-trained language based models? (\sec\ref{sec:rq6-bert-stand-alone})

\item Can a deep learning model for Sentisead outperform the BERT-based models? (\sec\ref{sec:rq-deep-learning-sentisead-of-shallow})

\item Can Sentisead based on an ensemble of all the individual models offer the best performance of all tools? (\sec\ref{sec:sentisead-with-all})
\end{enumerate}

\subsection{\texorpdfstring{RQ$_6$}{RQ6} Can Sentisead$_{RF}$ outperform the BERT-based advanced pre-trained language based models?}
\label{sec:rq6-bert-stand-alone}

\subsubsection{Motivation}

As we explained in \sec\ref{sec:intro}, we started our analysis of the feasibility of developing a hybrid tool of SE-specific sentiment detectors before deep learning models for sentiment detection in SE were described. Recent studies show that deep learning models, specially language-based pre-trained transformer models (PTMs) can be trained using SE-specific data to offer performance superior to the traditional shallow learning models~\cite{Zhang-SEBERTSentiment-ICSME2020,Biswas-ReliableSentiSEBERT-ICSME2020}.

The most recent paper on this topic is by Zhang et al.~\cite{Zhang-SEBERTSentiment-ICSME2020}, who reported that models like BERT and RoBERTa can outperform stand-alone rule-based and shallow learning models (e.g., Senti4SD, SentiCR). Therefore, we must analyze whether such PTMs could also outperform Sentisead, i.e., our hybrid of stand-alone rule-based 
and shallow learning sentiment detection tools.

\subsubsection{Approach} 

We investigate all the four PTMs from Zhang et al.~\cite{Zhang-SEBERTSentiment-ICSME2020}, i.e., BERT, RoBERTa, ALBERT, and XLNet. For each of the studied dataset, we train and test the performance of each model using the same 10-fold cross-validation that we used to answer RQ$_1$ (see \sec\ref{sec:approach-rq1-complementarity}). Similar to Zhang et al.~\cite{Zhang-SEBERTSentiment-ICSME2020}, we also fine-tune the hyperparameters for the PTMs. Following Delvin et al.~\cite{Delvin-BERTArch-Arxiv2018}, we investigate with the following hyperparameters: \begin{inparaenum} \item Batch size, \item Number of epochs, and \item Learning rate. \end{inparaenum} For all the PTMs, we find the following parameters after fine-tuning: batch size of 16, 4 epochs, and a learning rate of $2^{-5}$. The values are similar to those reported by Zhang et al.~\cite{Zhang-SEBERTSentiment-ICSME2020}.

\begin{table}
  \centering  \caption{Performance of stand-alone advanced pre-trained advanced language-based deep learning models across the datasets. 
  Performance analysis for each polarity class is provided in \tbl\ref{tab:performance-bert-stand-alone} of Appendix \ref{sec:detailsPerformanceAnalysis}.}
    \begin{tabular}{lrrrr|rrr}
    \toprule
          & & \multicolumn{3}{c}{\textbf{Macro}} & \multicolumn{3}{c}{\textbf{Micro}}\\
          \cmidrule{3-8}
         & \bf{$\kappa$} & {\textbf{F1}} & {\textbf{P}} & {\textbf{R}} & {\textbf{F1}} & {\textbf{P}} & {\textbf{R}}\\
          \midrule
    BERT & 0.70 & {0.793} & {0.793} & {0.792} & {0.81} & {0.81} & {0.81} \\
    Macro/Micro Change Over Sentisead$_{RF}$ & -- & 0.8\% & -2.8\% & 4.3\% & 0.6\% & 0.6\% & 0.6\% \\
    \midrule
    RoBERTa & 0.71 & {0.801} & {0.799} & {0.804} & {0.815} & {0.815} & {0.815} \\
    Macro/Micro Change Over Sentisead$_{RF}$ & -- & 1.8\% & -2.1\% & 5.9\% & 1.2\% & 1.2\% & 1.2\% \\
    \midrule
    ALBERT  & 0.69 & {0.787} & {0.795} & {0.78} & {0.806} & {0.806} & {0.806}\\
    Macro/Micro Change Over Sentisead$_{RF}$ & -- & 0.0\% & -2.6\% & 2.8\% & 0.1\% & 0.1\% & 0.1\% \\
    \midrule
    XLNet & 0.70 & {0.799} & {0.797} & {0.8} & {0.814} & {0.814} & {0.814} \\
    Macro/Micro Change Over Sentisead$_{RF}$ & -- & 1.5\% & -2.3\% & 5.4\% & 1.1\% & 1.1\% & 1.1\% \\
    \bottomrule
    \end{tabular}%
  \label{tab:performance-bert-stand-aloneShort}%
\end{table}%

\begin{table}[ht]
  \centering
  \caption{Distribution of error Categories by advanced pre-trained language-based models on dataset and categories from \tbl\ref{tab:misclassificationCategories159Units}}
    \begin{tabular}{lllll|l}\toprule
    \textbf{Category} & \textbf{BERT} & \textbf{RoBERTa} & \textbf{Albert} & \textbf{XLNet} & \bf{Sentisead$_B$} \\
    \midrule
    \textbf{Context/Pragmatics (62)} & 16\%  & 21\%  & 21\%  & 26\% & 20\% \\
    \textbf{Diverse Polarity (35)} & 11\%  & 9\%   & 17\%  & 17\% & 23\% \\
    \textbf{Domain Specificity (23)} & 9\%   & 9\%   & 13\%  & 13\% & 9\% \\
    \textbf{General Error (22)} & 9\%   & 14\%  & 14\%  & 9\% & 5\% \\
    \textbf{Politeness (17)} & 18\%  & 24\%  & 24\%  & 24\% & 24\% \\
    \bottomrule
    \end{tabular}%
  \label{tab:misclassification-bert-stand-alone}%
\end{table}%

\subsubsection{Results}

In \tbl\ref{tab:performance-bert-stand-aloneShort}, we present the performance of the four stand-alone PTMs across the six benchmark datasets. For each model, we also show how the performance of the model improved/decreased compared to Sentisead$_B$ from RQ$_4$ (i.e., the hybrid of the five rule-based and shallow learning models based on Random Forest).

In terms of Macro F1-score, all the PTMs offer slightly better performance than Sentisead$_B$. The increase in Macro F1-score in the PTMs compared to Sentisead$_B$ is due to the increase in recall but not in precision. RoBERTa shows the best Macro F1-score of 0.801, which outperforms Sentisead$_B$ by 1.8\%. It also shows the best recall of 0.804, which outperforms Sentisead$_B$ by 5.9\%. It also has the best precision of 0.799 among the four PTMs, which is 2.1\% less than the Macro precision of Sentisead$_B$.

A closer look at the class-based performance shows that the drop in precision is 7.1\% in RoBERTa compared to Sentisead$_B$ for both positive and negative polarity classes, while the increase in recall is 13.9\% and 15.3\% for the positive and negative classes, respectively. Thus, with PTMs like RoBERTa, we can expect to find more instances of polarity classes compared to Sentisead$_B$, but that also contributes to a loss of precision. In terms of F1-score, RoBERTa outperforms Sentisead$_B$ for all polarity by 0.5\% (neutral), 3.1\% (positive), and 4.1\% (negative).

\tbl\ref{tab:misclassification-bert-stand-alone} shows the distribution of error categories by the four PTMs in the dataset and categories from \tbl\ref{tab:misclassificationCategories159Units}. Among the four PTMs, BERT has the lowest number of errors related to contexts/pragmatics (16\%) and politeness (18\%), while RoBERTa has the lowest number of errors for two other categories (diverse polarity and domain specificity, both 9\%). Both BERT and XLNet show the lowest number of errors under the general category (9\%). The last column in \tbl\ref{tab:misclassification-bert-stand-alone} shows the distribution of error categories in the same dataset by Sentisead$_B$ (taken from \tbl\ref{tab:misclassificationCatCovSentiseadB}). 

Overall, the best performing PTM (in terms of Macro F1-score), RoBERTa improves on Sentisead$_B$ in this dataset in only one of the five error categories, diverse polarity (23\% in Sentisead$_B$ vs.\ 9\% in RoBERTa). For example, RoBERTa corrects the following misclassification of Sentisead$_B$ from neutral to positive, \emt{Seeing all the information and activity, it does look very constructive and \_perhaps\_ it shouldn't have been closed so eagerly!}. RoBERTa also corrects this misclassification of Sentisead$_B$ from neutral to negative: \emt{The problem is, the application is developed in Java not python} (in Lin et al.~\cite{Lin-SentimentDetectionNegativeResults-ICSE2018} app review dataset).

Therefore, when we can afford to use GPUs that are required to develop the PTMs and when we need
even slight performance improvement like 1.8\% (Macro F1-score) over a hybrid
of shallow and rule-based models in Sentisead$_B$, we could use the PTMs.
Otherwise, Sentisead$_{RF}$ might be sufficient to meet real-world needs for
sentiment detection in SE.

\subsection{\texorpdfstring{RQ$_7$}{RQ7} Can a deep learning model for Sentisead outperform the BERT-based models?}
\label{sec:rq-deep-learning-sentisead-of-shallow}


\subsubsection{Motivation}

As we observed in RQ$_6$, the stand-alone PTMs slightly outperform Sentisead$_B$, i.e., the Random Forest-based hybrid of five rule-based and shallow learning SE-specific sentiment detectors (Opiner, POME, Senti4SD, SentiCR, and SentistrengthSE). The PTMs offer better recall than Sentisead$_B$, but suffers from lower precision. The current ensembler in Sentisead$_B$ is a Random Forest, which is a shallow learning model. Therefore, we investigate whether we could further improve the performance of Sentisead$_B$ by introducing a PTM as ensembler.

\subsubsection{Approach}

Our goal is to replace the Random Forest algorithm in Sentisead$_B$ by the PTMs from RQ$_6$, as follows. For each studied dataset, we do a 10-fold cross validation using a PTM following an approach similar to RQ$_1$ in \sec\ref{sec:approach-rq1-complementarity}. Each iteration takes as input the following features for each unit: bag of words and sentiment polarity labels of the five rule-based and shallow models (i.e., OpinerPOME, Senti4SD, SentiCR, and SentistrengthSE). The target value is the manual label for each unit. Instead of Random Forest as ensembler, we use one of the PTM. Therefore, for each studied datasets, we do a total of $4 \times 10 = 40$ runs, 10 for each of the PTMs and therefore $6 \times 40 = 240$ iterations. We call the new ensemble models Sentisead$_{BERT}$, Sentisead$_{RoBERTa}$, Sentisead$_{ALBERT}$, and Sentisead$_{XLNet}$ when using BERT, RoBERTa, ALBERT, and XLNet, respectively.

\begin{table}[t]
\centering
  \caption{Performance of Sentisead using pre-trained langange based advanced deep learning models as ensembler. 
  Performance analysis for each polarity class is provided in \tbl\ref{tab:performance-sentisead-shallow-ensemble-bert} of Appendix \ref{sec:detailsPerformanceAnalysis}.}
    \begin{tabular}{lrrrr|rrr}\toprule
          & & \multicolumn{3}{c}{\textbf{Macro}} & \multicolumn{3}{c}{\textbf{Micro}}\\
          \cmidrule{3-8}
          &\bf{$\kappa$} & {\textbf{F1}} & {\textbf{P}} & {\textbf{R}} & {\textbf{F1}} & {\textbf{P}} & {\textbf{R}}\\
          \midrule
    Sentisead$_{BERT}$ & 0.69 & {0.791} & {0.791} & {0.79} & {0.807} & {0.807} & {0.807}\\
    Macro/Micro Change Over Sentisead$_{RF}$ & -- & 0.5\% & -3.1\% & 4.1\% & 0.2\% & 0.2\% & 0.2\%\\
    Macro/Micro Change Over BERT & -- & -0.3\% & -0.3\% & -0.3\% & -0.4\% & -0.4\% & -0.4\% \\
          \midrule
    Sentisead$_{RoBERTa}$ & 0.72 & {0.805} & {0.807} & {0.803} & {0.82} & {0.82} & {0.82} \\
    Macro/Micro Change Over Sentisead$_{RF}$ & -- & 2.3\% & -1.1\% & 5.8\% & 1.9\% & 1.9\% & 1.9\%\\
    Macro/Micro Change Over RoBERTa & -- & 0.5\% & 1.0\% & -0.1\% & 0.6\% & 0.6\% & 0.6\% \\
          \midrule
    Sentisead$_{ALBERT}$ & 0.68 & {0.779} & {0.779} & {0.779} & {0.798} & {0.798} & {0.798} \\
    Macro/Micro Change Over Sentisead$_{RF}$ & -- & -1.0\% & -4.5\% & 2.6\% & -0.9\% & -0.9\% & -0.9\%\\
    Macro/Micro Change Over ALBERT & -- & -1.0\% & -2.0\% & -0.1\% & -1.0\% & -1.0\% & -1.0\% \\
          \midrule
    Sentisead$_{XLNet}$ & 0.71 & {0.8} & {0.799} & {0.801} & {0.815} & {0.815} & {0.815}\\
    Macro/Micro Change Over Sentisead$_{RF}$ & -- & 1.7\% & -2.1\% & 5.5\% & 1.2\% & 1.2\% & 1.2\%\\
    Macro/Micro Change Over XLNet & -- & 0.1\% & 0.3\% & 0.1\% & 0.1\% & 0.1\% & 0.1\%\\
	\bottomrule    
	\end{tabular}%
  \label{tab:performance-sentisead-shallow-ensemble-bertShort}%
\end{table}%

\begin{table}[t]
  \centering
  \caption{Error Categories by Sentisead based on advanced pre-trained language-based models on dataset from \tbl\ref{tab:misclassificationCategories159Units}}
    \begin{tabular}{lllll}\toprule
    \textbf{Category} & \textbf{Sentisead$_{BERT}$} & \textbf{Sentisead$_{RoBERTa}$} & \textbf{Sentisead$_{ALBERT}$} & \textbf{Sentisead$_{XLnet}$} \\
    \midrule
    \textbf{Context/Pragmatics (62)} & 21\%  & 23\%  & 33\%  & 30\% \\
    \textbf{Diverse Polarity (35)} & 17\%  & 14\%  & 4\%   & 11\% \\
    \textbf{Domain Specificity (23)} & 9\%   & 4\%   & 17\%  & 4\% \\
    \textbf{General Error (22)} & 9\%   & 9\%   & 14\%  & 9\% \\
    \textbf{Politeness (17)} & 12\%  & 29\%  & 18\%  & 24\% \\
    \bottomrule
    \end{tabular}%
  \label{tab:error-categories-sentisead-shallow-ensemble-bert}%
\end{table}
\subsubsection{Results}

\tbl\ref{tab:performance-sentisead-shallow-ensemble-bertShort} shows the
performance of the four new PTM-based ensemblers Sentisead$_{RoBERTa}$,
Sentisead$_{BERT}$, Sentisead$_{ALBERT}$, and Sentisead$_{XLNet}$. It also
reports how their performances are different from those of Sentisead$_B$ and the
stand-alone PTMs.

In terms of Macro F1-score, Sentisead$_{RoBERTa}$ is the best performer with an
F1-score of 0.805. It outperforms Sentisead$_B$ by 2.3\% and the stand-alone
RoBERTa model by 0.5\%. The second best performer is Sentisead$_{XLNet}$ with a
Macro F1-score of 0.8, which outperforms Sentisead$_B$ by 1.7\% and the
stand-alone XLNet by 0.1\%. The other two ensemblers, Sentisead$_{BERT}$ and
Sentisead$_{ALBERT}$ perform worse than the corresponding stand-alone models
(i.e., BERT and ALBERT, respectively).

Similar to the stand-alone PTM, RoBERTa, the increase in F1-score of
Sentisead$_{RoBERTa}$ over Sentisead$_B$ is due to a higher recall. However,
Sentisead$_{RoBERTa}$ does have better precision than RoBERTa: the incorporation
of polarity labels from the five rule-based and shallow learning models has
reduced the loss of precision from Sentisead$_B$ from 7.1\% (in RoBERTa) to
6.2\% in Sentisead$_{RoBERTa}$.

\tbl\ref{tab:error-categories-sentisead-shallow-ensemble-bert} reports the
distribution of error categories per ensembler PTMs on the dataset and
categories from \tbl\ref{tab:misclassificationCategories159Units}. The best
Sentisead$_{RoBERTa}$ shows a reduction from stand-alone RoBERTa in
misclassifications for two error categories (domain specificity from 9\% to 4\%
and general error from 14\% to 9\%). For the other three error categories, the
performance of Sentisead$_{RoBERTa}$ is actually lower than the stand-alone
RoBERTa mode. For example, for errors related to lack of context/pragmatics, Sentisead$_{RoBERTa}$ shows the error in 23\% of all cases. In 
contrast, stand-alone RoBERTa shows the error in 21\% of all such cases. This means that the ensemble model was slightly more 
confused to correct such errors compared to the stand-alone model. This confusion could arise in the ensembler from the polarity labeling 
of stand-alone tools, which themselves can be confused with the exact polarity labeling of a sentence. This can happen for different reasons like 
the underlying context was implicit, or it was not annotated properly during the manual labeling. Indeed, subjectivity in annotation 
can be a problem with SE-specific datasets (see Novielli et al.~\cite{Novielli-BenchmarkStudySentiSE-MSR2018,Uddin-OpinionValue-TSE2019}).
A detailed understanding of the exact reasons of performance drop in a deep learning ensembler compared to the stand-alone PTMs 
requires the analysis of the PTMs and the ensemblers based on the principles of explainable/interpretable machine learning, which 
we leave as our future work. Nevertheless, it is arguable whether a performance improvement of
0.5\% over stand-alone PTM does warrant a PTM-based ensembler over a stand-alone
PTM.

\subsection{\texorpdfstring{RQ$_8$}{RQ8} Can Sentisead based on an ensemble of all the individual models offer the best performance?}
\label{sec:sentisead-with-all}

\subsubsection{Motivation}

Our findings from RQ$_6$ show that stand-alone PTMs, like RoBERTa, can outperform Sentisead$_B$ based on shallow learning. Our findings from RQ$_7$ show that an ensemble of the five rule-based and shallow learning models based on PTMs, like Sentisead$_{RoBERTa}$, can offer 0.5\% increase in Macro F1-score over the stand-alone RoBERTa model. Given the superiority of the stand-alone PTM over Sentisead$_B$, we want to learn whether an ensemble of all the available stand-alone models (i.e., rule-based, shallow, and PTMs) could offer the best performance of all the available ensemblers and stand-alone models.

\subsubsection{Approach}

We replicate the approach of RQ$_7$ for each dataset as follows. We perform a 10-fold cross validation similar to the approach in \sec\ref{sec:approach-rq1-complementarity}. For each unit in a dataset, we consider the following items as input features: textual contents of the unit as vectors in word embedding, sentiment polarity labels from the five rule-based and shallow learning models and the four PTMs. The target is the sentiment polarity label based on manual label. For the ensembler, we use one PTM at a time. As such, similar to RQ$_7$, for each PTM as an ensembler, we run 10 iterations per dataset. We report the performance of the ensemblers on the test dataset. The ensembler model is named after each PTM as follows: Sentisead$_{BERT}+$, Sentisead$_{RoBERTa}+$, Sentisead$_{ALBERT}+$, and Sentisead$_{ALBERT}+$.

\begin{table}[t]
  \centering
  \caption{Performance of Sentisead using pre-trained advanced deep learning models as ensembler and stand-alone detectors. 
  Performance analysis for each polarity class is provided in \tbl\ref{tab:performance-sentisead-shallow-ensemble-bert-plus} of Appendix \ref{sec:detailsPerformanceAnalysis}.}
    \begin{tabular}{lrrrr|rrr}\toprule
          & & \multicolumn{3}{c}{\textbf{Macro}} & \multicolumn{3}{c}{\textbf{Micro}} \\
          \cmidrule{3-8}
          & \bf{$\kappa$} & {\textbf{F1}} & {\textbf{P}} & {\textbf{R}} & {\textbf{F1}} & {\textbf{P}} & {\textbf{R}}\\
          \midrule
    Sentisead$_{BERT}$+ & 0.69 & {0.79} & {0.79} & {0.79} & {0.807} & {0.807} & {0.807}\\
    Macro/Micro Change Over Sentisead$_{RF}$ & -- & 0.4\% & -3.2\% & 4.1\% & 0.2\% & 0.2\% & 0.2\% \\
    Macro/Micro Change Over Sentisead$_{BERT}$ & -- & -0.1\% & -0.1\% & 0.0\% & 0.0\% & 0.0\% & 0.0\%\\
    Macro/Micro Change Over BERT & -- & -0.4\% & -0.4\% & -0.3\% & -0.4\% & -0.4\% & -0.4\% \\
          \midrule
    Sentisead$_{RoBERTa}$+ & 0.71 & {0.8} & {0.797} & {0.802} & {0.814} & {0.814} & {0.814}\\
    Macro/Micro Change Over Sentisead$_{RF}$ & -- & 1.7\% & -2.3\% & 5.7\% & 1.1\% & 1.1\% & 1.1\% \\
    Macro/Micro Change Over Sentisead$_{RoBERTa}$ & -- & -0.6\% & -1.2\% & -0.1\% & -0.7\% & -0.7\% & -0.7\% \\
    Macro/Micro Change Over RoBERTa & -- & -0.1\% & -0.3\% & -0.2\% & -0.1\% & -0.1\% & -0.1\%\\
          \midrule
    Sentisead$_{ALBERT}$+ & 0.68 & {0.783} & {0.784} & {0.782} & {0.801} & {0.801} & {0.801} \\
    Macro/Micro Change Over Sentisead$_{RF}$ & -- & -0.5\% & -3.9\% & 3.0\% & -0.5\% & -0.5\% & -0.5\% \\
    Macro/Micro Change Over Sentisead$_{ALBERT}$ & -- & 0.5\% & 0.6\% & 0.4\% & 0.4\% & 0.4\% & 0.4\%\\
    Macro/Micro Change Over Albert & -- & -0.5\% & -1.4\% & 0.3\% & -0.6\% & -0.6\% & -0.6\% \\
          \midrule
    Sentisead$_{XLNet}$+ & 0.70 & {0.797} & {0.795} & {0.799} & {0.811} & {0.811} & {0.811} \\
    Macro/Micro Change Over Sentisead$_{RF}$ & -- & 1.3\% & -2.6\% & 5.3\% & 0.7\% & 0.7\% & 0.7\%\\
    Macro/Micro Change Over Sentisead$_{XLNet}$ & -- & -0.4\% & -0.5\% & -0.2\% & -0.5\% & -0.5\% & -0.5\% \\
    Macro/Micro Change Over XLNet & -- & -0.3\% & -0.3\% & -0.1\% & -0.4\% & -0.4\% & -0.4\%\\
	\bottomrule
    \end{tabular}%
  \label{tab:performance-sentisead-shallow-ensemble-bert-plusShort}%
\end{table}%

\begin{table}[t]
  \centering
  \caption{Error Categories by Sentisead based on ensemble of all models and dataset from \tbl\ref{tab:misclassificationCategories159Units}}
    \begin{tabular}{lllll}\toprule
    \textbf{Category} & \textbf{Sentisead$_{BERT}$+} & \textbf{Sentisead$_{RoBERTa}$+} & \textbf{Sentisead$_{ALBERT}$+} & \textbf{Sentisead$_{XLNet}$+} \\
    \midrule
    \textbf{Context/Pragmatics (62)} & 21\%  & 26\%  & 25\%  & 25\% \\
    \textbf{Diverse Polarity (35)} & 14\%  & 14\%  & 14\%  & 11\% \\
    \textbf{Domain Specificity (23)} & 9\%   & 4\%   & 9\%   & 0\% \\
    \textbf{General Error (22)} & 5\%   & 14\%  & 9\%   & 14\% \\
    \textbf{Politeness (17)} & 18\%  & 18\%  & 24\%  & 24\% \\
    \bottomrule
    \end{tabular}%
  \label{tab:error-categories-sentisead-shallow-ensemble-bert-plus}%
\end{table}%
\subsubsection{Results}

\tbl\ref{tab:performance-sentisead-shallow-ensemble-bert-plusShort} reports the performance of the four new PTM-based ensemblers that produce the sentiment polarity label for a unit by taking as input the unit and the polarity labels from eight stand-alone sentiment detectors for SE (rule-based, shallow-learning, and four PTMs). For each model, like Sentisead$_{BERT}+$, we also compare how its performance has increased/decreased against previous tools like Sentisead$_B$, stand-alone PTM (i.e, BERT), and the ensembler from RQ$_7$ (i.e., BERT-based ensembler of five rule-based and shallow learning models).

The best performing model is Sentisead$_{RoBERTa}+$, with a Macro F1-score of 0.8. This F1-score is lower than the F1-score of Sentisead$_{RoBERTa}$, which is 0.805. The inclusion of polarity labels from the PTMs has not increased the performance of Sentisead$_{RoBERTa}+$. However, Sentisead$_{RoBERTa}+$ does show an increase of 1.7\% in Macro F1-score over Sentisead$_B$. Overall, among the four ensemblers in this RQ, only Sentisead$_{ALBERT}+$ shows a Macro F1-score lower than Sentisead$_B$. These findings are consistent with those from RQ$_7$ and RQ$_6$ and recent results from Zhang et al.~\cite{Zhang-SEBERTSentiment-ICSME2020}, which show that deep learning models offer better performance than rule-based and shallow learning models for SE-specific sentiment detection.

However, as shown in RQ$_6$ to RQ$_7$, the increase in performance is at most 2.3\% over the shallow learning-based ensembler (which we observed in Sentisead$_{RoBERTa}$). In addition, the lack of performance improvement in PTM-based ensembler of all stand-alone models in this RQ can serve as a guidance that a hybrid tool cannot outperform other tools by simply adding polarity labels from all available stand-alone SE-specific sentiment detectors.   

\section{Discussions}\label{sec:discussion}
In this section, we discuss major themes we observed during the analysis of our study results. 

\subsection{The Underwhelming Roles of Polarity Entropy Metrics in Sentiment Labeling} 
The findings from RQ$_5$ suggest that the three polarity entropy metrics were not effective to further improve the performance 
of our hybrid sentiment detection engine. To determine the root cause of this, we assessed the impact of the diversity metrics on the misclassification of 
the polarity labels from the three tools, Senti4SD, SentiCR and SentistrengthSE. The three tools are the three most dominant features in Sentisead. 
For each tool, the response variable is ``Misclassification'', i.e., for a given unit whether the tool output is wrong or not. 
The explanatory variables are the four diversity metrics above, i.e., entropy measurements using polarity, adjective, and verb. 
To fit and interpret the models, we follow standard practices in the literature~\cite{Patanamon-ReviewParticipationInCodeReview-JournalEMSE2017,Valiev-EcosystemLevelSurvivalPyPI-FSE2018}: 
\begin{inparaenum}
\item When fitting the models, we test for multicollinearity between the explanatory variables using the 
Variable Inflation Factor (VIF), and remove variables with VIF scores above the recommended maximum of 5~\cite{Cohen-AppliedMultipleRegression-Lawrence2002}.
\item When interpreting the models, we consider coefficient importance if they are statistically 
significant ($p-value \le 0.05$). We also estimate their effect sizes based on ANOVA type-2 analyses (column ``LR-Chisq'' of \tbl\ref{tab:regressionResultDiversityCaseStudy}). 
\item We report the goodness of fit using the McFadden's pseudo $R^2$. 
\end{inparaenum} In \tbl\ref{tab:regressionResultDiversityCaseStudy}, we show the results
of regression models of each of the three tools for the inputs for which the
tool was wrong but at least one of the other two tools was right, i.e., the
wrong tool could have benefited from the correct classifications from other
tools. The $Polarity_{Entropy}$ is statistically significant for each
tool, i.e., the misclassification of each tool increases with the increase in
the diversity of polarity words. The $Adjective_{Entropy}$ is significant for SentiCR and $Verb_{Entropy}$ is significant for SentistrengthSE. However, the $R^2$ is only between 3.06-3.86\%. That means that the entropy metrics can only capture around 3-4\% of variations in the data while correcting a tool. This low impact can become negligible in the presence of more dominant features, e.g., bag of words.

\begin{table*}[ht]
  \centering
  \caption{Regression models for the sentiment detectors on the misclassified units where at least one detector was right}
  \resizebox{\columnwidth}{!}{%
    \begin{tabular}{llll|lll|lll}\toprule
          & \multicolumn{3}{c}{\textbf{Senti4SD}} & \multicolumn{3}{c}{\textbf{SentiCR}} & \multicolumn{3}{c}{\textbf{SentistrengthSE}} \\
          & \multicolumn{3}{c}{{Pseudo $R^2$ = 3.06\%}}         & \multicolumn{3}{c}{{Pseudo $R^2$ = 3.86\%}}   & \multicolumn{3}{c}{{Pseudo $R^2$ = 3.74\%}}  \\
          \cmidrule{2-10}
          & \textbf{Coeffs} & \textbf{Std. Err} & \textbf{LR Chisq} & \textbf{Coeffs} & \textbf{Std. Err} & \textbf{LR Chisq} & \textbf{Coeffs} & \textbf{Std. Err} & \textbf{LR Chisq} \\
          \midrule
    (Intercept) & 2.66***  & 0.27  &       & -2.77*** & 0.27  &       & -1.64*** & 0.22  &  \\
    Polarity~$_{Entropy}$ & 0.45***  & 0.11  & 25.72*** & 0.52***  & 0.11  & 36.15*** & 0.64***  & 0.10  & 75.16*** \\
    Adjective~$_{Entropy}$ & 0.01  & 0.11  & 0.01  & 0.23*  & 0.11  & 4.57*  & 0.01  & 0.10  & 0.00 \\
    Verb~$_{Entropy}$ & 0.01  & 0.12  & 0.11  & -0.17 & 0.12  & 1.50  & 0.22*  & 0.10  & 6.07* \\
    \bottomrule
    \multicolumn{10}{l}{Signif. codes:  ***$p < 0.001$, **$p < 0.01$, *$p < 0.05$}
    \end{tabular}%
    }
  \label{tab:regressionResultDiversityCaseStudy}%
\end{table*}%

\subsection{\it{Issues} with Politeness in Sentiment Labeling} 17 of the 153 manually analyzed misclassified units in our complementary dataset (\sec\ref{subsec:misclassification-cats}) 
were related to the confusions among the sentiment tools to handle politeness. While around 76\% of those were corrected by Sentisead, the the rest 24\% remained misclassified across 
all the different settings of Sentisead and Sentisead+ (see \tbl\ref{tab:misclassificationCatCovSentiseadB}). There are two main reasons for this. First, the presence of politeness 
does not necessarily make a unit positive. Second, the manual annotation of units carrying politeness were judged differently by different human raters in the benchmarks.   
For example, the following is labeled as neutral in the Jira Ortu et al.~\cite{Ortu-BulliesMoreProductive-MSR2015} 
dataset, \emt{Jacopo  Thanks for diving in a pointing Scott at the appropriate screens.}. However, the following is labeled as positive in the same dataset \emt{Patch applied!  Thanks.}. 
This discrepancy can confuse a supervised classifier, including Sentisead. 

\subsection{The Problems with Contextual Inference Using \it{Surrounding} Sentences in Sentiment Detection} Among the six benchmark datasets in our study, Sentisead$_B$ achieves 0.83-0.98 F1-score to label both positive and 
negative classes for four datasets: Jira and App reviews from Lin et al.~\cite{Lin-SentimentDetectionNegativeResults-ICSE2018},  SO Calefato et al.~\cite{Calefato-Senti4SD-EMSE2017}, 
and Jira Ortu et al.~\cite{Ortu-BulliesMoreProductive-MSR2015}. Sentisead shows only 0.31-0.47 F1-scores for the positive and negative 
classes in other two datasets: SO Lin et al.~\cite{Lin-SentimentDetectionNegativeResults-ICSE2018} and SO Uddin et al.~\cite{Uddin-OpinionValue-TSE2019}, 
with the worst in SO Uddin et al.~\cite{Uddin-OpinionValue-TSE2019}. Both of these use data from Stack Overflow. The SO Uddin et al.~\cite{Uddin-OpinionValue-TSE2019} dataset 
is different from other benchmarks, because it is based on all the sentences from 1338 posts from Stack Overflow. Thus, when the raters were presented those sentences, they 
also saw other sentences immediately after and before the sentences. This \it{surrounding} contextual information influenced their annotation. This particular issue was also 
discussed in Uddin et al.~\cite{Uddin-OpinionValue-TSE2019}. The SO Lin et al.~\cite{Lin-SentimentDetectionNegativeResults-ICSE2018} benchmark seems to have the same problem. For example, the following sentence 
is rated as `negative' in the dataset \emt{I am not sure about Map implementation in HazelCast.} No sentiment detection tool is designed to check such domain specificity and \it{surrounding} sentences as contexts during 
the polarity classification. Therefore, further improvement in Sentisead for these two datasets will need the development of algorithms to handle such domain specificity 
and contextual information.

\subsection{Run-time Performance}
In \sec\ref{sec:intro} (Introduction), we write about different use cases for Sentisead: Sentisead is
useful for situations where we do not want to pick different tools for different
datasets. Instead, we simply run each stand-alone tool for a given dataset and
let Sentisead make the final decision on sentiment polarity. 

An analysis can be done offline or online. For example, when analyzing millions of 
code reviews from a large software system like OpenStack, 
we expect that the sentiment detection would be done offline and
not in real-time. Nevertheless, we were able to run all stand-alone tools within reasonable
time e.g, SentiCR for any dataset for one iteration (i.e., one of the
10-folds) took around 2 minutes (training + testing). Note that this was done on
a Laptop machine with only 2 cores and 4GB RAM. During testing, for a given
sentence, SentiCR~\cite{Ahmed-SentiCRNIER-ASE2017} was able to produce a polarity 
label in less than a second. Please note
that like any machine learning model, the training of a tool takes the most
amount of time (often around 95\%). This is true for any currently available
SE-specific sentiment detection tool. With advances in computing infrastructure,
parallel processing, and high-performance computing, we can expect to
further reduce such training and testing time. Indeed, we had some performance 
issues with Senti4SD~\cite{Calefato-Senti4SD-EMSE2017} initially, but
Senti4SD was recently updated to a new and faster version to support large-scale
data. Therefore, while the training
of such a tool takes the majority of time in Sentisead, the tool can make a polarity label on
a given unit at run-time in a short time (as noted above with SentiCR example).
Indeed, Sentisead can determine the polarity label of a given unit in less than
a second, given it is built on SentiCR - which as noted above is fast to produce
labels. Note that, we previously analyzed large number of OpenStack code reviews in a reasonably short time using some of the 
the studied tools in this paper
(e.g., Senti4SD, SentiCR) (please see Asri et al.~\cite{Ikram-SentimentCodeReview-IST2019}).

\subsection{The Role of Features in Sentisead Accuracy Improvement}
An important area of research in machine learning is the handling of overfitting of a machine learning model. 
Indeed, in machine learning literature, it is a standard practice to prune the
feature space of a model to improve model inference runtime. Models like Random
Forest are designed to handle overfitting efficiently by automatically pruning
the decision trees. In RQ$_4$ (\sec\ref{subsec:sentisead}), we find that a Random Forest-based hybrid
tool is the best performer in each individual dataset. In RQ$_5$ (\sec\ref{subsec:sentiseadplus}), 
we find that by simply adding extra features (e.g., partial polarity, polarity entropy),
we could not gain notable performance improvement in the new hybrid tool in RQ$_5$ (i.e.,
Sentisead$_B$+) over the the hybrid tool from RQ$_4$ (i.e, Sentisead$_B$). In RQ$_8$, 
we attempted to develop a hybrid Sentisead engine
by taking polarity labels of all 9 stand-alone sentiment detectors (5 rule-based
and shallow learning and 4 BERT-based PTMs from Zhang et al.~\cite{Zhang-SEBERTSentiment-ICSME2020}). We found
that this Sentisead engine performed a bit worse than the Sentisead engine that
only took polarity labels from 5 rule-based and shallow learning sentiment
detectors. These three observations offer insights that by simply adding
new features, we cannot not improve the performance of our hybrid sentiment detection
tool.

\subsection{Best Performers Per Dataset}\label{sec:bestPerformerPerDataset}
When we analyzed the results of each tool, we analyzed along two dimensions: Per
dataset and across all datasets. We provide a detailed explanation of how we
trained and tested each tool per dataset in \sec\ref{sec:approach-rq1-complementarity}. We also explain in
\sec\ref{sec:approach-rq1-complementarity}, how we combine the test results from each dataset to report an
overall performance score of a tool across the six datasets.

An overall performance score across the six datasets informs us of the overall effectiveness of 
a tool across the six datasets. This score is important given that these datasets are designed for different purposes. 
As noted in \sec\ref{sec:intro}, the major motivation behind designing our 
hybrid tool Sentisead was to obtain better results than the stand-alone sentiment detectors on diverse datasets. 
This overall performance score tells us whether the tool does indeed outperform each stand-alone tool across the datasets using a single metric.

Given that we studied six benchmark datasets, nine stand-alone sentiment
detection tools, various settings of our hybrid tool Sentisead by answering
eight research questions in the revised manuscript, it is a non-trivial to
report the performance of each studied tool and the hybrid tool for each
research question per studied dataset: doing so would be lengthy and difficult
to read. Nevertheless, we share the results of each tool per dataset and per
research question in our online replication package~\cite{website:senisead-online-appendix-ase2020}. In \tbl\ref{tab:bestPerformerPerDataset}, we
report the top three tools in each dataset of our study based in Macro F1 score.
\begin{table}[h]
  \centering
  \caption{The three best performing tools per dataset in our study (F1 = Macro F1 score)}
    \begin{tabular}{lllllll}\toprule
    \multicolumn{1}{c}{\multirow{2}[0]{*}{\textbf{Dataset}}} & \multicolumn{2}{c}{\textbf{First}} & \multicolumn{2}{c}{\textbf{Second}} & \multicolumn{2}{c}{\textbf{Third}} \\
    \cmidrule{2-7}
          & \textbf{Tool} & \textbf{F1} & \textbf{Tool} & \textbf{F1} & \textbf{Tool} & \textbf{F1} \\
          \midrule
    JIRA Lin et al.~\cite{Lin-SentimentDetectionNegativeResults-ICSE2018} & SentistrengthSE & 0.98  & Sentisead$_{BERT}$ & 0.97  & RoBERTa & 0.97 \\
    SO Uddin et al.~\cite{Uddin-OpinionValue-TSE2019} & Sentisead$_{XLNet}$ & 0.59  & Sentisead$_{XLNet+}$ & 0.59  & Sentisead$_{RoBERTa}$ & 0.59 \\
    Mobile App Lin et al.~\cite{Lin-SentimentDetectionNegativeResults-ICSE2018} & BERT  & 0.72  & Sentisead$_{RoBERTa}$ & 0.68  & RoBERTa & 0.62 \\
    SO Lin et al.~\cite{Lin-SentimentDetectionNegativeResults-ICSE2018} & Sentisead$_{RoBERTa+}$ & 0.78  & RoBERTa & 0.77  & Sentisead$_{XLNert+}$ & 0.74 \\
    SO Calefato et al.~\cite{Calefato-Senti4SD-EMSE2017} & Sentisead$_{RoBERTa}$ & 0.89  & Sentisead$_{XLNet}$ & 0.88  & RoBERTa & 0.88 \\
    JIRA Ortu et al..~\cite{Ortu-BulliesMoreProductive-MSR2015} & XLNet & 0.83  & Sentisead$_{RoBERTa}$ & 0.82  & Sentisead$_{RoBERTa+}$ & 0.82 \\
    \bottomrule
    \end{tabular}%
  \label{tab:bestPerformerPerDataset}%
\end{table}%

Around 61\% (i.e., 11 out of the 18) places of the best performing tools in \tbl\ref{tab:bestPerformerPerDataset} are occupied by the 
different ensemble combinations of Sentisead (e.g., Sentisead$_{RoBERTa}$), while 33\% of the tools (i.e., 6 out of 18) are  
the stand-alone PTMs, and the rest (i.e., 1 out of 18) of the tools is SentistrengthSE. Among the 61\% places occupied by Sentisead 
variations, all datasets contain different combinations of the PTMs as the ensembler. For the one dataset (JIRA Lin et al.~\cite{Lin-SentimentDetectionNegativeResults-ICSE2018}), 
where SentistrengthSE is the best performer with a Macro F1-score of 0.98, the second best are two tools with an F1-score of 0.97 (Sentisead$_BERT$ and RoBERTa). 
For the other five datasets, the topmost performer is Sentisead-based in three SO-related datasets: SO Uddin et al.~\cite{Uddin-OpinionValue-TSE2019},  
SO Lin et al.~\cite{Lin-SentimentDetectionNegativeResults-ICSE2018}, and SO Calefato et al.~\cite{Calefato-Senti4SD-EMSE2017}. 
For the other two datasets, stand-alone PTMs are the best performers: BERT for Mobile App Lin et al.~\cite{Lin-SentimentDetectionNegativeResults-ICSE2018} and XLNet for 
JIRA Ortu et al..~\cite{Ortu-BulliesMoreProductive-MSR2015}. The findings denote that PTMs as stand-alone tool or as an ensembler of existing tools 
can be the best candidates to detect sentiments in diverse SE-specific datasets.

\section{Threats to Validity}\label{sec:threats}
We discuss the threats to validity of our studies following common guidelines
for empirical studies~\cite{Woh00}.

\bf{Construct Validity Threats} concern the relation between theory and
observations. In our study, they could be due to measurement errors. For our manual labeling of error categories, we followed the coding guide and 
examples of Novielli et al.~\cite{Novielli-BenchmarkDatasetSentiSE-MSR2018}. This is not first time we did the manual labeling of such error categories. In our previously published 
paper~\cite{Uddin-OpinionValue-TSE2019}, we did similar exercise on a different dataset. Nevertheless, similar to Novieli et al.~\cite{Novielli-BenchmarkDatasetSentiSE-MSR2018}, we found 
three error categories: Contextual information, Politeness, and General Errors. In addition, we found two new error categories: Domain-specificity and presence of diverse 
polarity. We provide examples of each category in \sec\ref{sec:results} (RQ$_3$). In addition, we share the manually labeled dataset in our online appendix~\cite{website:senisead-online-appendix-ase2020}.

\bf{Internal validity} warrants the presence of systematic error (e.g., bias) in the study. This can happen due to the reporting of a performance metric that may not be standard or the 
due to the error in the computation of the standard metric. We mitigated the first bias by reporting standard metrics used in the literature. For example, we compute 
both Macro and Micro scores, following standard reporting practices in sentiment analysis for software engineering. We mitigated the second 
concern by confirming the computation of the metrics against the same metrics that are presently computed in widely used open source libraries (e.g., scikit-learn~\cite{website:scikitlearn}).  
In addition, we also report the details of how each metric parameters are computed (e.g., true positives, false positives, etc.).

\bf{External validity} concerns the \it{generalizability}
of the results presented in this paper, which are based on six publicly-available sentiment benchmarks. 
Together, these benchmarks consist of more than 17K units (sentences, posts, etc.), each manually annotated for polarity by multiple coders. 
Therefore, our evaluation corpus is bigger than any previous research on sentiment detection for software engineering. 

\bf{Reliability} concerns the possibility of replicating this evaluation. We provide the necessary details in an online appendix~\cite{website:senisead-online-appendix-ase2020}. 
\section{Related Work}
\label{sec:related-work}

Sentiment analysis in SE research has focused on two main areas: studying sentiments SE repositories/scenarios (\sec\ref{sec:related-work-study}) 
and developing tools to improve sentiment detection in SE (\sec\ref{sec:related-work-tool}).

\subsection{Studying Sentiments SE Repositories/Scenarios} \label{sec:related-work-study}
A significant number of studies are conducted to analyze the prevalence and
impact of sentiments in SE repositories and development scenarios, such as IT
tickets~\cite{Castaldi-SentimentAnalysisITSupport-MSR2016}, collaborative
distributed team communications~\cite{Guzman-SentimentAnalysisGithub-MSR2014},
twitter feeds~\cite{Guzman-NeedleInHaystackTwitterSoftware-RE2016},  commit
logs~\cite{Sinha-DeveloperSentimentCommitLog-MSR2016}, security-related
discussions in GitHub~\cite{Pletea-SecurityEmotionSE-MSR2014}, API
reviews from developer forums by resolving API mentions and by associating opinions to the API 
mentions~\cite{Uddin-OpinionValue-TSE2019,Uddin-OpinionSurvey-TSE2019,Uddin-OpinerReviewAlgo-ASE2017,Uddin-OpinerReviewToolDemo-ASE2017,Uddin-OpinerAPIUsageScenario-TOSEM2020,Uddin-OpinerAPIUsageScenario-IST2020,Uddin-APIAspectMining-TechReport2017,Uddin-ResolvingAPIMentions-TechReport2015a},
and code reviews~\cite{Ikram-SentimentCodeReview-IST2019}. Studies also combined
sentiment analysis with other techniques to improve domain-specific
classification, such as to classify app
reviews~\cite{Maalej-AutomaticClassificationAppReviews-RE2016,Panichella-ClassifyAppUserReview-ICSME2015}.

In parallel, a number of research efforts analyze emotions in SE repositories.
While sentiment analysis focuses on polarity (positive, negative, neutral),
emotion analysis focuses finer grained expressions such as anger, love, etc.
Emotion analysis is used to determine team cohesions, such as the impact of
bullies in an SE team~\cite{Ortu-AreBulliesMoreProductive-MSR205}, the relation
of VAD (Valence, Arousal, Dominance) scores~\cite{Warriner-VadLexicons-BRM2013}
with productivity and burn-out in SE
teams~\cite{Mika-MiningValenceBurnout-MSR2016}, and the usage of expressed
emotions to prioritize improvements or raise team
awareness~\cite{Guzman-EmotionalAwareness-FSE2013,Gachechiladze-AngerDirectionCollaborativeSE-ICSENIER2017}.
Overall, analysis of sentiments and emotions is an increasingly popular field in
SE~\cite{Novielli-IntroToSpIssueAffectSE-JSS2019,Novielli-SentimentEmotionSE-IEEESoftware2019}

Contrary to the above work, in this paper, we focus on the development of a sentiment tool for SE.

\subsection{Development of Sentiment Detection Tools in SE} \label{sec:related-work-tool}
SE-specific sentiment tools are needed, because off-the-shelf sentiment detection tools from other domains are not very accurate~\cite{Jongeling-SentimentAnalysisToolsSe-ICSME2015,Jongeling-SentimentNegative-EMSE2017}. Subsequently, a number of sentiment and emotion detectors and benchmarks were developed for SE~\cite{Islam-DevaEmotionSE-ACMSAC2018,Islam-SentistrengthSE-MSR2017, Lin-PatternBasedOpinionMining-ICSE2019,Uddin-OpinionValue-TSE2019,Calefato-Senti4SD-EMSE2017,Ahmed-SentiCRNIER-ASE2017,Murgia-DoDevelopersFeelEmotion-MSR2014,Ortu-EmotionalSideJira-MSR2016,Chen-SentiEmoji-FSE2019,Islam-Deva-ACMSAC2018,Calefato-EmoTxt-ACII2017,Maipradit-SentimentClassificationNGram-IEEESW2019}. Largely, sentiment detection efforts for SE can be divided into three categories: rule-based, shallow learning-based, and deep learning-based.

We started to build our Hybrid tool, Sentisead, by combining the polarity labels of five SE-specific rule-based and shallow learning sentiment detection tools (Opiner~\cite{Uddin-OpinionValue-TSE2019}, POME~\cite{Lin-PatternBasedOpinionMining-ICSE2019}, SentiCR~\cite{Ahmed-SentiCRNIER-ASE2017}, Senti4SD~\cite{Calefato-Senti4SD-EMSE2017}, and SentistrengthSE~\cite{Islam-SentistrengthSE-MSR2017}). SentistrengthSE~\cite{Islam-SentistrengthSE-MSR2017} was studied by Lin et al.~\cite{Lin-SentimentDetectionNegativeResults-ICSE2018} along with other tools. 
Recently, Novielli et al.~\cite{Novielli-BenchmarkStudySentiSE-MSR2018} showed that the two supervised detectors (Senti4SD~\cite{Calefato-Senti4SD-EMSE2017} and SentiCR~\cite{Ahmed-SentiCRNIER-ASE2017}) performed better once re-trained for a domain. Thus, we retrained those for each of our six benchmarks. In a subsequent study, Novielli et al.~\cite{Novielli-SEToolCrossPlatform-MSR2020} found that unsupervised sentiment tools for SE, such as SentistrengthSE work better than supervised tools when applied in a cross-platform setting. Maipradit et al.~\cite{Maipradit-SentimentClassificationNGram-IEEESW2019} reported superior performance using n-grams in shallow learning models, like SVM, in multiple datasets. However, it is not clear which value of $n$ in the n-grams provided the best result.

Recently, several deep learning models to detect sentiment in SE artifacts were described in the literature~\cite{Chen-SentiEmoji-FSE2019,Biswas-SentiSEWordEmbedding-MSR2019,Biswas-ReliableSentiSEBERT-ICSME2020,Zhang-SEBERTSentiment-ICSME2020}. In 2019, Chen et al.~\cite{Chen-SentiEmoji-FSE2019} proposed SentiEmoji, an emoji-powered sentiment detection tools for SE data. SentiEmoji uses posts containing emotional emojis from Tweets and GitHub posts as new data to automatically produce new sentiment benchmark dataset. SentiEmoji combines the dataset with existing SE sentiment benchmark dataset and proposes a representation learning approach to train/test sentiments in the dataset. The approach offered superior performance than shallow learning tools like Senti4SD. 

In 2020, several papers reported language-based pre-trained transformer models (PTMs)~\cite{Biswas-ReliableSentiSEBERT-ICSME2020,Zhang-SEBERTSentiment-ICSME2020}. Biswas et al.~\cite{Biswas-ReliableSentiSEBERT-ICSME2020} showed that the use of PTMs, like BERT, can offer superior performance to other deep learning models, like RNN, which they studied before~\cite{Biswas-ReliableSentiSEBERT-ICSME2020}. At the same time, Zhang et al.~\cite{Zhang-SEBERTSentiment-ICSME2020} investigated four PTMs: BERT, ALBERT, RoBERTa, XLNet. They found that the PTMs outperformed state-of-the-art shallow learning and rule based SE-specific sentiment detectors across multiple datasets. The findings of Zhang et al.~\cite{Zhang-SEBERTSentiment-ICSME2020} and Biswas et al.~\cite{Biswas-ReliableSentiSEBERT-ICSME2020} motivated our inclusion of the PTMs into our hybrid tool.
  
Unlike our paper, none of the above studies investigate the feasibility of combining the tools. Our study, for the first time in SE, shows that a hybrid of the stand-alone SE-specific sentiment detection tools can be developed and that such a tool can indeed outperform each of those stand-alone SE-specific sentiment tool.

\section{Conclusions}
\label{sec:conclusions}

Sentiment detection is important to understand and support diverse scenarios in
software engineering and software-engineering research, yet studies from Lin et
al.~\cite{Lin-SentimentDetectionNegativeResults-ICSE2018} and Jongeling et
al.~\cite{Jongeling-SentimentNegative-EMSE2017} showed that currently available
sentiment detection tools for software engineering may not accurately detect
polarity in software textual contents.

To determine whether and how the existing SE-specific sentiment tools could be
combined to increase their performance, we reported the results of an empirical study that
explores the feasibility of creating a hybrid sentiment detection tool. We
started with five stand-alone rule-based and shallow learning sentiment
detectors for SE. We reported that the stand-alone tools can complement each
other to offer better performance, but a majority voting-based unsupervised
classifier fails to increase performance. We found that a simple supervised
combination of polarity labels from individual tools and bag of words as
features could increase the performance (F1-scores) of stand-alone tools by 4\%
to more than 100\%. We named the developed supervised hybrid tool Sentisead. Therefore, Sentisead can offer better performance 
than stand-alone detection tool, but it was not more than 4\% against the best performing stand-alone tool (Senti4SD).

Our study was motivated by the negative results reported by Lin et
al.~\cite{Lin-SentimentDetectionNegativeResults-ICSE2018} in 2018 and then the
positive results reported by Lin et
al.~\cite{Lin-PatternBasedOpinionMining-ICSE2019} in 2019 based on a newly
developed tool. We wanted to understand whether we could further improve the
performance of the stand-alone sentiment detectors (which were publicly
available at the time of our analysis). Instead of creating a new stand-alone
sentiment detection tool for software engineering, we wanted to determine
whether we could leverage the strengths of each individual SE-specific
individual tool. Therefore, we studied the feasibility of developing an ensemble
of those stand-alone tools. With a comparison of the performance of SentistrengthSE reported in Lin et
al.~\cite{Lin-SentimentDetectionNegativeResults-ICSE2018} and Sentisead, we
found that Sentisead offers better performance in the three datasets summarized
in \tbl\ref{tab:summary}. Note that the performance of SentistrengthSE on Jira Lin et al.~\cite{Lin-SentimentDetectionNegativeResults-ICSE2018} dataset was reported 
as 0.91 by Lin et al.~\cite{Lin-SentimentDetectionNegativeResults-ICSE2018}, but we observed it to be 0.97 as shown in in \tbl\ref{tab:bestPerformerPerDataset} (\sec\ref{sec:bestPerformerPerDataset}). 
\begin{table}[ht]
  \centering
    \begin{tabular}{lrr|rr}\toprule
     & \multicolumn{2}{l}{\textbf{Positive Polarity}} & \multicolumn{2}{l}{\textbf{Negative Polarity}} \\
     \cmidrule{2-5}
    \textbf{Dataset} & {\textbf{Lin~\cite{Lin-SentimentDetectionNegativeResults-ICSE2018}}} & {\textbf{Sentisead}} & {\textbf{Lin~\cite{Lin-SentimentDetectionNegativeResults-ICSE2018}}} & {\textbf{Sentisead}} \\
    \midrule
    \textbf{SO} & 0.26  & 0.35  & 0.27  & 0.31 \\
    \textbf{App} & 0.78  & 0.89  & 0.45  & 0.83 \\
    \textbf{Jira} & 0.91  & 0.95  & 0.82  & 0.98 \\

    \bottomrule
    \end{tabular}%
    \caption{Sentisead$_{RF}$ vs.\ SentistrengthSE performance reported by Lin et al.~\cite{Lin-SentimentDetectionNegativeResults-ICSE2018}}
    \label{tab:summary}
\end{table}%

The recent promising results in SE-specific sentiment detection with
language-based pre-trained transformer models (PTMs, e.g., BERT), as reported in
several works, e.g., Zhang et al.\cite{Zhang-SEBERTSentiment-ICSME2020}, has
motivated the second phase of our study, where we compared Sentisead with four
stand-alone PTMs (BERT, RoBERTa, ALBERT, and XLNet). We found that a stand-alone
RoBERTa outperforms Sentisead by 1.8\% (in terms of Macro F1-score). When we
replaced the ensembler in Sentisead from a Random Forest to RoBERTa, we observed
a performance increase of 2.3\% over the Random Forest-based Sentisead (i.e., Sentisead$_{RF}$).
However, when we checked the performance per dataset, the findings were inconclusive against the stand-alone PTMs, i.e., 
Sentisead-based various ensemble marginally outperformed the stand-alone PTMs in several cases but were outperformed by those PTMs in other 
cases. In particular, we found in \sec\ref{sec:resultsDeep} that PTMs like RoBERTa
performed considerably well as stand-alone sentiment detectors.
Therefore, when such marginal improvment via Sentisead is not warranted, a stand-alone PTM might suffice for sentiment detection in SE datasets.

As noted in \sec\ref{subsec:misclassification-cats}, out of the five
misclassification categories, we observed that three categories were also
previously observed and reported by Novielli et
al.~\cite{Novielli-BenchmarkStudySentiSE-MSR2018}. While Novielli et
al.~\cite{Novielli-BenchmarkStudySentiSE-MSR2018} observed the error categories
in units where all tools were wrong, we observed the error categories for units
where at least one of the tools is correct. The purpose of our study to derive
the misclassification categories was twofold: first, we needed to know the
specific reasons for misclassification in one tool where another tool could
offer the correct result. Second, we wanted to use knowledge gained from the
error categories to develop our hybrid tool, Sentisead.

Around 50\% of all misclassifications were due to the inability of the wrong
tool to understand the underlying contexts, while the correct tool understood
the context. Based on this insight, we hypothesized that if we could combine the
polarity labels of all tools (where the correct tool is also included) with the
bag of words of the unit that would serve as contexts, we could offer better
results in our hybrid tool, Sentisead. Indeed, Sentisead corrects the
contexts-based error for 80\% of the cases (i.e., it was wrong in only 20\% of
the 62 units where at least one stand-alone tool is wrong), which is more than
any of the stand-alone tools. \tbl\ref{tab:misclassificationCatCovSentiseadB}
offers a qualitative breakdown of how Sentisead offered improvement over each
stand-alone tool based on the five error categories.

Sentisead infrastructure could be leveraged to add the polarity labels of any
newly introduced SE-specific stand-alone sentiment detection tool. We envision
that this flexibility of Sentisead will allow the co-creation of two new-breeds
of efforts in software engineering research: \begin{inparaenum}[(1)] \item More
efforts to create new SE-specific stand-alone sentiment detection tool focusing
on diverse SE datasets, and \item Studying \begin{inparaenum}[(a)] \item the
incorporation of diverse complementary SE-specific stand-alone sentiment
detection tools. \item The effectiveness of diverse features, such as
fine-grained emotions from various tools into Sentistead (or its extension).
\end{inparaenum}
\end{inparaenum} 

As we noted in \sec\ref{sec:intro}, recent advances in deep learning-based
SE-specific sentiment detection shows more promising results than those observed
by Lin et al.~\cite{Lin-SentimentDetectionNegativeResults-ICSE2018}. While
within the limited space of this paper, we attempted to offer a comprehensive
analysis by investigating the feasibility of developing a hybrid tool based on
rule-based, shallow and deep learning stand-alone classifiers, we leave
following extensions to Sentisead for future work:

\begin{inparaenum}[(1)]
\item SentiEmoji~\cite{Chen-SentiEmoji-FSE2019}, a tool developed by Chen et al.
shows that we can use emoji-powered data from Twitter and Facebook to further
improve sentiment classification in SE.
SentiEmoji~\cite{Chen-SentiEmoji-FSE2019} also shows that an embedding based on
the emoji information and the textual contents can be more useful. Sentisead
could leverage such emoji-powered information as features besides word-embedding
or bag of words.

\item Deep learning models, like PTMs, are found to offer good performance even
with low amount of data due to their design as pre-trained models. However, deep
learning models also benefit from more data. A problem with SE-specific
sentiment detection is the diversity of datasets. With emoji-powered labels like
those used by SentiEmoji~\cite{Chen-SentiEmoji-FSE2019}, we may be able to get
more data with automatic polarity labels, which then can help with
domain-specific nuances more. Therefore, Sentisead could benefit from such
large-scale data, especially for PTMs like a RoBERTa as an ensembler (see
RQ$_6$).
\end{inparaenum}

We conclude that our empirical study with Sentisead offers valuable insights to
the field of SE-specific sentiment detection research with empirical insights,
actionable results, and a hybrid tool Sentisead that shows marginal to promising performance
over stand-alone SE-specific sentiment detection tools.

\section*{Acknowledgment}

We would like to thank the anonymous reviewers, who helped improve our work with comments and suggestions. We would like to thank Nibir Mandal, Mohamed Raed El Aoun, and Houssem Ben Braiek, who helped run the experiments with the advanced language-based pre-trained models in \sec\ref{sec:resultsDeep}.

\bibliographystyle{abbrv}
\bibliography{main}

\begin{thebibliography}{10}

\bibitem{Ahmed-SentiCRNIER-ASE2017}
T.~Ahmed, A.~Bosu, A.~Iqbal, and S.~Rahimi.
\newblock Senticr: A customized sentiment analysis tool for code review
  interactions.
\newblock In {\em Proceedings of the 32nd International Conference on Automated
  Software Engineering}, pages 106--111, 2017.

\bibitem{Ikram-SentimentCodeReview-IST2019}
I.~E. Asri, N.~Kerzazi, G.~Uddin, F.~Khomh, and M.~J. Idrissi.
\newblock An empirical study of sentiments in code reviews.
\newblock {\em Information and Software Technology}, 114:37--54, 2019.

\bibitem{Biswas-ReliableSentiSEBERT-ICSME2020}
E.~Biswas, M.~E. Karabulut, L.~Pollock, and K.~Vijay-Shanker.
\newblock Achieving reliable sentiment analysis in the software engineering
  domain using bert.
\newblock In {\em IEEE International Conference on Software Maintenance and
  Evolution}, pages 162--173, 2020.

\bibitem{Biswas-SentiSEWordEmbedding-MSR2019}
E.~Biswas, K.~Vijay-Shanker, and L.~Pollock.
\newblock Exploring word embedding techniques to improve sentiment analysis of
  software engineering texts.
\newblock In {\em Proceedings of the 16th International Conference on Mining
  Software Repositories}, pages 68--78, 2019.

\bibitem{Calefato-Senti4SD-EMSE2017}
F.~Calefato, F.~Lanubile, F.~Maiorano, and N.~Novielli.
\newblock Sentiment polarity detection for software development.
\newblock {\em Journal Empirical Software Engineering}, pages 2543--2584, 2017.

\bibitem{Calefato-EmoTxt-ACII2017}
F.~Calefato, F.~Lanubile, and N.~Novielli.
\newblock Emotxt: A toolkit for emotion recognition from text.
\newblock In {\em Proc. 7th Affective Computing and Intelligent Interaction},
  page~2, 2017.

\bibitem{Castaldi-SentimentAnalysisITSupport-MSR2016}
C.~Castaldi, A.~Blaz, and K.~Becker.
\newblock Sentiment analysis in tickets for {IT} support.
\newblock In {\em IEEE/ACM 13th Working Conference on Mining Software
  Repositories}, pages 235--246, 2016.

\bibitem{Chawla-SMOTE-JournalAIResearch2002}
N.~V. Chawla, K.~W. Bowyer, L.~O. Hall, and W.~P. Kegelmeyer.
\newblock Smote: synthetic minority over-sampling technique.
\newblock {\em Journal of Artificial Intelligence Research}, 16(1):321--357,
  2002.

\bibitem{Chen-SentiEmoji-FSE2019}
Z.~Chen, Y.~Cao, X.~Lu, Q.~Mei, and X.~Liu.
\newblock Sentimoji: an emoji-powered learning approach for sentiment analysis
  in software engineering.
\newblock In {\em 27th ACM Joint Meeting on European Software Engineering
  Conference and Symposium on the Foundations of Software Engineering}, pages
  841--852, 2019.

\bibitem{Cohen-AppliedMultipleRegression-Lawrence2002}
J.~Cohen, S.~G. West, L.~Aiken, and P.~Cohen.
\newblock {\em Applied Multiple Regression/Correlation Analysis for the
  Behavioral Sciences}.
\newblock Lawrence Erlbaum Associates, 3rd edition, 2002.

\bibitem{Delvin-BERTArch-Arxiv2018}
J.~Devlin, M.-W. Chang, K.~Lee, and K.~Toutanova.
\newblock {BERT}: Pre-training of deep bidirectional transformers for language
  understanding.
\newblock Technical report, \url{https://arxiv.org/abs/1810.04805}, 2018.

\bibitem{Balage-HybridSentimentDetectorTwitter-SemEval2014}
P.~P.~B. Filho, L.~Avanco, T.~A.~S. Pardo, and M.~G.~V. Nunes.
\newblock Nilc\_usp: An improved hybrid system for sentiment analysis in
  twitter messages.
\newblock In {\em Proceedings of the 8th International Workshop on Semantic
  Evaluation}, pages 428--432, 2014.

\bibitem{Gachechiladze-AngerDirectionCollaborativeSE-ICSENIER2017}
D.~Gachechiladze, F.~Lanubile, N.~Novielli, and A.~Serebrenik.
\newblock Anger and its direction in collaborative software development.
\newblock In {\em 39th International Conference on Software Engineering: New
  Ideas and Emerging Results Track}, pages 11--14, 2017.

\bibitem{Ghotra-RevisitingImpactClassificationDefectPrediction-ICSE2015}
B.~Ghotra, S.~McIntosh, and A.~E. Hassan.
\newblock Revisiting the impact of classification techniques on the performance
  of defect prediction models.
\newblock In {\em Proceedings of the 37th International Conference on Software
  Engineering}, pages 789--800, 2015.

\bibitem{Goncalves-ComparingCombiningSentimentAnalysisTools-COSN2013}
P.~Goncalves, M.~Araujo, F.~Benevenuto, and M.~Cha.
\newblock Comparing and combining sentiment analysis methods.
\newblock In {\em Proceedings of the first ACM conference on Online social
  networks}, pages 27--38, 2013.

\bibitem{Guzman-NeedleInHaystackTwitterSoftware-RE2016}
E.~Guzman, R.~Alkadhi, , and N.~Seyf.
\newblock A needle in a haystack: What do twitter users say about software?
\newblock In {\em 24th IEEE International Requirements Engineering Conference},
  pages 96--105, 2016.

\bibitem{Guzman-SentimentAnalysisGithub-MSR2014}
E.~Guzman, D.~Az\'{o}car, and Y.~Li.
\newblock Sentiment analysis of commit comments in github: an empirical study.
\newblock In {\em Proceedings of the 11th Working Conference on Mining Software
  Repositories}, pages 352--355, 2014.

\bibitem{Guzman-EmotionalAwarenessSoftwareDevelopmentTeams-FSE2013}
E.~Guzman and B.~Bruegge.
\newblock Towards emotional awareness in software development teams.
\newblock In {\em Proceedings of the 2013 9th Joint Meeting on Foundations of
  Software Engineering}, pages 671--674, 2013.

\bibitem{Guzman-EmotionalAwareness-FSE2013}
E.~Guzman and B.~Bruegge.
\newblock Towards emotional awareness in software development teams.
\newblock In {\em Proceedings of the 7th Joint Meeting on Foundations of
  Software Engineering}, pages 671--674, 2013.

\bibitem{Hassan-PredictingFaultsUsingCodeComplexity-ICSE2009}
A.~E. Hassan.
\newblock Predicting faults using the complexity of code changes.
\newblock In {\em Proc. 31st International Conference on Software Engineering},
  pages 78--89, 2009.

\bibitem{Hu-MiningSummarizingCustomerReviews-KDD2004}
M.~Hu and B.~Liu.
\newblock Mining and summarizing customer reviews.
\newblock In {\em ACM SIGKDD International Conference on Knowledge Discovery
  and Data Mining}, pages 168--177, 2004.

\bibitem{Islam-SentistrengthSE-MSR2017}
M.~R. Islam and M.~F. Zibran.
\newblock Leveraging automated sentiment analysis in software engineering.
\newblock In {\em Proc. 14th International Conference on Mining Software
  Repositories}, pages 203--214, 2017.

\bibitem{Islam-DevaEmotionSE-ACMSAC2018}
M.~R. Islam and M.~F. Zibran.
\newblock Deva: Sensing emotions in the valence arousal space in software
  engineering text.
\newblock In {\em 33rd Annual ACM Symposium on Applied Computing}, page
  1536–1543, 2018.

\bibitem{Islam-Deva-ACMSAC2018}
M.~R. Islam and M.~F. Zibran.
\newblock Deva: Sensing emotions in the valence arousal space in software
  engineering text.
\newblock In {\em 33rd Annual ACM Symposium on Applied Computing}, page
  1536–1543, 2018.

\bibitem{Jongeling-SentimentAnalysisToolsSe-ICSME2015}
R.~Jongeling, S.~Datta, and A.~Serebrenik.
\newblock Choosing your weapons: On sentiment analysis tools for software
  engineering research.
\newblock In {\em Proceedings of the 31st International Confernece on Software
  Maintenance and Evolution}, 2015.

\bibitem{Jongeling-SentimentNegative-EMSE2017}
R.~Jongeling, P.~Sarkar, S.~Datta, and A.~Serebrenik.
\newblock On negative results when using sentiment analysis tools for software
  engineering research.
\newblock {\em Journal Empirical Software Engineering}, 22(5):2543--2584, 2017.

\bibitem{Khomh-EntropyEvaluationTriageMozillaCrash-WCRE2011}
F.~Khomh, B.~Chan, Y.~Zou, and A.~E. Hassan.
\newblock An entropy evaluation approach for triaging field crashes: A case
  study of mozilla firefox.
\newblock In {\em Proceedings of the 2011 18th Working Conference on Reverse
  Engineering}, pages 261--270, 2011.

\bibitem{Ko-LinguisticAnalysis-VLHCC2005a}
A.~J. Ko, B.~A. Myers, and D.~H. Chau.
\newblock A linguistic analysis of how people describe software problems.
\newblock In {\em 2005 IEEE Symposium on Visual Languages and Human-Centric
  Computing}, pages 127--134, 2005.

\bibitem{website:senisead-online-appendix-ase2020}
D.~Lab.
\newblock {\em An Empirical Study of the Effectiveness of an Ensemble of
  Stand-alone Sentiment Detection Tools for Software Engineering Datasets
  (Online Appendix)}.
\newblock \url{https://github.com/disa-lab/HybridSESentimentTOSEM}, 21 April
  2021 (last accessed).

\bibitem{Lan-Albert-Arxiv2020}
Z.~Lan, M.~Chen, S.~Goodman, K.~Gimpel, P.~Sharma, and R.~Soricut.
\newblock {ALBERT}: A lite bert for self-supervised learning of language
  representations.
\newblock Technical report, \url{https://arxiv.org/abs/1909.11942}, 2020.

\bibitem{Lin-PatternBasedOpinionMining-ICSE2019}
B.~Lin, F.~Zampetti, G.~Bavota, M.~D. Penta, and M.~Lanza.
\newblock Pattern-based mining of opinions in {Q\&A} websites.
\newblock In {\em Proc. 41st International Conference on Software Engineering},
  pages 548--559, 2019.

\bibitem{Lin-SentimentDetectionNegativeResults-ICSE2018}
B.~Lin, F.~Zampetti, G.~Bavota, M.~D. Penta, M.~Lanza, and R.~Oliveto.
\newblock Sentiment analysis for software engineering: how far can we go?
\newblock In {\em Proc. 40th International Conference on Software Engineering},
  pages 94--104, 2018.

\bibitem{Liu-SentimentAnalysisOpinionMining-MorganClaypool2012}
B.~Liu.
\newblock {\em Sentiment Analysis and Opinion Mining}.
\newblock Morgan \& Claypool Publishers, 1st edition, May 2012.

\bibitem{Liu-Roberta-Arxiv2019}
Y.~Liu, M.~Ott, N.~Goyal, J.~Du, M.~Joshi, D.~Chen, O.~Levy, M.~Lewis,
  L.~Zettlemoyer, and V.~Stoyanov.
\newblock {RoBERTa}: A robustly optimized bert pretraining approach.
\newblock Technical report, \url{https://arxiv.org/abs/1907.11692}, 2019.

\bibitem{Maalej-AutomaticClassificationAppReviews-RE2016}
W.~Maalej, Z.~Kurtanovi\'{c}, H.~Nabil, and C.~Stanik.
\newblock On the automatic classification of app reviews.
\newblock In {\em International Requirements Engineering Conference}, page
  311–331, 2016.

\bibitem{Maipradit-SentimentClassificationNGram-IEEESW2019}
R.~Maipradit, H.~Hata, and K.~Matsumoto.
\newblock Sentiment classification using n-gram inverse document frequency and
  automated machine learning.
\newblock {\em IEEE Software}, 36(5):65--70, 2019.

\bibitem{Manning-IRIntroBook-Cambridge2009}
C.~D. Manning, P.~Raghavan, and H.~Sch\"{u}tze.
\newblock {\em An Introduction to Information Retrieval}.
\newblock Cambridge Uni Press, 2009.

\bibitem{Mika-MiningValenceBurnout-MSR2016}
M.~M\.{a}ntyl\.{a}, B.~Adams, G.~Destefanis, D.~Graziotin, and M.~Ortu.
\newblock Mining valence, arousal, and dominance -- possibilities for detecting
  burnout and productivity?
\newblock In {\em Proceedings of the 13th Working Conference on Mining Software
  Repositories}, pages 247--258, 2016.

\bibitem{Misrili-IndustrialCaseStudyEnsembleSoftwareDefects-SQJ2011}
A.~T. Misirli, A.~B. Bener, and B.~Turhan.
\newblock An industrial case study of classifier ensembles for locating
  software defects.
\newblock {\em Software Quality Journal}, 19(3):515--536, 2011.

\bibitem{Murgia-DoDevelopersFeelEmotion-MSR2014}
A.~Murgia, P.~Tourani, B.~Adams, and M.~Ortu.
\newblock Do developers feel emotions? an exploratory analysis of emotions in
  software artifacts.
\newblock In {\em Proceedings of the 11th Working Conference on Mining Software
  Repositories}, 2014.

\bibitem{Novielli-IntroToSpIssueAffectSE-JSS2019}
N.~Novielli, A.~Begel, and W.~Maalej.
\newblock Introduction to the special issue on affect awareness in software
  engineering.
\newblock {\em Journal of Systems and Software}, 148:180--182, 2019.

\bibitem{Novielli-SEToolCrossPlatform-MSR2020}
N.~Novielli, F.~Calefato, D.~Dongiovanni, D.~Girardi, and F.~Lanubile.
\newblock Can we use se-specific sentiment analysis tools in a cross-platform
  setting?
\newblock In {\em 17th International Conference on Mining Software
  Repositories}, page~11, 2020.

\bibitem{Novielli-ChallengesSentimentDetectionProgrammerEcosystem-SSE2015}
N.~Novielli, F.~Calefato, and F.~Lanubile.
\newblock The challenges of sentiment detection in the social programmer
  ecosystem.
\newblock In {\em Proceedings of the 7th International Workshop on Social
  Software Engineering}, pages 33--40, 2015.

\bibitem{Novielli-BenchmarkDatasetSentiSE-MSR2018}
N.~Novielli, F.~Calefato, and F.~Lanubile.
\newblock A gold standard for emotion annotation in stack overflow.
\newblock In {\em Proceedings of the 15th International Conference on Mining
  Software Repositories (Data Showcase)}, page~4, 2018.

\bibitem{Novielli-BenchmarkStudySentiSE-MSR2018}
N.~Novielli, D.~Girardi, and F.~Lanubile.
\newblock A benchmark study on sentiment analysis for software engineering
  research.
\newblock In {\em Proceedings of the 15th International Conference on Mining
  Software Repositories}, page~12, 2018.

\bibitem{Novielli-SentimentEmotionSE-IEEESoftware2019}
N.~Novielli and A.~Serebrenik.
\newblock Sentiment and emotion in software engineering.
\newblock {\em IEEE Software}, 36(5):6--23, 2019.

\bibitem{Nucci-DynamicSelectionOfClassifiersBugPrediction-IEEETranEmergingTopics2017}
D.~D. Nucci, F.~Palomba, R.~Oliveto, and A.~D. Lucia.
\newblock Dynamic selection of classifiers in bug prediction: An adaptive
  method.
\newblock {\em IEEE Transactions on Emerging Topics in Computational
  Intelligence}, 1(3):202--212, 2017.

\bibitem{Ortu-BulliesMoreProductive-MSR2015}
M.~Ortu, B.~Adams, G.~Destefanis, P.~Tourani, M.~Marchesi, and R.~Tonelli.
\newblock Are bullies more productive? empirical study of affectiveness vs.
  issue fixing time.
\newblock In {\em Proceedings of the 12th Working Conference on Mining Software
  Repositories}, pages 303--313, 2015.

\bibitem{Ortu-AreBulliesMoreProductive-MSR205}
M.~Ortu, B.~Adams, G.~Destefanis, P.~Tourani, M.~Marchesi, and R.~Tonelli.
\newblock Are bullies more productive? empirical study of affectiveness vs.
  issue fixing time.
\newblock In {\em Proceedings of the 12th Working Conference on Mining Software
  Repositories}, 2015.

\bibitem{Ortu-EmotionalSideJira-MSR2016}
M.~Ortu, A.~Murgia, G.~Destefanis, P.~Tourani, R.~Tonelli, M.~L. Marchesi, and
  B.~Adams.
\newblock The emotional side of software developers in jira.
\newblock In {\em 13th International Conference on Mining Software
  Repositories}, pages 480--483, 2016.

\bibitem{Pang-SentimentClassificationMachineLearning-EMNLP2002}
B.~Pang, L.~Lee, and S.~Vaithyanathan.
\newblock Thumbs up? sentiment classification using machine learning
  techniques.
\newblock In {\em Conference on Empirical Methods in Natural Language
  Processing}, pages 79--86, 2002.

\bibitem{Panichella-ClassifyAppUserReview-ICSME2015}
S.~Panichella, A.~D. Sorbo, E.~Guzman, C.~A. Visaggio, G.~Canfora, and H.~C.
  Gall.
\newblock How can i improve my app? classifying user reviews for software
  maintenance and evolution.
\newblock In {\em IEEE International Conf. on Software Maintenance and
  Evolution}, pages 281--290, 2015.

\bibitem{Petric-EnsembleDefectDiversity-ESEM2016}
J.~Petric, D.~Bowes, T.~Hall, B.~Christianson, and N.~Baddoo.
\newblock Building an ensemble for software defect prediction based on
  diversity selection.
\newblock In {\em Proceedings of the 10th ACM/IEEE International Symposium on
  Empirical Software Engineering and Measurement}, page Article No. 46, 2016.

\bibitem{Pletea-SecurityEmotionSE-MSR2014}
D.~Pletea, B.~Vasilescu, and A.~Serebrenik.
\newblock Security and emotion: sentiment analysis of security discussions on
  github.
\newblock In {\em Proceedings of the 11th Working Conference on Mining Software
  Repositories}, pages 348--351, 2014.

\bibitem{website:scikitlearn}
scikit learn.
\newblock {\em Machine Learning in Python}.
\newblock \url{http://scikit-learn.org/stable/index.html#}, 2017.

\bibitem{Sebastiani-MachineLearningTextCategorization-ACMSurveys2002}
F.~Sebastiani.
\newblock Machine learning in automated text categorization.
\newblock {\em Journal of ACM Computing Surveys}, 34(1):1--47, 2002.

\bibitem{Shannon-MathematicTheoryOfCommunication-BellJournal1948}
C.~E. Shannon.
\newblock A mathematical theory of communication.
\newblock {\em The Bell System Technical Journal}, 27(3):379--423, 1948.

\bibitem{Sinha-DeveloperSentimentCommitLog-MSR2016}
V.~Sinha, A.~Lazar, and B.~Sharif.
\newblock Analyzing developer sentiment in commit logs.
\newblock In {\em 13th International Conference on Mining Software
  Repositories}, pages 281--290, 2016.

\bibitem{Socher-Sentiment-EMNLP2013}
R.~Socher, A.~Perelygin, J.~Wu, C.~Manning, A.~Ng, and J.~Chuang.
\newblock Recursive models for semantic compositionality over a sentiment
  treebank.
\newblock In {\em Proc. Conference on Empirical Methods in Natural Language
  Processing (EMNLP)}, page~12, 2013.

\bibitem{Thelwall-Sentistrength-ASICT2010}
M.~Thelwall, K.~Buckley, G.~Paltoglou, D.~Cai, and A.~Kappas.
\newblock Sentiment in short strength detection informal text.
\newblock {\em Journal of the American Society for Information Science and
  Technology}, 61(12):2544--2558, 2010.

\bibitem{Patanamon-ReviewParticipationInCodeReview-JournalEMSE2017}
P.~Thongtanunam, S.~Mcintosh, A.~E. Hassan, and H.~Iida.
\newblock Review participation in modern code review: An empirical study of the
  android, qt, and openstack projects.
\newblock {\em Journal of Empirical Software Engineering}, 22(2):768--817,
  2017.

\bibitem{Uddin-OpinionSurvey-TSE2019}
G.~Uddin, O.~Baysal, L.~Guerroj, and F.~Khomh.
\newblock Understanding how and why developers seek and analyze api related
  opinions.
\newblock {\em IEEE Transactions on Software Engineering}, page~40, 2019.

\bibitem{Uddin-OpinerReviewAlgo-ASE2017}
G.~Uddin and F.~Khomh.
\newblock Automatic summarization of {API} reviews.
\newblock In {\em Proc. 32nd IEEE/ACM International Conference on Automated
  Software Engineering}, pages 159--170, 2017.

\bibitem{Uddin-APIAspectMining-TechReport2017}
G.~Uddin and F.~Khomh.
\newblock Mining api aspects in api reviews.
\newblock Technical report,
  \url{https://swat.polymtl.ca/data/opinionvalue-technical-report.pdf}, 2017.

\bibitem{Uddin-OpinerReviewToolDemo-ASE2017}
G.~Uddin and F.~Khomh.
\newblock Opiner: A search and summarization engine for {API} reviews.
\newblock In {\em Proc. 32nd IEEE/ACM International Conference on Automated
  Software Engineering}, pages 978--983, 2017.

\bibitem{Uddin-OpinionValue-TSE2019}
G.~Uddin and F.~Khomh.
\newblock Automatic opinion mining from {API} reviews from stack overflow.
\newblock {\em IEEE Transactions on Software Engineering}, page~35, 2019.

\bibitem{Uddin-OpinerAPIUsageScenario-TOSEM2020}
G.~Uddin, F.~Khomh, and C.~K. Roy.
\newblock Automatic api usage scenario documentation from technical q\&a sites.
\newblock {\em ACM Transactions on Software Engineering and Methodology},
  page~43, 2020.

\bibitem{Uddin-OpinerAPIUsageScenario-IST2020}
G.~Uddin, F.~Khomh, and C.~K. Roy.
\newblock Automatic mining of api usage scenarios from stack overflow.
\newblock {\em Information and Software Technology (IST)}, page~16, 2020.

\bibitem{Uddin-ResolvingAPIMentions-TechReport2015a}
G.~Uddin and M.~P. Robillard.
\newblock Resolving api mentions in forum texts.
\newblock Technical report, McGill University, 2015.

\bibitem{Valiev-EcosystemLevelSurvivalPyPI-FSE2018}
M.~Valiev, B.~Vasilescu, and J.~Herbsleb.
\newblock Ecosystem-level determinants of sustained activity in open-source
  projects: a case study of the pypi ecosystem.
\newblock In {\em Proceedings of the 2018 26th ACM Joint Meeting on European
  Software Engineering Conference and Symposium on the Foundations of Software
  Engineering}, pages 644--655, 2018.

\bibitem{Villarroel-ReleasePlanningMobileAppReviews-ICSE2016}
L.~Villarroel, G.~Bavota, B.~Russo, R.~Oliveto, and M.~D. Penta.
\newblock Release planning of mobile apps based on user reviews.
\newblock In {\em Proceedings of the 38th International Conference on Software
  Engineering}, pages 14--24, 2016.

\bibitem{Wang-BaselinesBigrams-ACL2012}
S.~Wang and C.~D. Manning.
\newblock Baselines and bigrams: simple, good sentiment and topic
  classification.
\newblock In {\em Proceedings of the 50th Annual Meeting of the Association for
  Computational Linguistics}, pages 90--94, 2012.

\bibitem{Warriner-VadLexicons-BRM2013}
A.~B. Warriner, V.~Kuperman, and M.~Brysbaert.
\newblock Norms of valence, arousal, and dominance for 13,915 english lemmas.
\newblock {\em Behavior Research Methods}, 45(4):1191--1207, 2013.

\bibitem{Woh00}
C.~Wohlin, P.~Runeson, M.~H\"{o}st, M.~C. Ohlsson, B.~Regnell, and
  A.~Wessl\'{e}n.
\newblock {\em Experimentation in software engineering: an introduction}.
\newblock Kluwer Academic Publishers, Norwell, MA, USA, 2000.

\bibitem{Yang-TwoLayerEnsembleJustInTimeDefect-IST2017}
X.~Yang, D.~Lo, X.~Xia, and J.~Sun.
\newblock {TLEL}: A two-layer ensemble learning approach for just-in-time
  defect prediction.
\newblock {\em Information and Software Technology}, 87(2):206 -- 220, 2017.

\bibitem{Yang-Xlnet-Arxiv2020}
Z.~Yang, Z.~Dai, Y.~Yang, J.~Carbonell, R.~Salakhutdinov, and Q.~V. Le.
\newblock {XLNet}: Generalized autoregressive pretraining for language
  understanding.
\newblock Technical report, \url{https://arxiv.org/abs/1906.08237}, 2020.

\bibitem{Zhang-SEBERTSentiment-ICSME2020}
T.~Zhang, B.~Xu, F.~Thung, S.~A. Haryono, D.~Lo, and L.~Jiang.
\newblock Sentiment analysis for software engineering: How far can pre-trained
  transformer models go?
\newblock In {\em IEEE International Conference on Software Maintenance and
  Evolution}, pages 70--80, 2020.

\bibitem{Zhang-CombinedClassifierCrossDefect-FrontierComputerScience2018}
Y.~Zhang, D.~Lo, X.~Xia, and J.~Sun.
\newblock Combined classifier for cross-project defect prediction: an extended
  empirical study.
\newblock {\em Frontiers of Computer Science}, 22(2):280 -- 296, 2018.

\end{thebibliography}


\begin{thebibliography}{10}
\providecommand{\url}[1]{#1}
\csname url@samestyle\endcsname
\providecommand{\newblock}{\relax}
\providecommand{\bibinfo}[2]{#2}
\providecommand{\BIBentrySTDinterwordspacing}{\spaceskip=0pt\relax}
\providecommand{\BIBentryALTinterwordstretchfactor}{4}
\providecommand{\BIBentryALTinterwordspacing}{\spaceskip=\fontdimen2\font plus
\BIBentryALTinterwordstretchfactor\fontdimen3\font minus
  \fontdimen4\font\relax}
\providecommand{\BIBforeignlanguage}[2]{{%
\expandafter\ifx\csname l@#1\endcsname\relax
\typeout{** WARNING: IEEEtran.bst: No hyphenation pattern has been}%
\typeout{** loaded for the language `#1'. Using the pattern for}%
\typeout{** the default language instead.}%
\else
\language=\csname l@#1\endcsname
\fi
#2}}
\providecommand{\BIBdecl}{\relax}
\BIBdecl

\bibitem{Liu-SentimentAnalysisOpinionMining-MorganClaypool2012}
B.~Liu, \emph{Sentiment Analysis and Opinion Mining}, 1st~ed.\hskip 1em plus
  0.5em minus 0.4em\relax Morgan \& Claypool Publishers, May 2012.

\bibitem{Pang-SentimentClassificationMachineLearning-EMNLP2002}
B.~Pang, L.~Lee, and S.~Vaithyanathan, ``Thumbs up? sentiment classification
  using machine learning techniques,'' in \emph{Conference on Empirical Methods
  in Natural Language Processing}, 2002, pp. 79--86.

\bibitem{Uddin-OpinerReviewAlgo-ASE2017}
G.~Uddin and F.~Khomh, ``Automatic summarization of {API} reviews,'' in
  \emph{Proc. 32nd IEEE/ACM International Conference on Automated Software
  Engineering}, 2017, pp. 159--170.

\bibitem{Guzman-EmotionalAwarenessSoftwareDevelopmentTeams-FSE2013}
E.~Guzman and B.~Bruegge, ``Towards emotional awareness in software development
  teams,'' in \emph{Proceedings of the 2013 9th Joint Meeting on Foundations of
  Software Engineering}, 2013, pp. 671--674.

\bibitem{Mika-MiningValenceBurnout-MSR2016}
M.~M\.{a}ntyl\.{a}, B.~Adams, G.~Destefanis, D.~Graziotin, and M.~Ortu,
  ``Mining valence, arousal, and dominance -- possibilities for detecting
  burnout and productivity?'' in \emph{Proceedings of the 13th Working
  Conference on Mining Software Repositories}, 2016, pp. 247--258.

\bibitem{Ortu-BulliesMoreProductive-MSR2015}
M.~Ortu, B.~Adams, G.~Destefanis, P.~Tourani, M.~Marchesi, and R.~Tonelli,
  ``Are bullies more productive? empirical study of affectiveness vs. issue
  fixing time,'' in \emph{Proceedings of the 12th Working Conference on Mining
  Software Repositories}, 2015, pp. 303--313.

\bibitem{Pletea-SecurityEmotionSE-MSR2014}
D.~Pletea, B.~Vasilescu, and A.~Serebrenik, ``Security and emotion: sentiment
  analysis of security discussions on github,'' in \emph{Proceedings of the
  11th Working Conference on Mining Software Repositories}, 2014, pp. 348--351.

\bibitem{Uddin-OpinerReviewToolDemo-ASE2017}
G.~Uddin and F.~Khomh, ``Opiner: A search and summarization engine for {API}
  reviews,'' in \emph{Proc. 32nd IEEE/ACM International Conference on Automated
  Software Engineering}, 2017, pp. 978--983.

\bibitem{Novielli-ChallengesSentimentDetectionProgrammerEcosystem-SSE2015}
N.~Novielli, F.~Calefato, and F.~Lanubile, ``The challenges of sentiment
  detection in the social programmer ecosystem,'' in \emph{Proceedings of the
  7th International Workshop on Social Software Engineering}, 2015, pp. 33--40.

\bibitem{Socher-Sentiment-EMNLP2013}
R.~Socher, A.~Perelygin, J.~Wu, C.~Manning, A.~Ng, and J.~Chuang, ``Recursive
  models for semantic compositionality over a sentiment treebank,'' in
  \emph{Proc. Conference on Empirical Methods in Natural Language Processing
  (EMNLP)}, 2013, p.~12.

\bibitem{Islam-SentistrengthSE-MSR2017}
M.~R. Islam and M.~F. Zibran, ``Leveraging automated sentiment analysis in
  software engineering,'' in \emph{Proc. 14th International Conference on
  Mining Software Repositories}, 2017, pp. 203--214.

\bibitem{Calefato-Senti4SD-EMSE2017}
F.~Calefato, F.~Lanubile, F.~Maiorano, and N.~Novielli, ``Sentiment polarity
  detection for software development,'' \emph{Journal Empirical Software
  Engineering}, pp. 2543--2584, 2017.

\bibitem{Ahmed-SentiCRNIER-ASE2017}
T.~Ahmed, A.~Bosu, A.~Iqbal, and S.~Rahimi, ``Senticr: A customized sentiment
  analysis tool for code review interactions,'' in \emph{Proceedings of the
  32nd International Conference on Automated Software Engineering}, 2017, pp.
  106--111.

\bibitem{Lin-SentimentDetectionNegativeResults-ICSE2018}
B.~Lin, F.~Zampetti, G.~Bavota, M.~D. Penta, M.~Lanza, and R.~Oliveto,
  ``Sentiment analysis for software engineering: how far can we go?'' in
  \emph{Proc. 40th International Conference on Software Engineering}, 2018, pp.
  94--104.

\bibitem{Novielli-BenchmarkStudySentiSE-MSR2018}
N.~Novielli, D.~Girardi, and F.~Lanubile, ``A benchmark study on sentiment
  analysis for software engineering research,'' in \emph{Proceedings of the
  15th International Conference on Mining Software Repositories}, 2018, p.~12.

\bibitem{Balage-HybridSentimentDetectorTwitter-SemEval2014}
P.~P.~B. Filho, L.~Avanco, T.~A.~S. Pardo, and M.~G.~V. Nunes, ``Nilc\_usp: An
  improved hybrid system for sentiment analysis in twitter messages,'' in
  \emph{Proceedings of the 8th International Workshop on Semantic Evaluation},
  2014, pp. 428--432.

\bibitem{Goncalves-ComparingCombiningSentimentAnalysisTools-COSN2013}
P.~Goncalves, M.~Araujo, F.~Benevenuto, and M.~Cha, ``Comparing and combining
  sentiment analysis methods,'' in \emph{Proceedings of the first ACM
  conference on Online social networks}, 2013, pp. 27--38.

\bibitem{Ghotra-RevisitingImpactClassificationDefectPrediction-ICSE2015}
B.~Ghotra, S.~McIntosh, and A.~E. Hassan, ``Revisiting the impact of
  classification techniques on the performance of defect prediction models,''
  in \emph{Proceedings of the 37th International Conference on Software
  Engineering}, 2015, pp. 789--800.

\bibitem{Nucci-DynamicSelectionOfClassifiersBugPrediction-IEEETranEmergingTopics2017}
D.~D. Nucci, F.~Palomba, R.~Oliveto, and A.~D. Lucia, ``Dynamic selection of
  classifiers in bug prediction: An adaptive method,'' \emph{IEEE Transactions
  on Emerging Topics in Computational Intelligence}, vol.~1, no.~3, pp.
  202--212, 2017.

\bibitem{Uddin-OpinionValue-TSE2019}
G.~Uddin and F.~Khomh, ``Automatic opinion mining from {API} reviews,''
  \emph{IEEE Transactions on Software Engineering}, p.~37, 2019.

\bibitem{Lin-PatternBasedOpinionMining-ICSE2019}
B.~Lin, F.~Zampetti, G.~Bavota, M.~D. Penta, and M.~Lanza, ``Pattern-based
  mining of opinions in {Q\&A} websites,'' in \emph{Proc. 41st International
  Conference on Software Engineering}, 2019, pp. 548--559.

\bibitem{Chawla-SMOTE-JournalAIResearch2002}
N.~V. Chawla, K.~W. Bowyer, L.~O. Hall, and W.~P. Kegelmeyer, ``Smote:
  synthetic minority over-sampling technique,'' \emph{Journal of Artificial
  Intelligence Research}, vol.~16, no.~1, pp. 321--357, 2002.

\bibitem{Thelwall-Sentistrength-ASICT2010}
M.~Thelwall, K.~Buckley, G.~Paltoglou, D.~Cai, and A.~Kappas, ``Sentiment in
  short strength detection informal text,'' \emph{Journal of the American
  Society for Information Science and Technology}, vol.~61, no.~12, pp.
  2544--2558, 2010.

\bibitem{Hu-MiningSummarizingCustomerReviews-KDD2004}
M.~Hu and B.~Liu, ``Mining and summarizing customer reviews,'' in \emph{ACM
  SIGKDD International Conference on Knowledge Discovery and Data Mining},
  2004, pp. 168--177.

\bibitem{BlairGoldensohn-SentimentSummarizerLocalReviews-NLPIX2008}
S.~Blair-Goldensohn, K.~Hannan, R.~McDonald, T.~Neylon, G.~A. Reis, and
  J.~Reyner, ``Building a sentiment summarizer for local search reviews,'' in
  \emph{WWW Workshop on NLP in the Information Explosion Era}, 2008, p.~10.

\bibitem{Villarroel-ReleasePlanningMobileAppReviews-ICSE2016}
L.~Villarroel, G.~Bavota, B.~Russo, R.~Oliveto, and M.~D. Penta, ``Release
  planning of mobile apps based on user reviews,'' in \emph{Proceedings of the
  38th International Conference on Software Engineering}, 2016, pp. 14--24.

\bibitem{Misrili-IndustrialCaseStudyEnsembleSoftwareDefects-SQJ2011}
A.~T. Misirli, A.~B. Bener, and B.~Turhan, ``An industrial case study of
  classifier ensembles for locating software defects,'' \emph{Software Quality
  Journal}, vol.~19, no.~3, pp. 515--536, 2011.

\bibitem{Petric-EnsembleDefectDiversity-ESEM2016}
J.~Petric, D.~Bowes, T.~Hall, B.~Christianson, and N.~Baddoo, ``Building an
  ensemble for software defect prediction based on diversity selection,'' in
  \emph{Proceedings of the 10th ACM/IEEE International Symposium on Empirical
  Software Engineering and Measurement}, 2016, p. Article No. 46.

\bibitem{Zhang-CombinedClassifierCrossDefect-FrontierComputerScience2018}
Y.~Zhang, D.~Lo, X.~Xia, and J.~Sun, ``Combined classifier for cross-project
  defect prediction: an extended empirical study,'' \emph{Frontiers of Computer
  Science}, vol.~22, no.~2, pp. 280 -- 296, 2018.

\bibitem{Yang-TwoLayerEnsembleJustInTimeDefect-IST2017}
X.~Yang, D.~Lo, X.~Xia, and J.~Sun, ``{TLEL}: A two-layer ensemble learning
  approach for just-in-time defect prediction,'' \emph{Information and Software
  Technology}, vol.~87, no.~2, pp. 206 -- 220, 2017.

\bibitem{Bowes-SoftwareDefectPredictionDoDifferentClassifiersFindSameDefects-SQJ2018}
D.~Bowes, T.~Hall, and J.~Petric, ``Software defect prediction: do different
  classifiers find the same defects?'' \emph{Software Quality Journal},
  vol.~26, no.~2, pp. 525--552, 2018.

\bibitem{Menzies-LocalVsGlobalDefectPrediction-TSE2013}
T.~Menzies, A.~Butcher, D.~Cok, A.~Marcus, L.~Layman, F.~S.~B. Turhan, and
  T.~Zimmermann, ``Local versus global lessons for defect prediction and effort
  estimation,'' \emph{IEEE Transactions on Software Engineering}, vol.~39,
  no.~6, pp. 822 -- 834, 2013.

\bibitem{miles_1994}
M.~Miles and A.~Huberman, \emph{Qualitative Data Analysis: {An} Expanded
  Sourcebook}.\hskip 1em plus 0.5em minus 0.4em\relax SAGE Publications, 1994.

\bibitem{Patanamon-ReviewParticipationInCodeReview-JournalEMSE2017}
P.~Thongtanunam, S.~Mcintosh, A.~E. Hassan, and H.~Iida, ``Review participation
  in modern code review: An empirical study of the android, qt, and openstack
  projects,'' \emph{Journal of Empirical Software Engineering}, vol.~22, no.~2,
  pp. 768--817, 2017.

\bibitem{Valiev-EcosystemLevelSurvivalPyPI-FSE2018}
M.~Valiev, B.~Vasilescu, and J.~Herbsleb, ``Ecosystem-level determinants of
  sustained activity in open-source projects: a case study of the pypi
  ecosystem,'' in \emph{Proceedings of the 2018 26th ACM Joint Meeting on
  European Software Engineering Conference and Symposium on the Foundations of
  Software Engineering}, 2018, pp. 644--655.

\bibitem{Cohen-AppliedMultipleRegression-Lawrence2002}
J.~Cohen, S.~G. West, L.~Aiken, and P.~Cohen, \emph{Applied Multiple
  Regression/Correlation Analysis for the Behavioral Sciences}, 3rd~ed.\hskip
  1em plus 0.5em minus 0.4em\relax Lawrence Erlbaum Associates, 2002.

\bibitem{Ortu-AreBulliesMoreProductive-MSR205}
M.~Ortu, B.~Adams, G.~Destefanis, P.~Tourani, M.~Marchesi, and R.~Tonelli,
  ``Are bullies more productive? empirical study of affectiveness vs. issue
  fixing time,'' in \emph{Proceedings of the 12th Working Conference on Mining
  Software Repositories}, 2015.

\bibitem{Warriner-VadLexicons-BRM2013}
A.~B. Warriner, V.~Kuperman, and M.~Brysbaert, ``Norms of valence, arousal, and
  dominance for 13,915 english lemmas,'' \emph{Behavior Research Methods},
  vol.~45, no.~4, pp. 1191--1207, 2013.

\bibitem{Guzman-SentimentAnalysisGithub-MSR2014}
E.~Guzman, D.~Az\'{o}car, and Y.~Li, ``Sentiment analysis of commit comments in
  github: an empirical study,'' in \emph{Proceedings of the 11th Working
  Conference on Mining Software Repositories}, 2014, pp. 352--355.

\bibitem{Guzman-EmotionalAwareness-FSE2013}
E.~Guzman and B.~Bruegge, ``Towards emotional awareness in software development
  teams,'' in \emph{Proceedings of the 7th Joint Meeting on Foundations of
  Software Engineering}, 2013, pp. 671--674.

\bibitem{Blei-LDA-JournalMachineLearning2003}
D.~M. Blei, A.~Y. Ng, and M.~I. Jordan, ``Latent dirichlet allocation,''
  \emph{Journal of Machine Learning Research}, vol.~3, no. 4-5, pp. 993--1022,
  2003.

\bibitem{Jongeling-SentimentAnalysisToolsSe-ICSME2015}
R.~Jongeling, S.~Datta, and A.~Serebrenik, ``Choosing your weapons: On
  sentiment analysis tools for software engineering research,'' in
  \emph{Proceedings of the 31st International Confernece on Software
  Maintenance and Evolution}, 2015.

\bibitem{website:alchemy}
IBM, \emph{Alchemy sentiment detection},
  \url{http://www.alchemyapi.com/products/alchemylanguage/sentiment-analysis},
  2016.

\bibitem{website:nltk}
NLTK, \emph{Sentiment Analysis},
  \url{http://www.nltk.org/howto/sentiment.html}, 2016.

\end{thebibliography}


\begin{thebibliography}{10}

\bibitem{website:senisead-online-appendix-ase2020}
{\em Hybrid Sentiment Detection for Software Artifacts: A Case Study (Online
  Appendix)}.
\newblock \url{https://github.com/anonsubmissions2/HybridSentimentSEStudy}, 8
  May 2020 (last accessed).

\bibitem{Ahmed-SentiCRNIER-ASE2017}
T.~Ahmed, A.~Bosu, A.~Iqbal, and S.~Rahimi.
\newblock Senticr: A customized sentiment analysis tool for code review
  interactions.
\newblock In {\em Proceedings of the 32nd International Conference on Automated
  Software Engineering}, pages 106--111, 2017.

\bibitem{Ikram-SentimentCodeReview-IST2019}
I.~E. Asri, N.~Kerzazi, G.~Uddin, F.~Khomh, and M.~J. Idrissi.
\newblock An empirical study of sentiments in code reviews.
\newblock {\em Information and Software Technology}, 114:37--54, 2019.

\bibitem{Calefato-Senti4SD-EMSE2017}
F.~Calefato, F.~Lanubile, F.~Maiorano, and N.~Novielli.
\newblock Sentiment polarity detection for software development.
\newblock {\em Journal Empirical Software Engineering}, pages 2543--2584, 2017.

\bibitem{Castaldi-SentimentAnalysisITSupport-MSR2016}
C.~Castaldi, A.~Blaz, and K.~Becker.
\newblock Sentiment analysis in tickets for {IT} support.
\newblock In {\em IEEE/ACM 13th Working Conference on Mining Software
  Repositories}, pages 235--246, 2016.

\bibitem{Chawla-SMOTE-JournalAIResearch2002}
N.~V. Chawla, K.~W. Bowyer, L.~O. Hall, and W.~P. Kegelmeyer.
\newblock Smote: synthetic minority over-sampling technique.
\newblock {\em Journal of Artificial Intelligence Research}, 16(1):321--357,
  2002.

\bibitem{Cohen-AppliedMultipleRegression-Lawrence2002}
J.~Cohen, S.~G. West, L.~Aiken, and P.~Cohen.
\newblock {\em Applied Multiple Regression/Correlation Analysis for the
  Behavioral Sciences}.
\newblock Lawrence Erlbaum Associates, 3rd edition, 2002.

\bibitem{Balage-HybridSentimentDetectorTwitter-SemEval2014}
P.~P.~B. Filho, L.~Avanco, T.~A.~S. Pardo, and M.~G.~V. Nunes.
\newblock Nilc\_usp: An improved hybrid system for sentiment analysis in
  twitter messages.
\newblock In {\em Proceedings of the 8th International Workshop on Semantic
  Evaluation}, pages 428--432, 2014.

\bibitem{Gachechiladze-AngerDirectionCollaborativeSE-ICSENIER2017}
D.~Gachechiladze, F.~Lanubile, N.~Novielli, and A.~Serebrenik.
\newblock Anger and its direction in collaborative software development.
\newblock In {\em 39th International Conference on Software Engineering: New
  Ideas and Emerging Results Track}, pages 11--14, 2017.

\bibitem{Ghotra-RevisitingImpactClassificationDefectPrediction-ICSE2015}
B.~Ghotra, S.~McIntosh, and A.~E. Hassan.
\newblock Revisiting the impact of classification techniques on the performance
  of defect prediction models.
\newblock In {\em Proceedings of the 37th International Conference on Software
  Engineering}, pages 789--800, 2015.

\bibitem{Goncalves-ComparingCombiningSentimentAnalysisTools-COSN2013}
P.~Goncalves, M.~Araujo, F.~Benevenuto, and M.~Cha.
\newblock Comparing and combining sentiment analysis methods.
\newblock In {\em Proceedings of the first ACM conference on Online social
  networks}, pages 27--38, 2013.

\bibitem{Guzman-NeedleInHaystackTwitterSoftware-RE2016}
E.~Guzman, R.~Alkadhi, , and N.~Seyf.
\newblock A needle in a haystack: What do twitter users say about software?
\newblock In {\em 24th IEEE International Requirements Engineering Conference},
  pages 96--105, 2016.

\bibitem{Guzman-SentimentAnalysisGithub-MSR2014}
E.~Guzman, D.~Az\'{o}car, and Y.~Li.
\newblock Sentiment analysis of commit comments in github: an empirical study.
\newblock In {\em Proceedings of the 11th Working Conference on Mining Software
  Repositories}, pages 352--355, 2014.

\bibitem{Guzman-EmotionalAwarenessSoftwareDevelopmentTeams-FSE2013}
E.~Guzman and B.~Bruegge.
\newblock Towards emotional awareness in software development teams.
\newblock In {\em Proceedings of the 2013 9th Joint Meeting on Foundations of
  Software Engineering}, pages 671--674, 2013.

\bibitem{Guzman-EmotionalAwareness-FSE2013}
E.~Guzman and B.~Bruegge.
\newblock Towards emotional awareness in software development teams.
\newblock In {\em Proceedings of the 7th Joint Meeting on Foundations of
  Software Engineering}, pages 671--674, 2013.

\bibitem{Hassan-PredictingFaultsUsingCodeComplexity-ICSE2009}
A.~E. Hassan.
\newblock Predicting faults using the complexity of code changes.
\newblock In {\em Proc. 31st International Conference on Software Engineering},
  pages 78--89, 2009.

\bibitem{Hu-MiningSummarizingCustomerReviews-KDD2004}
M.~Hu and B.~Liu.
\newblock Mining and summarizing customer reviews.
\newblock In {\em ACM SIGKDD International Conference on Knowledge Discovery
  and Data Mining}, pages 168--177, 2004.

\bibitem{Islam-SentistrengthSE-MSR2017}
M.~R. Islam and M.~F. Zibran.
\newblock Leveraging automated sentiment analysis in software engineering.
\newblock In {\em Proc. 14th International Conference on Mining Software
  Repositories}, pages 203--214, 2017.

\bibitem{Islam-DevaEmotionSE-ACMSAC2018}
M.~R. Islam and M.~F. Zibran.
\newblock Deva: Sensing emotions in the valence arousal space in software
  engineering text.
\newblock In {\em 33rd Annual ACM Symposium on Applied Computing}, page
  1536–1543, 2018.

\bibitem{Jongeling-SentimentAnalysisToolsSe-ICSME2015}
R.~Jongeling, S.~Datta, and A.~Serebrenik.
\newblock Choosing your weapons: On sentiment analysis tools for software
  engineering research.
\newblock In {\em Proceedings of the 31st International Confernece on Software
  Maintenance and Evolution}, 2015.

\bibitem{Jongeling-SentimentNegative-EMSE2017}
R.~Jongeling, P.~Sarkar, S.~Datta, and A.~Serebrenik.
\newblock On negative results when using sentiment analysis tools for software
  engineering research.
\newblock {\em Journal Empirical Software Engineering}, 22(5):2543--2584, 2017.

\bibitem{Khomh-EntropyEvaluationTriageMozillaCrash-WCRE2011}
F.~Khomh, B.~Chan, Y.~Zou, and A.~E. Hassan.
\newblock An entropy evaluation approach for triaging field crashes: A case
  study of mozilla firefox.
\newblock In {\em Proceedings of the 2011 18th Working Conference on Reverse
  Engineering}, pages 261--270, 2011.

\bibitem{Ko-LinguisticAnalysis-VLHCC2005a}
A.~J. Ko, B.~A. Myers, and D.~H. Chau.
\newblock A linguistic analysis of how people describe software problems.
\newblock In {\em 2005 IEEE Symposium on Visual Languages and Human-Centric
  Computing}, pages 127--134, 2005.

\bibitem{Lin-PatternBasedOpinionMining-ICSE2019}
B.~Lin, F.~Zampetti, G.~Bavota, M.~D. Penta, and M.~Lanza.
\newblock Pattern-based mining of opinions in {Q\&A} websites.
\newblock In {\em Proc. 41st International Conference on Software Engineering},
  pages 548--559, 2019.

\bibitem{Lin-SentimentDetectionNegativeResults-ICSE2018}
B.~Lin, F.~Zampetti, G.~Bavota, M.~D. Penta, M.~Lanza, and R.~Oliveto.
\newblock Sentiment analysis for software engineering: how far can we go?
\newblock In {\em Proc. 40th International Conference on Software Engineering},
  pages 94--104, 2018.

\bibitem{Liu-SentimentAnalysisOpinionMining-MorganClaypool2012}
B.~Liu.
\newblock {\em Sentiment Analysis and Opinion Mining}.
\newblock Morgan \& Claypool Publishers, 1st edition, May 2012.

\bibitem{Maalej-AutomaticClassificationAppReviews-RE2016}
W.~Maalej, Z.~Kurtanovi\'{c}, H.~Nabil, and C.~Stanik.
\newblock On the automatic classification of app reviews.
\newblock In {\em International Requirements Engineering Conference}, page
  311–331, 2016.

\bibitem{Manning-IRIntroBook-Cambridge2009}
C.~D. Manning, P.~Raghavan, and H.~Sch\"{u}tze.
\newblock {\em An Introduction to Information Retrieval}.
\newblock Cambridge Uni Press, 2009.

\bibitem{Mika-MiningValenceBurnout-MSR2016}
M.~M\.{a}ntyl\.{a}, B.~Adams, G.~Destefanis, D.~Graziotin, and M.~Ortu.
\newblock Mining valence, arousal, and dominance -- possibilities for detecting
  burnout and productivity?
\newblock In {\em Proceedings of the 13th Working Conference on Mining Software
  Repositories}, pages 247--258, 2016.

\bibitem{Misrili-IndustrialCaseStudyEnsembleSoftwareDefects-SQJ2011}
A.~T. Misirli, A.~B. Bener, and B.~Turhan.
\newblock An industrial case study of classifier ensembles for locating
  software defects.
\newblock {\em Software Quality Journal}, 19(3):515--536, 2011.

\bibitem{Murgia-DoDevelopersFeelEmotion-MSR2014}
A.~Murgia, P.~Tourani, B.~Adams, and M.~Ortu.
\newblock Do developers feel emotions? an exploratory analysis of emotions in
  software artifacts.
\newblock In {\em Proceedings of the 11th Working Conference on Mining Software
  Repositories}, 2014.

\bibitem{Novielli-IntroToSpIssueAffectSE-JSS2019}
N.~Novielli, A.~Begel, and W.~Maalej.
\newblock Introduction to the special issue on affect awareness in software
  engineering.
\newblock {\em Journal of Systems and Software}, 148:180--182, 2019.

\bibitem{Novielli-SEToolCrossPlatform-MSR2020}
N.~Novielli, F.~Calefato, D.~Dongiovanni, D.~Girardi, and F.~Lanubile.
\newblock Can we use se-specific sentiment analysis tools in a cross-platform
  setting?
\newblock In {\em 17th International Conference on Mining Software
  Repositories}, page~11, 2020.

\bibitem{Novielli-ChallengesSentimentDetectionProgrammerEcosystem-SSE2015}
N.~Novielli, F.~Calefato, and F.~Lanubile.
\newblock The challenges of sentiment detection in the social programmer
  ecosystem.
\newblock In {\em Proceedings of the 7th International Workshop on Social
  Software Engineering}, pages 33--40, 2015.

\bibitem{Novielli-BenchmarkStudySentiSE-MSR2018}
N.~Novielli, D.~Girardi, and F.~Lanubile.
\newblock A benchmark study on sentiment analysis for software engineering
  research.
\newblock In {\em Proceedings of the 15th International Conference on Mining
  Software Repositories}, page~12, 2018.

\bibitem{Novielli-SentimentEmotionSE-IEEESoftware2019}
N.~Novielli and A.~Serebrenik.
\newblock Sentiment and emotion in software engineering.
\newblock {\em IEEE Software}, 36(5):6--23, 2019.

\bibitem{Nucci-DynamicSelectionOfClassifiersBugPrediction-IEEETranEmergingTopics2017}
D.~D. Nucci, F.~Palomba, R.~Oliveto, and A.~D. Lucia.
\newblock Dynamic selection of classifiers in bug prediction: An adaptive
  method.
\newblock {\em IEEE Transactions on Emerging Topics in Computational
  Intelligence}, 1(3):202--212, 2017.

\bibitem{Ortu-BulliesMoreProductive-MSR2015}
M.~Ortu, B.~Adams, G.~Destefanis, P.~Tourani, M.~Marchesi, and R.~Tonelli.
\newblock Are bullies more productive? empirical study of affectiveness vs.
  issue fixing time.
\newblock In {\em Proceedings of the 12th Working Conference on Mining Software
  Repositories}, pages 303--313, 2015.

\bibitem{Ortu-AreBulliesMoreProductive-MSR205}
M.~Ortu, B.~Adams, G.~Destefanis, P.~Tourani, M.~Marchesi, and R.~Tonelli.
\newblock Are bullies more productive? empirical study of affectiveness vs.
  issue fixing time.
\newblock In {\em Proceedings of the 12th Working Conference on Mining Software
  Repositories}, 2015.

\bibitem{Ortu-EmotionalSideJira-MSR2016}
M.~Ortu, A.~Murgia, G.~Destefanis, P.~Tourani, R.~Tonelli, M.~L. Marchesi, and
  B.~Adams.
\newblock The emotional side of software developers in jira.
\newblock In {\em 13th International Conference on Mining Software
  Repositories}, pages 480--483, 2016.

\bibitem{Pang-SentimentClassificationMachineLearning-EMNLP2002}
B.~Pang, L.~Lee, and S.~Vaithyanathan.
\newblock Thumbs up? sentiment classification using machine learning
  techniques.
\newblock In {\em Conference on Empirical Methods in Natural Language
  Processing}, pages 79--86, 2002.

\bibitem{Panichella-ClassifyAppUserReview-ICSME2015}
S.~Panichella, A.~D. Sorbo, E.~Guzman, C.~A. Visaggio, G.~Canfora, and H.~C.
  Gall.
\newblock How can i improve my app? classifying user reviews for software
  maintenance and evolution.
\newblock In {\em IEEE International Conf. on Software Maintenance and
  Evolution}, pages 281--290, 2015.

\bibitem{Petric-EnsembleDefectDiversity-ESEM2016}
J.~Petric, D.~Bowes, T.~Hall, B.~Christianson, and N.~Baddoo.
\newblock Building an ensemble for software defect prediction based on
  diversity selection.
\newblock In {\em Proceedings of the 10th ACM/IEEE International Symposium on
  Empirical Software Engineering and Measurement}, page Article No. 46, 2016.

\bibitem{Pletea-SecurityEmotionSE-MSR2014}
D.~Pletea, B.~Vasilescu, and A.~Serebrenik.
\newblock Security and emotion: sentiment analysis of security discussions on
  github.
\newblock In {\em Proceedings of the 11th Working Conference on Mining Software
  Repositories}, pages 348--351, 2014.

\bibitem{Sebastiani-MachineLearningTextCategorization-ACMSurveys2002}
F.~Sebastiani.
\newblock Machine learning in automated text categorization.
\newblock {\em Journal of ACM Computing Surveys}, 34(1):1--47, 2002.

\bibitem{Shannon-MathematicTheoryOfCommunication-BellJournal1948}
C.~E. Shannon.
\newblock A mathematical theory of communication.
\newblock {\em The Bell System Technical Journal}, 27(3):379--423, 1948.

\bibitem{Sinha-DeveloperSentimentCommitLog-MSR2016}
V.~Sinha, A.~Lazar, and B.~Sharif.
\newblock Analyzing developer sentiment in commit logs.
\newblock In {\em 13th International Conference on Mining Software
  Repositories}, pages 281--290, 2016.

\bibitem{Socher-Sentiment-EMNLP2013}
R.~Socher, A.~Perelygin, J.~Wu, C.~Manning, A.~Ng, and J.~Chuang.
\newblock Recursive models for semantic compositionality over a sentiment
  treebank.
\newblock In {\em Proc. Conference on Empirical Methods in Natural Language
  Processing (EMNLP)}, page~12, 2013.

\bibitem{Thelwall-Sentistrength-ASICT2010}
M.~Thelwall, K.~Buckley, G.~Paltoglou, D.~Cai, and A.~Kappas.
\newblock Sentiment in short strength detection informal text.
\newblock {\em Journal of the American Society for Information Science and
  Technology}, 61(12):2544--2558, 2010.

\bibitem{Patanamon-ReviewParticipationInCodeReview-JournalEMSE2017}
P.~Thongtanunam, S.~Mcintosh, A.~E. Hassan, and H.~Iida.
\newblock Review participation in modern code review: An empirical study of the
  android, qt, and openstack projects.
\newblock {\em Journal of Empirical Software Engineering}, 22(2):768--817,
  2017.

\bibitem{Uddin-OpinerReviewAlgo-ASE2017}
G.~Uddin and F.~Khomh.
\newblock Automatic summarization of {API} reviews.
\newblock In {\em Proc. 32nd IEEE/ACM International Conference on Automated
  Software Engineering}, pages 159--170, 2017.

\bibitem{Uddin-OpinerEval-ASE2017}
G.~Uddin and F.~Khomh.
\newblock Automatic summarization of api reviews.
\newblock In {\em Submitted to 32nd IEEE/ACM International Conference on
  Automated Software Engineering}, page~12, 2017.

\bibitem{Uddin-OpinerReviewToolDemo-ASE2017}
G.~Uddin and F.~Khomh.
\newblock Opiner: A search and summarization engine for {API} reviews.
\newblock In {\em Proc. 32nd IEEE/ACM International Conference on Automated
  Software Engineering}, pages 978--983, 2017.

\bibitem{Uddin-OpinionValue-TSE2019}
G.~Uddin and F.~Khomh.
\newblock Automatic opinion mining from {API} reviews.
\newblock {\em IEEE Transactions on Software Engineering}, page~37, 2019.

\bibitem{Valiev-EcosystemLevelSurvivalPyPI-FSE2018}
M.~Valiev, B.~Vasilescu, and J.~Herbsleb.
\newblock Ecosystem-level determinants of sustained activity in open-source
  projects: a case study of the pypi ecosystem.
\newblock In {\em Proceedings of the 2018 26th ACM Joint Meeting on European
  Software Engineering Conference and Symposium on the Foundations of Software
  Engineering}, pages 644--655, 2018.

\bibitem{Villarroel-ReleasePlanningMobileAppReviews-ICSE2016}
L.~Villarroel, G.~Bavota, B.~Russo, R.~Oliveto, and M.~D. Penta.
\newblock Release planning of mobile apps based on user reviews.
\newblock In {\em Proceedings of the 38th International Conference on Software
  Engineering}, pages 14--24, 2016.

\bibitem{Warriner-VadLexicons-BRM2013}
A.~B. Warriner, V.~Kuperman, and M.~Brysbaert.
\newblock Norms of valence, arousal, and dominance for 13,915 english lemmas.
\newblock {\em Behavior Research Methods}, 45(4):1191--1207, 2013.

\bibitem{Yang-TwoLayerEnsembleJustInTimeDefect-IST2017}
X.~Yang, D.~Lo, X.~Xia, and J.~Sun.
\newblock {TLEL}: A two-layer ensemble learning approach for just-in-time
  defect prediction.
\newblock {\em Information and Software Technology}, 87(2):206 -- 220, 2017.

\bibitem{Zhang-CombinedClassifierCrossDefect-FrontierComputerScience2018}
Y.~Zhang, D.~Lo, X.~Xia, and J.~Sun.
\newblock Combined classifier for cross-project defect prediction: an extended
  empirical study.
\newblock {\em Frontiers of Computer Science}, 22(2):280 -- 296, 2018.

\end{thebibliography}

\begin{appendices}
\newpage
\section{Details of Performance Analysis of Tools}\label{sec:detailsPerformanceAnalysis}

\begin{table}[h]
  \centering
  \caption{Performance comparison of majority voting-based hybrid classifiers with the baselines. Extension of \tbl\ref{tab:majorityVotingShort}}
    \begin{tabular}{lrrr|rrr|rrr}\toprule
          & \multicolumn{3}{c}{\textbf{Positive}} & \multicolumn{3}{c}{\textbf{Negative}} & \multicolumn{3}{c}{\textbf{Neutral}} \\
          \cmidrule{2-10}
          & \textbf{F1} & \textbf{P} & \textbf{R} & \textbf{F1} & \textbf{P} & \textbf{R} & \textbf{F1} & \textbf{P} & \textbf{R} \\
    \midrule
    \textbf{Majority-All} & 0.721 & 0.825 & 0.640 & 0.645 & 0.859 & 0.516 & 0.819 & 0.734 & 0.925 \\
    \textbf{Majority-Supervised} & 0.706 & 0.818 & 0.621 & 0.645 & 0.879 & 0.510 & 0.825 & 0.735 & 0.939 \\
    \textbf{Majority-Unsupervised} & 0.536 & 0.736 & 0.422 & 0.490 & 0.762 & 0.361 & 0.755 & 0.647 & 0.907 \\
     \midrule
    Baseline - \textbf{Senti4SD} & 0.731 & 0.773 & 0.692 & 0.690 & 0.748 & 0.640 & 0.822 & 0.783 & 0.865 \\
    Baseline - \textbf{SentiCR} & 0.728 & 0.803 & 0.667 & 0.687 & 0.804 & 0.599 & 0.830 & 0.767 & 0.903 \\
    Baseline - \textbf{SentistrengthSE} & 0.720 & 0.752 & 0.690 & 0.640 & 0.736 & 0.566 & 0.795 & 0.750 & 0.847 \\
    Baseline - \textbf{Opiner} & 0.494 & 0.467 & 0.524 & 0.512 & 0.577 & 0.460 & 0.676 & 0.668 & 0.684 \\
    Baseline - \textbf{POME} & 0.127 & 0.389 & 0.076 & 0.065 & 0.415 & 0.035 & 0.697 & 0.550 & 0.950 \\
    \bottomrule
    \end{tabular}%
  \label{tab:majorityVoting}%
\end{table}

\begin{table}[h]
  \centering
 \caption{Performance comparison of the developed hybrid classifier (Sentisead) with the baselines. Extension of \tbl\ref{tab:ensembleBoWShort}}
    \begin{tabular}{l|rrr|rrr|rrr}\toprule
          & \multicolumn{3}{c}{\textbf{Positive}} & \multicolumn{3}{c}{\textbf{Negative}} & \multicolumn{3}{c}{\textbf{Neutral}} \\
          \cmidrule{2-10}
          &{\textbf{F1}} &{\textbf{P}} &{\textbf{R}} &{\textbf{F1}} &{\textbf{P}} &{\textbf{R}} &{\textbf{F1}} &{\textbf{P}} &{\textbf{R}} \\
     \midrule
    \textbf{Sentisead$_{RF-B}$} &{0.764} &{0.832} &{0.706} &{0.738} &{0.827} &{0.666} &{0.843} &{0.790} &{0.905} \\
    \bf{Sentisead$_{RF-N}$} & {0.759} & {0.797} & {0.724} & {0.72} & {0.777} & {0.671} & {0.837} & {0.801} & {0.877} \\

     \midrule
    \multicolumn{10}{c}{\textbf{Performance Increase of  \textbf{Sentisead$_B$ (i.e., Hybrid with Bag of Words)}}} \\
    \multicolumn{10}{c}{\textbf{Over Baseline (SentiSE = SentistrengthSE)}} \\
     \midrule
    \textbf{Majority} & 6.0\% & 0.8\% & 10.3\% & 14.4\% & -3.7\% & 29.1\% & 2.9\% & 7.6\% & -2.2\% \\
    \textbf{Senti4SD} & 4.5\% & 7.6\% & 2.0\% & 7.0\% & 10.6\% & 4.1\% & 2.6\% & 0.9\% & 4.6\% \\
    \textbf{SentiCR} & 4.9\% & 3.6\% & 5.8\% & 7.4\% & 2.9\% & 11.2\% & 1.6\% & 3.0\% & 0.2\% \\
    \textbf{SentiSE} & 6.1\% & 10.6\% & 2.3\% & 15.3\% & 12.4\% & 17.7\% & 6.0\% & 5.3\% & 6.8\% \\
    \textbf{Opiner} & 54.7\% & 78.2\% & 34.7\% & 44.1\% & 43.3\% & 44.8\% & 24.7\% & 18.3\% & 32.3\% \\
    \textbf{POME} & 501.6\% & 113.9\% & 828.9\% & 1035.4\% & 99.3\% & 1802.9\% & 20.9\% & 43.6\% & -4.7\% \\
    \bottomrule
    \end{tabular}%
  \label{tab:ensembleBoW}%
\end{table}

\begin{table}[h]
  \centering
  
  \caption{Performance comparison of the designed Hybrid detector (Sentisead) with added features from \tbl\ref{tab:sentiseadPlusStudiedFeatures}. Extension of \tbl\ref{tab:performanceSentiseadPlusShort}}
    \begin{tabular}{l|rrr|rrr|rrr}\toprule
          & \multicolumn{3}{c}{\textbf{Positive}} & \multicolumn{3}{c}{\textbf{Negative}} & \multicolumn{3}{c}{\textbf{Neutral}} \\
           \cmidrule{2-10}
          & {\textbf{F1}} & {\textbf{P}} & {\textbf{R}} & {\textbf{F1}} & {\textbf{P}} & {\textbf{R}} & {\textbf{F1}} & {\textbf{P}} & {\textbf{R}} \\
          \midrule
      \textbf{Sentisead$_{RF-B}$+} & 0.767 & 0.832 & 0.712 & 0.74  & 0.818 & 0.675 & 0.844 & 0.794 & 0.901 \\
  
    \textbf{Sentisead$_{RF-BNE}$+} & 0.761 & 0.831 & 0.702 & 0.738 & 0.832 & 0.663 & 0.843 & 0.788 & 0.906 \\
    \textbf{Sentisead$_{RF-BNP}$+} & 0.765 & 0.829 & 0.711 & 0.737 & 0.813 & 0.674 & 0.843 & 0.793 & 0.899 \\
    \midrule
    \textbf{SentiSead$_{RF-N}$+} & 0.747 & 0.777 & 0.719 & 0.698 & 0.741 & 0.66  & 0.822 & 0.793 & 0.854 \\
    \textbf{Sentisead$_{RF-NNE}$+} & 0.755 & 0.791 & 0.722 & 0.708 & 0.772 & 0.653 & 0.834 & 0.796 & 0.876 \\
    \textbf{Sentisead$_{RF-NNP}$+} & 0.746 & 0.773 & 0.721 & 0.698 & 0.741 & 0.66  & 0.821 & 0.793 & 0.852 \\
    \midrule
    \multicolumn{10}{l}{Sentisead$_{RF-B}*$: With bag of words. Sentisead$_{RF-N}*$: No bag of words. }\\
    \multicolumn{10}{l}{NP = No partial polarity. NE = No polarity entropy}
    \\
    \bottomrule
    \end{tabular}%
  \label{tab:performanceSentiseadPlus}%
  
\end{table}

\begin{table}[h]
  \centering  \caption{Performance of stand-alone advanced pre-trained advanced language-based deep learning models across the datasets. Extension of \tbl\ref{tab:performance-bert-stand-aloneShort}}
    \begin{tabular}{l|rrr|rrr|rrr}
    \toprule
          & \multicolumn{3}{c}{\textbf{Positive}} & \multicolumn{3}{c}{\textbf{Negative}} & \multicolumn{3}{c}{\textbf{Neutral}} \\
          \cmidrule{2-10}
          & {\textbf{F1}} & {\textbf{P}} & {\textbf{R}} & {\textbf{F1}} & {\textbf{P}} & {\textbf{R}} & {\textbf{F1}} & {\textbf{P}} & {\textbf{R}} \\
          \midrule
    BERT  & {0.772} & {0.768} & {0.775} & {0.759} & {0.767} & {0.752} & {0.847} & {0.845} & {0.848} \\
    Change Over Sentisead$_{RF}$ & 1.0\% & -7.7\% & 9.8\% & 2.8\% & -7.3\% & 12.9\% & 0.5\% & 7.0\% & -6.3\% \\
    \midrule
    RoBERTa & {0.788} & {0.773} & {0.804} & {0.768} & {0.768} & {0.768} & {0.847} & {0.855} & {0.839} \\
    Change Over Sentisead$_{RF}$ & 3.1\% & -7.1\% & 13.9\% & 4.1\% & -7.1\% & 15.3\% & 0.5\% & 8.2\% & -7.3\% \\
    \midrule
    ALBERT & {0.767} & {0.792} & {0.743} & {0.747} & {0.764} & {0.731} & {0.846} & {0.828} & {0.865} \\
    Change Over Sentisead$_{RF}$ & 0.4\% & -4.8\% & 5.2\% & 1.2\% & -7.6\% & 9.8\% & 0.4\% & 4.8\% & -4.4\% \\
    \midrule
    XLNet & {0.779} & {0.749} & {0.812} & {0.767} & {0.786} & {0.749} & {0.848} & {0.857} & {0.84} \\
    Change Over Sentisead$_{RF}$ & 2.0\% & -10.0\% & 15.0\% & 3.9\% & -5.0\% & 12.5\% & 0.6\% & 8.5\% & -7.2\% \\
    \bottomrule
    \end{tabular}%
  \label{tab:performance-bert-stand-alone}%
\end{table}%

\begin{table}[h]
\centering
  \caption{Performance of Sentisead using pre-trained langange based advanced deep learning models as ensembler. Extension of \tbl\ref{tab:performance-sentisead-shallow-ensemble-bertShort}}
    \begin{tabular}{l|rrr|rrr|rrr}\toprule
          & \multicolumn{3}{c}{\textbf{Positive}} & \multicolumn{3}{c}{\textbf{Negative}} & \multicolumn{3}{c}{\textbf{Neutral}} \\
          \cmidrule{2-10}
          & {\textbf{F1}} & {\textbf{P}} & {\textbf{R}} & {\textbf{F1}} & {\textbf{P}} & {\textbf{R}} & {\textbf{F1}} & {\textbf{P}} & {\textbf{R}} \\
          \midrule
    Sentisead$_{BERT}$ & {0.77} & {0.771} & {0.768} & {0.76} & {0.761} & {0.759} & {0.843} & {0.842} & {0.844} \\
    Change Over Sentisead$_{RF}$ & 0.8\% & -7.3\% & 8.8\% & 3.0\% & -8.0\% & 14.0\% & 0.0\% & 6.6\% & -6.7\% \\
    Change Over BERT & -0.3\% & 0.4\% & -0.9\% & 0.1\% & -0.8\% & 0.9\% & -0.5\% & -0.4\% & -0.5\% \\
          \midrule
    Sentisead$_{RoBERTa}$ & {0.787} & {0.78} & {0.794} & {0.774} & {0.792} & {0.757} & {0.853} & {0.849} & {0.857} \\
    Change Over Sentisead$_{RF}$ & 3.0\% & -6.2\% & 12.5\% & 4.9\% & -4.2\% & 13.7\% & 1.2\% & 7.5\% & -5.3\% \\
    Change Over RoBERTa & -0.1\% & 0.9\% & -1.2\% & 0.8\% & 3.1\% & -1.4\% & 0.7\% & -0.7\% & 2.1\% \\
          \midrule
    Sentisead$_{ALBERT}$ & {0.764} & {0.769} & {0.76} & {0.735} & {0.731} & {0.738} & {0.839} & {0.838} & {0.84} \\
    Change Over Sentisead$_{RF}$ & 0.0\% & -7.6\% & 7.6\% & -0.4\% & -11.6\% & 10.8\% & -0.5\% & 6.1\% & -7.2\% \\
    Change Over ALBERT & -0.4\% & -2.9\% & 2.3\% & -1.6\% & -4.3\% & 1.0\% & -0.8\% & 1.2\% & -2.9\% \\
          \midrule
    Sentisead$_{XLNet}$ & {0.785} & {0.774} & {0.796} & {0.767} & {0.772} & {0.763} & {0.848} & {0.852} & {0.844} \\
    Change Over Sentisead$_{RF}$ & 2.7\% & -7.0\% & 12.7\% & 3.9\% & -6.7\% & 14.6\% & 0.6\% & 7.8\% & -6.7\% \\
    Change Over XLNet & 0.8\% & 3.3\% & -2.0\% & 0.0\% & -1.8\% & 1.9\% & 0.0\% & -0.6\% & 0.5\% \\
	\bottomrule    
	\end{tabular}%
  \label{tab:performance-sentisead-shallow-ensemble-bert}%
\end{table}%

\begin{table}[h]
  \centering
  \caption{Performance of Sentisead using pre-trained advanced deep learning models as ensembler and stand-alone detectors. Extension of \tbl\ref{tab:performance-sentisead-shallow-ensemble-bert-plusShort}}
    \begin{tabular}{l|rrr|rrr|rrr}\toprule
          & \multicolumn{3}{c}{\textbf{Positive}} & \multicolumn{3}{c}{\textbf{Negative}} & \multicolumn{3}{c}{\textbf{Neutral}} \\
          \cmidrule{2-10}
          & {\textbf{F1}} & {\textbf{P}} & {\textbf{R}} & {\textbf{F1}} & {\textbf{P}} & {\textbf{R}} & {\textbf{F1}} & {\textbf{P}} & {\textbf{R}} \\
          \midrule
    Sentisead$_{BERT}$+ & {0.769} & {0.766} & {0.771} & {0.757} & {0.76} & {0.755} & {0.844} & {0.844} & {0.844} \\
    Change Over Sentisead$_{RF}$ & 0.7\% & -7.9\% & 9.2\% & 2.6\% & -8.1\% & 13.4\% & 0.1\% & 6.8\% & -6.7\% \\
    Change Over Sentisead$_{BERT}$ & -0.1\% & -0.6\% & 0.4\% & -0.4\% & -0.1\% & -0.5\% & 0.1\% & 0.2\% & 0.0\% \\
    Change Over BERT & -0.4\% & -0.3\% & -0.5\% & -0.3\% & -0.9\% & 0.4\% & -0.4\% & -0.1\% & -0.5\% \\
          \midrule
    Sentisead$_{RoBERTa}$+ & {0.783} & {0.758} & {0.809} & {0.768} & {0.776} & {0.761} & {0.847} & {0.857} & {0.837} \\
    Change Over Sentisead$_{RF}$ & 2.5\% & -8.9\% & 14.6\% & 4.1\% & -6.2\% & 14.3\% & 0.5\% & 8.5\% & -7.5\% \\
    Change Over Sentisead$_{RoBERTa}$ & -0.5\% & -2.8\% & 1.9\% & -0.8\% & -2.0\% & 0.5\% & -0.7\% & 0.9\% & -2.3\% \\
    Change Over RoBERTa & -0.6\% & -1.9\% & 0.6\% & 0.0\% & 1.0\% & -0.9\% & 0.0\% & 0.2\% & -0.2\% \\
          \midrule
    Sentisead$_{ALBERT}$+ & {0.766} & {0.763} & {0.768} & {0.744} & {0.752} & {0.736} & {0.84} & {0.838} & {0.843} \\
    Change Over Sentisead$_{RF}$ & 0.3\% & -8.3\% & 8.8\% & 0.8\% & -9.1\% & 10.5\% & -0.4\% & 6.1\% & -6.9\% \\
    Change Over Sentisead$_{ALBERT}$ & 0.3\% & -0.8\% & 1.1\% & 1.2\% & 2.9\% & -0.3\% & 0.1\% & 0.0\% & 0.4\% \\
    Change Over Albert & -0.1\% & -3.7\% & 3.4\% & -0.4\% & -1.6\% & 0.7\% & -0.7\% & 1.2\% & -2.5\% \\
          \midrule
    Sentisead$_{XLNet}$+ & {0.777} & {0.752} & {0.803} & {0.768} & {0.781} & {0.756} & {0.845} & {0.852} & {0.837} \\
    Change Over Sentisead$_{RF}$ & 1.7\% & -9.6\% & 13.7\% & 4.1\% & -5.6\% & 13.5\% & 0.2\% & 7.8\% & -7.5\% \\
    Change Over Sentisead$_{XLNet}$ & -1.0\% & -2.8\% & 0.9\% & 0.1\% & 1.2\% & -0.9\% & -0.4\% & 0.0\% & -0.8\% \\
    Change Over XLNet & -0.3\% & 0.4\% & -1.1\% & 0.1\% & -0.6\% & 0.9\% & -0.4\% & -0.6\% & -0.4\% \\
	\bottomrule
    \end{tabular}%
  \label{tab:performance-sentisead-shallow-ensemble-bert-plus}%
\end{table}%
\end{appendices}
\end{document}